\documentclass[%
  reprint,
  showpacs,preprintnumbers,
  amsmath,amssymb,
  aps,
  prb,
]{revtex4-1}

\usepackage{graphicx}% Include figure files
\usepackage{dcolumn}% Align table columns on decimal point
\usepackage{bm}% bold math
\usepackage{multirow}
\usepackage{longtable}
\usepackage{braket}
\usepackage{color}
\usepackage{ulem}

\usepackage{times}
\usepackage{booktabs}

\begin{document}

\preprint{APS/123-QED}

\title{
Bottom-up design of spin-split and reshaped electronic band structures\\ in spin-orbit-coupling free antiferromagnets: \\
Procedure on the basis of augmented multipoles 
}

\author{Satoru Hayami$^1$, Yuki Yanagi$^2$, and Hiroaki Kusunose$^3$}
\affiliation{
$^1$Department of Applied Physics, The University of Tokyo, Bunkyo, Tokyo 113-8656, Japan \\
$^2$Center for Computational Materials Science, Institute for Materials Research, Tohoku University, Sendai, Miyagi, 950-8577, Japan \\
$^3$Department of Physics, Meiji University, Kawasaki 214-8571, Japan 
}
 
\begin{abstract}
We propose an efficient microscopic design procedure of electronic band structures having intrinsic spin and momentum dependences in spin-orbit-coupling free antiferromagnets. 
Our bottom-up design approach to creating desired spin-split and reshaped electronic band structures could result in further findings of practical spin-orbit-coupling free materials exhibiting a giant spin-dependent and/or nonreciprocal transport, magneto-electric and elastic responses, and so on, as a consequence of such band structures. 
We establish a systematic guideline to construct symmetric/antisymmetric spin-split and antisymmetrically deformed spin-independent band structures in spin-orbit-coupling free systems by using two polar multipole degrees of freedom, i.e., electric and magnetic toroidal multipoles. 
The two polar multipoles constitute a complete set and describe arbitrary degrees of freedom in the hopping Hamiltonian, whose onsite and offsite degrees of freedom in a cluster are described as the so-called cluster and bond multipoles, respectively, and another degree of freedom connecting between clusters is expressed as momentum multipoles. 
By using these multipole descriptions, we elucidate simple microscopic conditions to realize intrinsic band deformations in magnetically ordered states: 
The symmetric spin splitting is realized in collinear magnets when cluster and bond multipoles contain the same symmetry of multipoles.  
The antisymmetric spin splitting occurs in noncollinear antiferromagnets when a bond-type magnetic toroidal multipole is present. 
Furthermore, the antisymmetric band deformation with spin degeneracy is realized in noncoplanar antiferromagnets. 
We exemplify three lattice systems formed by a triangle unit, triangular, kagome, and breathing kagome structures, in order to demonstrate the band deformations under the magnetic ordering. 
On the basis of the proposed procedure, we list up various candidate materials showing intrinsic band deformations in accordance with MAGNDATA, magnetic structures database. 
\end{abstract}
\maketitle

\section{Introduction}
\label{sec:Introduction}

The electronic band structures in solids play an important role in determining fundamental physical properties. 
In general, the electronic band dispersions $\varepsilon_{\sigma}(\bm{k})$, which are characterized by the wave vector $\bm{k}$ and the spin $\sigma$, are classified according to the presence and absence of space-time inversion symmetry, where the spatial inversion operation, $\mathcal{P}$, transforms $\varepsilon_{\sigma}(\bm{k})$ as $\mathcal{P}\varepsilon_{\sigma}(\bm{k})=\varepsilon_{\sigma}(-\bm{k})$ and the time-reversal operation, $\mathcal{T}$, transforms $\varepsilon_{\sigma}(\bm{k})$ as $\mathcal{T}\varepsilon_{\sigma}(\bm{k})=\varepsilon_{-\sigma}(-\bm{k})$. 
In the presence of $\mathcal{P}$ and $\mathcal{T}$, i.e., in the centrosymmetric paramagnetic state, the system has a twofold degeneracy, $\varepsilon_{\sigma}(\bm{k})= \varepsilon_{\sigma}(-\bm{k})=\varepsilon_{-\sigma}(\bm{k})$ in the entire Brillouin zone. 
The spin-split band structure is realized once either $\mathcal{P}$ or $\mathcal{T}$ is broken: 
The breaking of $\mathcal{T}$ ($\mathcal{P}$) results in the (anti)symmetric spin splitting in momentum space, provided that the spin and momentum degrees of freedom are coupled with each other. This is refereed as the spin-momentum locking~\cite{rashba1960properties,Sinova_PhysRevLett.92.126603,Dresselhaus_Dresselhaus_Jorio}.  

One of the microscopic key ingredients to connect the spin degree of freedom with kinetic motion of electrons is the spin-orbit coupling (SOC). 
For example, the relatively large SOC brings about the large antisymmetric spin splitting in the noncentrosymmetric materials, such as a polar semiconductor BiTeI~\cite{rashba1960properties,ishizaka2011giant,Bahramy_PhysRevB.84.041202} and monolayer transition-metal dichalcogenides, $MX_2$ ($M=$ Mo, W and $X=$ S, Se)~\cite{Zhu_PhysRevB.84.153402,wang2012electronics,ugeda2014giant,Andor_PhysRevX.4.011034}.
Although materials with the large SOC give rise to intriguing physical phenomena, such as the magnetoelectric effect~\cite{kimura2003magnetic,Fiebig0022-3727-38-8-R01,KhomskiiPhysics.2.20,furukawa2017observation,saito2018evidence}, spin Hall effect~\cite{hirsch1999spin,Sinova_PhysRevLett.92.126603,bernevig2006quantum,liu2012spin,sinova2015spin}, and nonreciprocal optics~\cite{kezsmarki2014one,Toyoda_PhysRevB.93.201109,Morimoto_PhysRevLett.117.146603,ideue2017bulk,tokura2018nonreciprocal,Aoki_PhysRevLett.122.057206}, it is usually nontrivial to control them microscopically, since 
the SOC is predominant in the complicated atomic orbitals and chemical composition. 
It prevents us engineering large spin splittings by tuning the built-in SOC of the materials constituted of moderately heavier elements.

On the contrary, recent studies indicate that even without relying on the SOC, a change of electronic state by magnetic orderings leads to a similar spin splitting depending on the crystal momentum~\cite{Ahn_PhysRevB.99.184432,naka2019spin,hayami2019momentum,Hayami2020b,Hayami_PhysRevB.101.220403,Yuan_PhysRevB.102.014422}. 
It has been discussed that the symmetric spin splitting with respect to momentum is realized in a nonsymmorphic organic compound, $\kappa$-(BETD-TTF)$_2$Cu[N(CN)$_2$]Cl~\cite{naka2019spin,Hayami2020b} and a distorted tetragonal compound, RuO$_2$~\cite{Berlijn_PhysRevLett.118.077201,Ahn_PhysRevB.99.184432} with collinear-type antiferromagnetic (AFM) structures.
The subsequent similar works also discuss the spin splitting based on the band calculation in one of the candidate materials, MnF$_{2}$~\cite{Yuan_PhysRevB.102.014422}. 
Moreover, antisymmetric spin splitting can be realized as well in a trigonal oxide Ba$_3$MnNb$_2$O$_9$ with a noncollinear AFM structure~\cite{Hayami_PhysRevB.101.220403}. 

The spin splittings driven by the magnetic phase transition induce interesting physical phenomena through the anisotropic spin-dependent kinetic motions of electrons, for instance, it is proposed the spin current generation by an electric field~\cite{Ahn_PhysRevB.99.184432,naka2019spin,hayami2019momentum} in collinear magnets and by a shear-type strain in noncollinear magnets~\cite{Hayami_PhysRevB.101.220403}. 
It is also shown that nonreciprocal transport arising from the antisymmetrically reshaped band structure is expected in noncoplanar magnets~\cite{Hayami_PhysRevB.101.220403,hayami2020phase}. 
Such a magnetic-order driven band deformation has an advantage in its flexible controllability, i.e.,  
it can be accessible by external fields, pressure and temperature. 
Furthermore, due to their kinetic origin, large spin splittings can be expected for the materials even with the negligibly small SOC.
This aspect is significant to extend the scope of materials and explore further efficient functional materials in the field of AFM spintronics~\cite{jungwirth2016antiferromagnetic,Baltz_RevModPhys.90.015005}.

In the present study, we further develop the above scenario, and we provide a complete microscopic guideline to engineer spin- and momentum-dependent band structures in SOC free AFMs. 
Our guideline is essentially based on local symmetry, which is embodied by the concept of augmented multipoles, especially with the electric and magnetic toroidal multipoles~\cite{suzuki2018first,hayami2018microscopic,Hayami_PhysRevB.98.165110,Watanabe_PhysRevB.98.245129,kusunose2020complete}.  
The analysis of couplings among these multipoles in the given Hamiltonian provides necessary ingredients for the band deformations, instead of performing band calculations.  
By introducing cluster-, bond-, and momentum-type electric and magnetic toroidal multipoles in a magnetic cluster, we can analyze which effective multipole coupling realizes the spin splitting and/or band deformation. 
Specifically, the symmetric spin splitting occurs in collinear AFMs when the Hamiltonian contains cluster and bond multipoles with the same symmetry. 
Similarly, the antisymmetric spin splitting is realized in noncollinear AFMs when a bond-type magnetic toroidal multipole is activated through the magnetic phase transition. 
Furthermore, the antisymmetric band deformation with keeping spin degeneracy is realized in noncoplanar AFMs. 
We exemplify three lattice systems consisting of a triangle unit, triangular, kagome, and breathing kagome structures, in order to demonstrate how the (spin-dependent) band deformations occur. 
Our analysis provides a simple prescription of bottom-up design for arbitrary electronic band structures from a microscopic viewpoint.
This simple procedure promotes further findings of materials exhibiting a giant spin splitting and related physical responses in SOC free AFMs.
As a fruitful outcome, we list up candidate materials showing intrinsic spin splitting and/or band deformations in accordance with MAGNDATA~\cite{gallego2016magndata}, magnetic structures database.

\begin{figure*}[htb!]
\begin{center}
\includegraphics[width=1.0 \hsize]{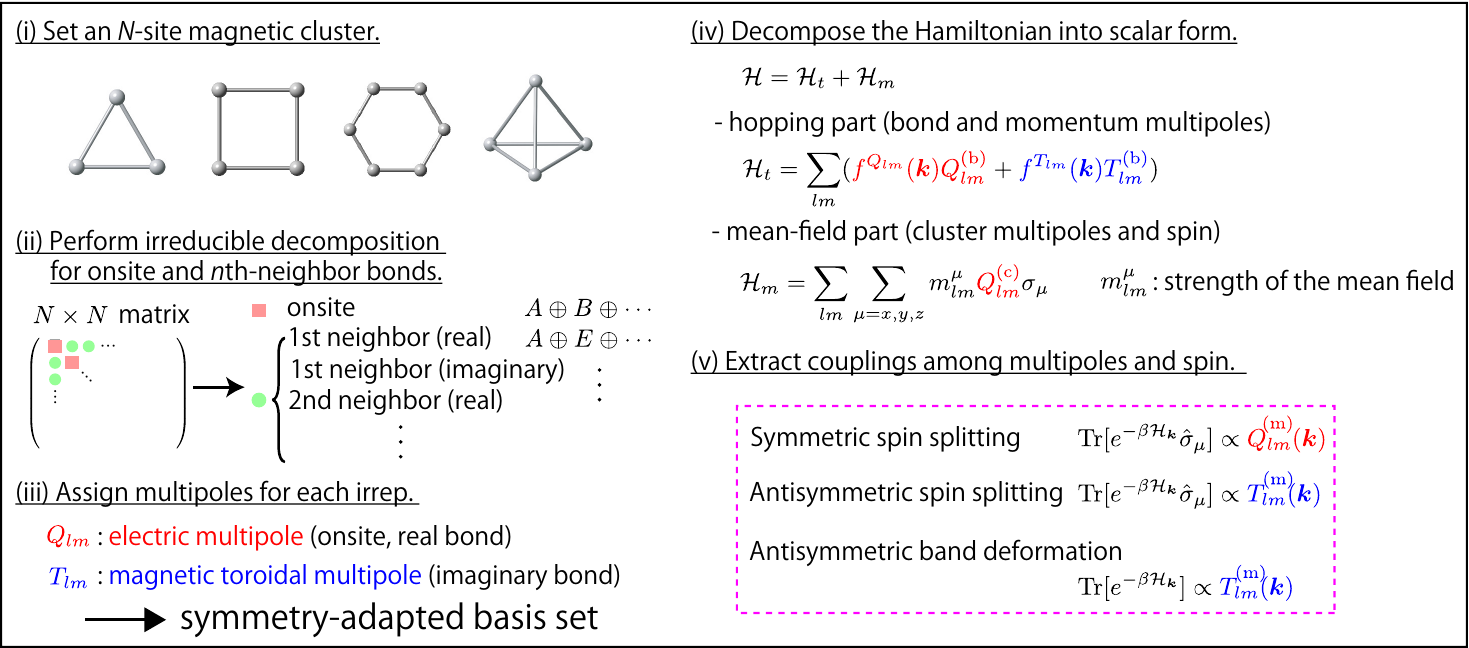} 
\caption{
\label{Fig:outline}
Outline of the engineering procedure of the spin-split and reshaped band structures in terms of augmented multipoles. 
}
\end{center}
\end{figure*}

\section{Outline}
\label{sec:Outline}
 
First, let us show the abstract procedure to engineer the spin-split and/or reshaped band structures by AFM. 
The overall guideline is summarized in Fig.~\ref{Fig:outline}, which consists of the following five parts:  
\begin{enumerate}
\item Set an $N$-site magnetic cluster to describe AFM structures, such as triangle, square, hexagon, and tetrahedron, in accordance with the crystallographic point groups.
\item Perform irreducible decomposition of an arbitrary hermitian matrix in the cluster for onsite and $n$th-neighbor bonds according to the point group. 
The independent $N\times N$ degrees of freedom are decoupled into the symmetry-adapted $N$ onsite degrees of freedom, and $N(N-1)$ off-diagonal ones in which half of them are for real part, and another half of them are imaginary part. 
\item Assign the augmented multipoles to each decomposed irreducible representation (irrep.), which gives intuitive view of microscopic degrees of freedom. 
In the decoupled spin and orbital basis, an introduction of two types of multipoles, electric and magnetic toroidal multipoles, is sufficient, which describe polar tensors with time-reversal even and odd, respectively. The onsite and real bond degrees of freedom are represented by the electric multipoles, whereas the imaginary bond degrees of freedom are represented by the magnetic toroidal multipoles. 
These multipoles are used to span the given Hamiltonian as the symmetry-adapted basis set. 
\item Decompose the hopping and mean-field Hamiltonians into a ``scalar-product'' form in terms of electric and magnetic toroidal multipoles. The hopping Hamiltonian is described by a linear combination of products between bond and momentum multipoles, while the mean-field Hamiltonian is described by a linear combination of products between cluster multipoles and Pauli matrices of spins. 
\item Extract effective spin-multipole couplings by evaluating momentum-dependent spin moments.
An effective coupling between cluster (molecular field) and bond multipoles induces momentum multipoles, which yields symmetric, antisymmetric spin splittings, and antisymmetric band deformations depending on the type of multipole couplings: 
The symmetric spin splitting is represented by momentum electric multipoles, and the antisymmetric spin splitting or spin-independent band deformation is represented by momentum magnetic toroidal multipoles. 
\end{enumerate}
Through the above procedure, the microscopic conditions (e.g., which part of hopping element is indispensable, or significant to obtain large splitting, deformation, etc.) for emergent symmetric and antisymmetric spin splittings and antisymmetric band deformations are systematically derived. 

The rest of the paper is organized as follows. 
In Sec.~\ref{sec:Magnetic cluster}, we set a magnetic cluster and perform irreducible decomposition for onsite and bond degrees of freedom, which corresponds to the procedures (i) and (ii). 
The remaining procedures (iii) to (v) are explained in Secs.~\ref{sec:Multipole description} and \ref{sec:Momentum-dependent spin splitting and band deformation}. 
In Sec.~\ref{sec:Multipole description}, we introduce the concept of three kinds of multipoles, cluster, bond, and momentum multipoles representing different electronic degrees of freedom. 
We describe a general condition of the spin-split band structure and asymmetric spin-degenerate band deformation in Sec.~\ref{sec:Momentum-dependent spin splitting and band deformation}. 
In Sec.~\ref{sec:Application to lattice systems}, we show three examples by considering the periodic lattice systems comprised of the triangle unit. 
We discuss potential candidate materials to exhibit spin splittings and band deformations driven by the magnetic order and summarize the paper 
in Sec.~\ref{sec:Summary}. 
In two Appendices, we show the explicit expressions of the electric multipoles in Appendix~\ref{appen:Expressions of electric multipoles}, and classification of multipoles under eleven Laue classes in Appendix~\ref{appen:Multipole notations under 11 Laue classes}. 

Throughout this paper, we focus on the limit of negligibly small SOC in order to extract intrinsic role of the multipole-spin couplings, and then we adopt the spin-orbital decoupled basis to express the electronic degrees of freedom.

\section{Magnetic cluster and irreducible decomposition of electronic degrees of freedom}
\label{sec:Magnetic cluster}

\begin{figure}[htb!]
\begin{center}
\includegraphics[width=1.0 \hsize]{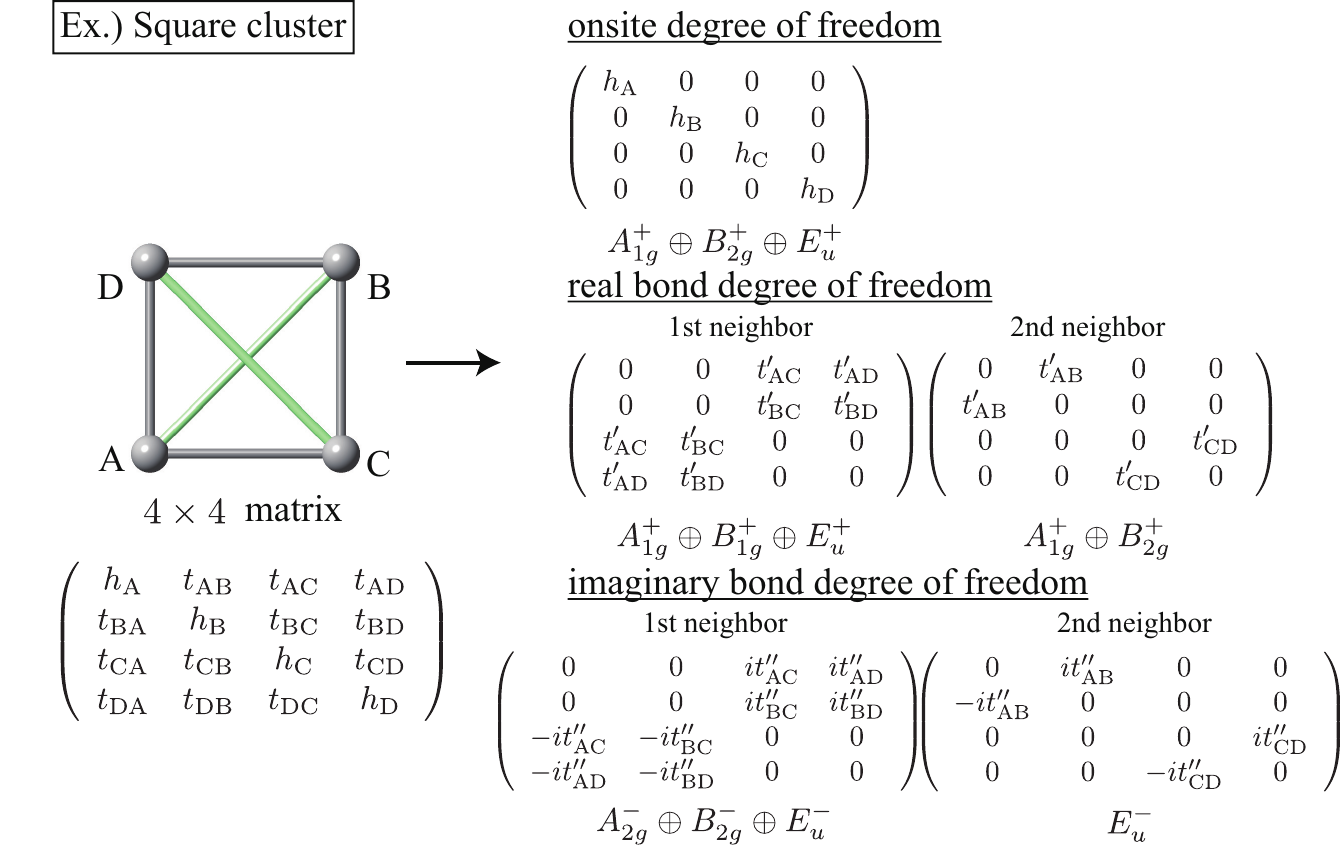} 
\caption{
\label{Fig:cluster}
Schematic picture of a four-site square cluster. 
The sixteen independent matrix elements are decomposed into four onsite degrees of freedom $h_{i}$, six plus six bond degrees of freedom $t_{ij}=t_{ij}'+it_{ij}''$ for the real and imaginary components. 
The corresponding matrix elements and irrep. under the point group $D_{4 \rm h}$ are also shown. 
}
\end{center}
\end{figure}

Before introducing the multipole descriptions, we perform the irreducible decomposition of the internal electronic degrees of freedom by using the group theory.
A magnetic cluster is introduced as a unit to represent the AFM structure. 
In other words, the magnetic cluster represents the minimal magnetic unit cell we focus on. 
In the following, we consider the single-orbital model and ignore the atomic orbital degree of freedom, although the extension to the multi-orbital system is straightforward. 
Then, a spinless basis wave function is represented by $\phi=(\phi_1, \phi_2, \cdots, \phi_N)$ where $\phi_i$ is the wave function at sublattice $i$ and $N$ is the number of sublattices. 
The Hamiltonian for one-body electronic state is represented by the $N \times N$ matrix except for the spin degree of freedom. 
As the Hamiltonian is hermitian matrix, its matrix elements are decomposed into the $N$ diagonal part, the $N(N-1)/2$ real and imaginary off-diagonal parts. 
The $N$ diagonal elements correspond to the onsite degrees of freedom such as charge and spin densities, while the $N(N-1)$ off-diagonal elements correspond to the bond degrees of freedom representing off-site kinetic motion of electrons. 
For each part, the irreducible decomposition can be performed according to the point group symmetry. 
It is noted that the $N(N-1)$ bond degrees of freedom can be further decomposed into $n$th-neighbor bond degree of freedom of the real and imaginary components. 

\begin{table*}
\caption{
Irreducible decomposition of the Hamiltonian matrix for the representative clusters~\cite{hayami2019momentum}. 
\# represents the number of sublattice. 
The parentheses in the fifth and sixth columns represent the irrep. for each neighbor bond. 
}
\label{tab_IRREP}
\centering
\begin{tabular}{cccccccccccccc|ccc|ccc} \hline\hline
Cluster & \#  & PG &  Onsite & Real bond & Imaginary bond  \\  \hline
Triangle  &  3 & $D_{3\rm h}$ & $A'^+_1 \oplus E'^+$ & $A'^+_1 \oplus E'^+$ & $A'^-_2 \oplus E'^-$\\
Rectangle & 4  & $D_{2\rm h}$ & $A^+_{g} \oplus B^+_{1g}\oplus B^+_{2u}\oplus B^+_{3u}$ & $(A^+_{g} \oplus B^+_{2u})\oplus (A^+_{g} \oplus B^+_{3u})\oplus (A^+_{g} \oplus B^+_{1g})$ &$(B^-_{1g} \oplus B^-_{3u})\oplus (B^-_{1g} \oplus B^-_{2u})\oplus (B^-_{2u} \oplus B^-_{3u})$  \\
Square &  4 & $D_{4\rm h}$ & 
$A^+_{1g} \oplus B^+_{2g}\oplus E^+_{u}$ & 
$(A^+_{1g} \oplus B^+_{1g}\oplus E^+_{u})\oplus (A^+_{1g} \oplus B^+_{2g})$ & 
$(A^-_{2g} \oplus B^-_{2g}\oplus E^-_{u})\oplus (E^-_{u})$ \\ 
Hexagon & 6 & $D_{6\rm h}$ &  
$A^+_{1g} \oplus B^+_{1u}\oplus E^+_{1u}\oplus E^+_{2g}$ & 
$(A^+_{1g} \oplus B^+_{1u}\oplus E^+_{1u}\oplus E^+_{2g})$& 
$(A^-_{2g} \oplus B^-_{2u}\oplus E^-_{1u}\oplus E^-_{2g})$ 
\\
 & & &  &$\oplus (A^+_{1g} \oplus B^+_{2u} \oplus E^+_{1u} \oplus E^+_{2g}) \oplus (A^+_{1g} \oplus E^+_{2g} )$ & 
 $\oplus (A^-_{2g} \oplus B^-_{1u} \oplus E^-_{1u} \oplus E^-_{2g}) \oplus (B^-_{1u} \oplus E^-_{1u} )$
 \\
Tetrahedron & 4 & $T_{\rm d}$ &  
$A^+_{1} \oplus T^+_{2}$ & 
$A^+_{1} \oplus E^+\oplus T^+_{2}$ & 
$T^-_{1} \oplus T^-_{2}$ \\ 
Octahedron & 6 & $O_{\rm h}$ & 
$A^+_{1g} \oplus E^+_{g}\oplus T^+_{1u}$ & 
$(A^+_{1g} \oplus E^+_{g}\oplus T^+_{1u}\oplus T^+_{2g}\oplus T^+_{2u})$ & 
$(A^-_{2g} \oplus E^-_{g}\oplus T^-_{1g}\oplus T^-_{1u}\oplus T^-_{2u})\oplus (T^-_{1u})$   \\
& & & & $\oplus (A^+_{1g} \oplus E^+_{g})$&  \\
Cube & 8 &$O_{\rm h}$ &  
$A^+_{1g} \oplus A^+_{2u}\oplus T^+_{1u}\oplus T^+_{2g}$ & 
$(A^+_{1g} \oplus E^+_{g}\oplus T^+_{1u}\oplus T^+_{2g}\oplus T^+_{2u})$ & 
$(A^-_{2u} \oplus E^-_{u} \oplus T^-_{1g} \oplus T^-_{1u} \oplus T^-_{2g})$
 \\
&&&& $\oplus (A^+_{1g} \oplus A^+_{2u} \oplus E^+_{g} \oplus E^+_{u} \oplus T^+_{1u} \oplus T^+_{2g})$ &$\oplus ( T^-_{1g} \oplus T^-_{1u} \oplus T^-_{2g}\oplus T^-_{2u})$ \\
&&&& $\oplus (A^+_{1g} \oplus T^+_{2g})$ & $\oplus (A^-_{2u} \oplus T^-_{1u})$
\\
\hline\hline
\end{tabular}
\end{table*}

As an example, let us consider a square cluster consisting of four sublattice as shown in Fig.~\ref{Fig:cluster}, 
which belongs to the point group $D_{4 \rm h}$. 
The Hamiltonian matrix in spinless space is generally represented by the $4\times 4$ matrix as
\begin{align}
\mathcal{H} &= \sum_{i,j={\rm A, B, C, D}} c_{i}^{\dagger} H_{ij} c_{j}, \\ 
\label{eq:Hmat}
H&=
\left(
\begin{array}{cccc}
h_{\rm A} & t_{\rm AB}& t_{\rm AC}& t_{\rm AD} \\
t_{\rm BA} &h_{\rm B} & t_{\rm BC}& t_{\rm BD} \\
t_{\rm CA} & t_{\rm CB}& h_{\rm C} & t_{\rm CD} \\
t_{\rm DA} & t_{\rm DB}& t_{\rm DC} &h_{\rm D} \\
\end{array}
\right), 
\end{align}
where $c_i^{\dagger}$ ($c_i$) is the creation (annihilation) operator at site $i$. 
$h_i$ and $t_{ij}=t_{ji}^{*}$ ($i,j={\rm A, B, C, D}$) are real and complex numbers corresponding to onsite and hopping terms, respectively. 
The Hamiltonian matrix $H$ in Eq.~(\ref{eq:Hmat}) is specified by giving sixteen independent model parameters consisting of four $h_{i}$ and twelve $t_{ij}$. 

The matrix $H$ is decomposed into the onsite potential and the hopping parts. 
The sublattice-basis wave function $\left\{\phi_{\rm A}, \phi_{\rm B}, \phi_{\rm C}, \phi_{\rm D}\right\}$ is decomposed into the irrep. of the point group $D_{\rm 4h}$ as $A_{1g} \oplus B_{2g}\oplus E_{u} $, each of which represents the molecular orbital belonging to its irrep.
Then, the internal degrees of freedom is given by its direct product as $(A_{1g} \oplus B_{2g}\oplus E_{u})\otimes (A_{1g} \oplus B_{2g}\oplus E_{u}) =2{A}_{1g}^+ \oplus {B}_{1g}^+ \oplus {B}_{2g}^+\oplus 2{E}_{u}^+ \oplus {A}_{2g}^- \oplus {B}_{2g}^-\oplus 2{E}_{u}^-$, where the subscript represents the spatial parity (even: $g$, odd: $u$) and the superscript represents the time-reversal parity (even: $+$, odd: $-$).
Among them, for the onsite part, four parameters $(h_{\rm A}, h_{\rm B}, h_{\rm C}, h_{\rm D})$ are decomposed into ${A}_{1g}^+ \oplus {B}_{2g}^+\oplus {E}_{u}^+$. 
By applying the symmetry operation of $D_{\rm 4h}$ to the degrees of freedom in the magnetic cluster, one can find that the matrix elements in each irrep. are given by 
\begin{align}
{A}_{1g}^+&:  h_{\rm A}=h_{\rm B}=h_{\rm C}=h_{\rm D},  \nonumber  \\
\label{eq:B2g_square}
{B}_{2g}^+&:  h_{\rm A}=h_{\rm B}=-h_{\rm C}=-h_{\rm D},  \nonumber   \\
{E}_{u}^+&:   h_{\rm A}=-h_{\rm B}=-h_{\rm C}=h_{\rm D},    \nonumber  \\
                    &:  h_{\rm A}=-h_{\rm B}=h_{\rm C}=-h_{\rm D}.
\end{align}

For the hopping part, the hopping parameters $t_{ij}$ are divided into the real $(t_{ij}'=t_{ji}')$ and imaginary $(t_{ij}''=-t_{ji}'')$ components. 
By performing the irreducible decomposition for each $n$th-neighbor bond, the real part is decomposed into 
\begin{align}
{A}_{1g}^+&:  
t_{\rm AC}'=t_{\rm AD}'=t_{\rm BC}'=t_{\rm BD}',   \nonumber  \\
{B}_{1g}^+&: 
 -t_{\rm AC}'=t_{\rm AD}'=t_{\rm BC}'=-t_{\rm BD}',   \nonumber \\
{E}_{u}^+&: 
  t_{\rm AD}'=-t_{\rm BC}', \ t_{\rm AC}'=t_{\rm BD}'=0,   \nonumber  \\
                    &: 
t_{\rm AC}'=-t_{\rm BD}', \ t_{\rm AD}'=t_{\rm BC}'=0, 
\end{align}
for the first-neighbor bond, and 
\begin{align}
{A}_{1g}^+&:  t_{\rm AB}'=t_{\rm CD}',    \nonumber \\
{B}_{1g}^+&:  t_{\rm AB}'=-t_{\rm CD}', 
 \end{align}
for the second-neighbor bond. 

Similarly, the imaginary bond degree of freedom is decomposed into 
\begin{align}
{A}_{2g}^-&: -t_{\rm AC}''=t_{\rm AD}''=t_{\rm BC}''=-t_{\rm BD}'',   \nonumber  \\
 {B}_{2g}^-&: t_{\rm AC}''=t_{\rm AD}''=t_{\rm BC}''=t_{\rm BD}'',  \nonumber  \\
 {E}_{u}^-  &:  -t_{\rm AC}''=t_{\rm BD}'', \ t_{\rm AD}''=t_{\rm BC}''=0,   \nonumber  \\
                    &: -t_{\rm AD}''=t_{\rm BC}'', \ t_{\rm AC}''=t_{\rm BD}''=0, 
\end{align}
for the first-neighbor bond, and 
\begin{align}
E_{u}^-&: -t_{\rm AB}''=t_{\rm CD}'',  \nonumber \\
                  &:  t_{\rm AB}''=t_{\rm CD}'', 
 \end{align}
for the second-neighbor bond. 

In general, the $N\times N$ matrix elements in $N$ sublattice cluster are also represented by the irreps. of the given point group.
It is noted that such an irreducible decomposition is performed much more intuitively by using the multipole description, as will be shown in the next section.
We summarize the irreducible decomposition for onsite and bond degrees of freedom in the representative clusters in Table~\ref{tab_IRREP}~\cite{hayami2019momentum}.

\section{Multipole description}
\label{sec:Multipole description}

In this section, we describe the concept of multipole. 
We introduce three kinds of multipole notations, which are necessary to describe the distinct electronic degrees of freedom in the tight-binding model.  
The cluster multipole is used to describe the onsite degree of freedom in Sec.~\ref{sec:Cluster multipole}, the bond multipole is for the bond degree of freedom in Sec.~\ref{sec:Bond multipole}, and the momentum multipole is for a wave-vector-dependent form factor in periodic lattice systems in Sec.~\ref{sec:Momentum multipole}. 
Then, we show the correspondence between the irreps. explained in the previous section and these multipoles in Sec.~\ref{sec:Irreducible representation of multipoles}.

\subsection{Cluster multipole}
\label{sec:Cluster multipole}

\begin{figure}[htb!]
\begin{center}
\includegraphics[width=1.0 \hsize]{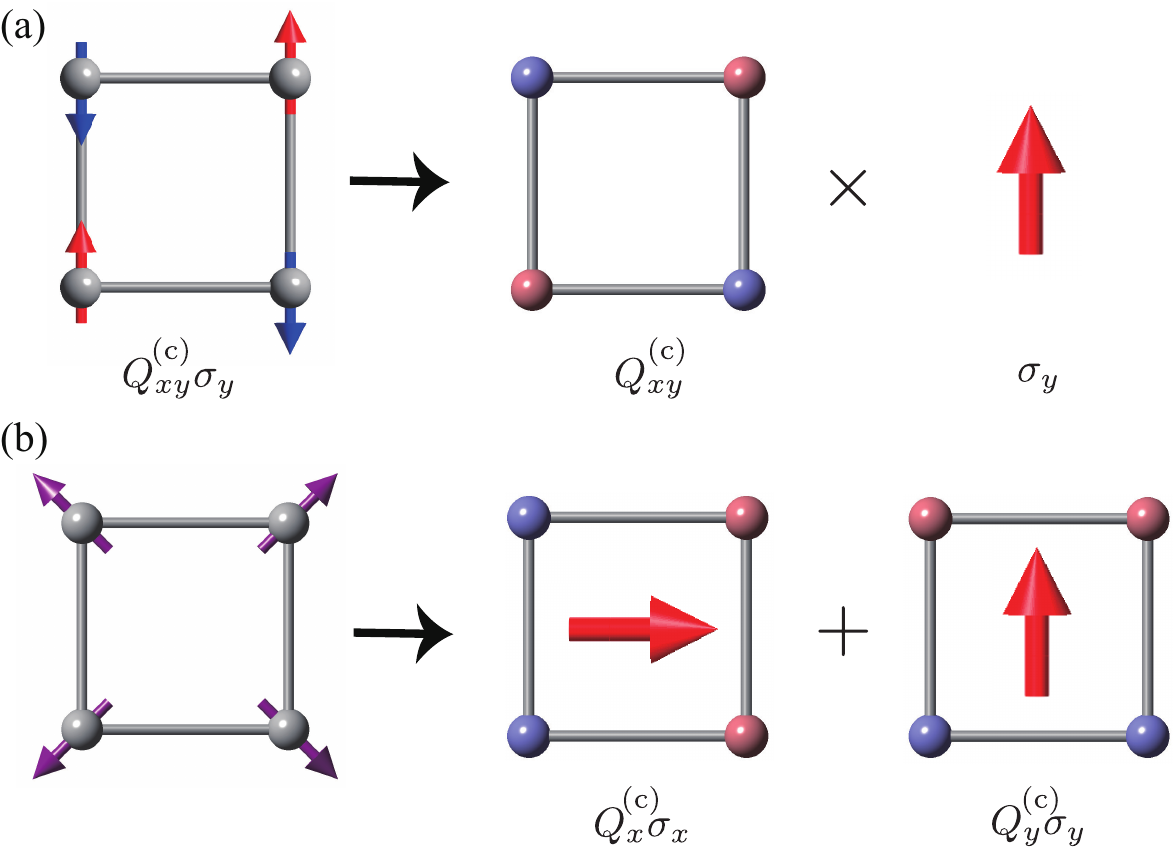} 
\caption{
\label{Fig:ponti_cluster}
(a) Collinear AFM order in a square cluster under the point group $D_{4{\rm h}}$. 
The collinear order parameter ($Q^{\rm (c)}_{xy}\sigma_y$) is decomposed into the electric-quadrupole-type alignment of point charges ($Q^{\rm (c)}_{xy}$) and the spin along the $y$ direction ($\sigma_y$). 
(b) Coplanar magnetic order in a square cluster, which is regarded as a superposition of two  collinear spin components with $Q^{\rm (c)}_{x}\sigma_x$ and $Q^{\rm (c)}_{y}\sigma_y$. 
}
\end{center}
\end{figure}

The cluster multipole is defined to describe the onsite degree of freedom in the tight-binding Hamiltonian. 
As arbitrary onsite degrees of freedom are represented by a superposition of local potentials at each atomic site, all the anisotropic charge distributions in a cluster are systematically represented by using the spherical harmonics with the origin at the center of the cluster, which is related to the electric multipole degree of freedom~\cite{Kusunose_JPSJ.77.064710,kuramoto2009multipole,Santini_RevModPhys.81.807}. 
Eventually, the anisotropic charge distributions on cluster sites are described as a cluster electric multipole $\tilde{Q}^{(c)}_{lm}$, which is given by 
\begin{align}
\label{eq:Qc}
\tilde{Q}^{\rm (c)}_{lm} = \sum_{i=1}^N q^{(lm)}_i O_{lm}(\bm{R}_i),  
\end{align}
where $O_{lm}(\bm{R}_i)= \sqrt{4\pi/(2l+1)}R_i^l Y^*_{lm}(\hat{\bm{R}}_i)$, $\bm{R}_i=(X_i, Y_i, Z_i)$ is the position vector of $i$th cluster site, $q^{(lm)}_i$ is the local electric charge of $i$th cluster site, and $N$ is the number of sites in a cluster. 
We omit electric charge unit $-e$ for notational simplicity. 
$Y_{lm}(\hat{\bm{R}}_i)$ is the spherical harmonics as a function of angle $\hat{\bm{R}}_i= \bm{R}_i/|\bm{R}_i|$ with the azimuthal and magnetic quantum numbers, $l$ and $m$ ($-l \leq m \leq l $). 
In the following, we regard the symbol $O_{lm}(\bm{r})$ as the harmonics of the point group such as cubic and hexagonal ones instead of the spherical harmonics, which are real functions given by linear combinations of $O_{lm}$ and $O_{l-m}$ as shown in Appendix~\ref{appen:Expressions of electric multipoles}~\cite{Kusunose_JPSJ.77.064710}. 
Through this expression, we define $q^{(lm)}_i$ for the specified electric multipole, and the corresponding matrix (operator) expression is given by $Q^{\rm (c)}_{lm}=\sum_i  q^{(lm)}_i \ket{i} \bra{i}$ where $\ket{i}$ is the atomic site basis. 

Such a cluster multipole can also describe magnetic ordering patterns in a cluster~\cite{Suzuki_PhysRevB.95.094406,Suzuki_PhysRevB.99.174407}. 
In the spin-orbital decoupled basis, it is useful to express the magnetic structure as a linear combination of direct products of $Q^{\rm (c)}_{lm}$ and the Pauli matrices of spin $\bm{\sigma}=(\sigma_x,\sigma_y,\sigma_z)$.  
Then, any types of magnetic orderings coupled with the corresponding molecular fields are expressed  by a linear combination of $Q^{\rm (c)}_{lm} \sigma_\mu$. 
Namely, the mean-field Hamiltonian of the AFM ordering is represented by 
\begin{align}
\label{eq:Hm}
H_m= \sum_{lm} \sum_{\mu=x,y,z} m^{\mu}_{lm} Q^{\rm (c)}_{lm} \sigma_\mu, 
\end{align}
where the coefficient $m^{\mu}_{lm}$ is a conjugate field of an order parameter in the AFM state. 
As $Q^{\rm (c)}_{lm}$ is time-reversal-even and $\sigma_\mu$ is time-reversal-odd, $H_m$ is time-reversal-odd as it is a symmetry breaking term. 
From the multipole viewpoint, the ordering pattern is characterized by the type of the emergent multipole: 
The ferromagnetic structure corresponds to the isotropic electric monopole and the AFM structure corresponds to the anisotropic electric multipoles for $l \geq 1$.

In the case of the collinear AFM order, the mean-field Hamiltonian matrix is represented by the single component of $\sigma_\mu$ where $\mu$ denotes the ordered moment direction, although it is taken to be arbitrary in the absence of the SOC. 
We show an example of the staggered AFM ordering in a square cluster under the point group $D_{\rm 4h}$ in Fig.~\ref{Fig:ponti_cluster}(a). 
By decomposing this magnetic structure into the alignment of point charges and spin as shown in Fig.~\ref{Fig:ponti_cluster}(a), and then, evaluating $Q^{\rm (c)}_{lm}$ via Eq.~(\ref{eq:Qc}), one can find that the corresponding multipole is $Q^{\rm (c)}_{xy}$. 

Such a multipole description is also understood from a symmetry viewpoint. 
The mean-field matrix in a square cluster in Eq.~(\ref{eq:Hmat}) is represented by 
\begin{align}
H_m&=
h \sigma \left(
\begin{array}{cccc}
1 & 0 & 0& 0 \\
0 & 1 & 0& 0 \\
0 & 0& -1 & 0 \\
0 & 0& 0 & -1 \\
\end{array}
\right), 
\end{align}
where $\sigma= \pm1$ for up and down spins. 
From Eq.~(\ref{eq:B2g_square}), this matrix element except for $\sigma$ belongs to the irrep. ${B}_{2g}^+$ under the point group $D_{4{\rm h}}$, which is the same irrep. of the $Q^{\rm (c)}_{xy}$-type electric quadrupole (See also the correspondence between the irrep. under the point group and multipoles in Sec.~\ref{sec:Irreducible representation of multipoles})~\cite{Hayami_PhysRevB.98.165110}. 
More intuitively, the real-space point charge alignment in a square cluster clearly indicates the presence of $xy$-type electric quadrupole; the positive charges are in the $[110]$ direction, while the negative ones are in the $[\bar{1}10]$ direction. 

In a similar manner, coplanar and noncoplanar magnetic structures are described by a linear combination of two and three components of $\sigma_\mu$, respectively.  
Figure~\ref{Fig:ponti_cluster}(b) shows an example of the coplanar spin structure in a square cluster where each spin points to the $\langle 110 \rangle$ radial direction. 
Also in this case, the plane including spins is taken to be arbitrary due to spin rotational symmetry.   
By using Eq.~(\ref{eq:Qc}) for two spin components, one can find that the spin pattern in Fig.~\ref{Fig:ponti_cluster}(b) is proportional to $Q^{\rm (c)}_{x} \sigma_x + Q^{\rm (c)}_{y} \sigma_y$. 
The mean-field matrix is given in the form of 
\begin{align}
H_m&=
h \left[\sigma_x \left(
\begin{array}{cccc}
-1 & 0 & 0& 0 \\
0 & 1 & 0& 0 \\
0 & 0& 1 & 0 \\
0 & 0& 0 & -1 \\
\end{array}
\right)+
\sigma_y \left(
\begin{array}{cccc}
-1 & 0 & 0& 0 \\
0 & 1 & 0& 0 \\
0 & 0& -1 & 0 \\
0 & 0& 0 & 1 \\
\end{array}
\right)
\right].  
\end{align}
It is apparent from Fig.~\ref{Fig:ponti_cluster}(b) that $Q_{x}^{\rm (c)}$ and $Q_{y}^{\rm (c)}$ represent the $x$ and $y$ electric dipoles in the multipole language.

\subsection{Bond multipole}
\label{sec:Bond multipole}

\begin{figure}[h!]
\begin{center}
\includegraphics[width=1.0 \hsize]{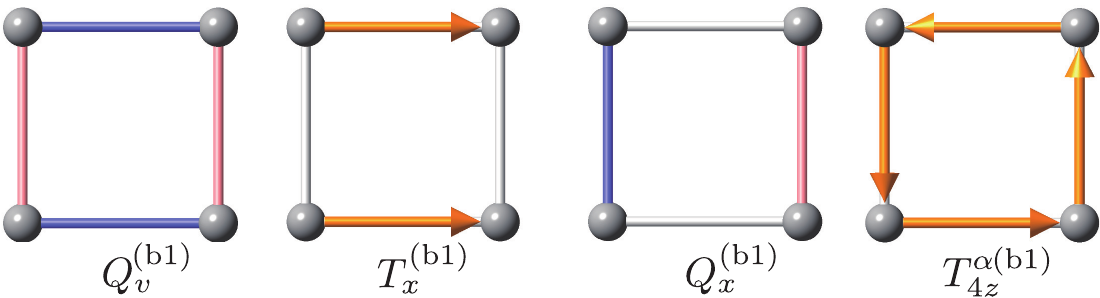} 
\caption{
\label{Fig:ponti_bond}
The examples of bond multipoles in a square unit. 
The real and imaginary hoppings correspond to the presence of the electric monopole on the bond center and the magnetic toroidal dipole $\bm{t}_{(ij)}$ along the bond direction, respectively. 
See also Eqs.~(\ref{eq:Qb}) and (\ref{eq:Tb}). 
From the left, the electric quadrupole $Q^{\rm (b1)}_v$, magnetic toroidal dipole $T^{\rm (b1)}_x$, electric dipole $Q^{\rm (b1)}_x$, and magnetic toroidal hexadecapole $T^{\alpha{\rm (b1)}}_{4z}$ are presented. 
}
\end{center}
\end{figure}

The bond multipole is introduced to describe the bond degree of freedom in the system, which corresponds to the off-diagonal hopping part in the tight-binding Hamiltonian~\cite{Hayami_PhysRevLett.122.147602}. 

First, in order to get an intuitive insight into the relation between bond multipoles and hoppings, let us consider a two-site problem connected by the complex hopping $t=t' + i t''$. 
The two sites are denoted as A and B, which are separated by the distance $a$ in the $x$ direction.
By using the molecular-orbital basis $\{\ket{\phi_1}, \ket{\phi_2}\}$ instead of the atomic site basis $\{\ket{{\rm A}}, \ket{{\rm B}}\}$, the real and imaginary hopping matrices are transformed as 
\begin{align}
{\rm Re}[H_t]=\left(
\begin{array}{cc}
t' & 0 \\
0 & -t' \\
\end{array}
\right), \ \
i{\rm Im}[H_t]=\left(
\begin{array}{cc}
0 & -it''
\\
it'' & 0 \\
\end{array}
\right), 
\end{align}
where $\ket{\phi_1}=(1/\sqrt{2})(\ket{{\rm A}}+ \ket{{\rm B}})$ and $\ket{\phi_2}=(1/\sqrt{2})(\ket{{\rm A}}- \ket{{\rm B}})$. 
As the anisotropy of the molecular orbitals $\{\ket{\phi_1}, \ket{\phi_2}\}$ is the same as the $s$- and $p_x$-orbital wave functions, the ordinary atomic-scale 
multipole description in Ref.~\onlinecite{hayami2018microscopic} can be applied. 
Then, by comparing the matrix elements in $s$-$p_x$ orbital basis, ${\rm Re}[H_t]$ corresponds to the electric monopole $Q_0$, while ${\rm Im}[H_t]$ corresponds to the magnetic toroidal dipole $T_x$~\cite{hayami2018microscopic}. 
This result indicates that the real hopping is expressed as the electric monopole on the bond center, while the imaginary hopping is expressed as the magnetic toroidal dipole along the bond direction. 
This assignment of multipole moments on the bond center is reasonable from a symmetry viewpoint, since the real (imaginary) hopping is equivalent with the time-reversal-even scalar (time-reversal-odd polar vector), which corresponds to the electric monopole (magnetic toroidal dipole). 

This result is generalized for arbitrary cluster systems. 
Any types of hoppings are represented by bond electric and magnetic toroidal multipoles, $\tilde{Q}^{\rm (b)}_{lm}$ and $\tilde{T}^{\rm (b)}_{lm}$, which are expressed as 
\begin{align}
\label{eq:Qb}
\tilde{Q}^{\rm (b)}_{lm}&= \sum_{(ij)}^{N_{\rm bond}} q^{(lm)}_{(ij)} O_{lm}(\bm{R}_{(ij)}),  \\
\label{eq:Tb}
\tilde{T}^{\rm (b)}_{lm}&= \sum_{(ij)}^{N_{\rm bond}} \bm{t}^{(lm)}_{(ij)}\cdot \bm{\nabla}O_{lm}(\bm{R}_{(ij)}),  
\end{align}
where $N_{\rm bond}$ is the number of bonds in a cluster and $\bm{R}_{(ij)}$ is the position vector at the $i$-$j$ bond center.
$q^{(lm)}_{(ij)}$ and $\bm{t}^{(lm)}_{(ij)}$ are the local electric charge and local magnetic toroidal dipole at $\bm{R}_{(ij)}$, respectively, where $\bm{t}^{(lm)}_{(ij)}=i t''^{(lm)}_{ij} \bm{n}_{ij}$ with $\bm{n}_{ij}$ being the unit vector connecting between $i$ and $j$ sites. 
The multipole assignment is independently performed per $n$th-neighbor bond.
The corresponding matrices (operators) of bond electric and magnetic toroidal multipoles are given by $Q^{\rm (b)}_{lm}=\sum_{(ij)}  q^{(lm)}_{(ij)} \ket{j}\bra{i}$ and $T^{\rm (b)}_{lm}=\sum_{(ij)}  (\bm{t}^{(lm)}_{(ij)} \cdot \bm{n}_{ij}) \ket{j}\bra{i}$. 

By using these matrices, we express any bond modulations in terms of bond multipoles.
In particular, the bond magnetic toroidal multipoles, $T^{\rm (b)}_{lm}$, represent the anisotropic current distribution including a loop-current distribution discussed in cuprates and iridates~\cite{Shekhter_PhysRevB.80.214501, zhao2016evidence, murayama2020bond}, as the imaginary hopping represents a local current along the bond. 

Let us again consider an example in a square cluster under the point group $D_{4{\rm h}}$, whose Hamiltonian is shown in Eq.~(\ref{eq:Hmat}). 
There are six real and imaginary bond degrees of freedom, which are assigned as six electric and magnetic toroidal multipoles, respectively. 
By using Eqs.~(\ref{eq:Qb}) and (\ref{eq:Tb}), the first-neighbor four real (imaginary) bonds correspond to electric monopole $Q^{\rm (b1)}_0$, electric quadrupole $Q^{\rm (b1)}_v$, and electric dipoles $(Q^{\rm (b1)}_x, Q^{\rm (b1)}_y)$ [magnetic toroidal hexadecapole $T^{\alpha {\rm (b1)}}_{4x}$, magnetic toroidal quadrupole $T^{\rm (b1)}_{xy}$, and magnetic toroidal dipoles $(T^{\rm (b1)}_x, T^{\rm (b1)}_y)$], while the second-neighbor two real (imaginary) bonds correspond to electric monopole $Q^{\rm (b2)}_0$ and electric quadrupole $Q^{\rm (b2)}_{xy}$ [magnetic toroidal dipoles $(T^{\rm (b2)}_x, T^{\rm (b2)}_y)$] where the integer $(n=1,2)$ in superscript represents the $n$th-neighbor bond. 
The specific examples of $Q^{\rm (b1)}_v$, $T^{\rm (b1)}_x$, $Q^{\rm (b1)}_x$, and $T^{\alpha{\rm (b1)}}_{4z}$ are shown in Fig.~\ref{Fig:ponti_bond}. 
It is noted that the magnetic toroidal hexadecapole $T_{4z}^{\rm \alpha(b1)}$ with the vortex-like alignment of $\bm{t}_{(ij)}$ in Fig.~\ref{Fig:ponti_bond} is equivalent to the magnetic dipole along $z$ direction.
Nevertheless, we use the higher-rank hexadecapole $T_{4z}^{\rm \alpha(b1)}$ since we use the convention in this paper that all bond degrees of freedom are described by the electric or magnetic toroidal multipoles.

\subsection{Momentum multipole}
\label{sec:Momentum multipole}

Finally, we introduce momentum multipoles to represent the momentum dependence in crystals.
In the $\bm{k} \to \bm{0}$ limit, the spherical harmonics as a function of $\hat{\bm{k}}= \bm{k}/|\bm{k}|$, $Y_{lm}(\hat{\bm{k}})$, gives the anisotropic momentum distribution. 
As $\bm{k}$ is a polar vector with time-reversal odd, the even-(odd-)rank component in $Y_{lm}(\hat{\bm{k}})$ corresponds to the electric (magnetic toroidal) multipoles, which are defined as 
\begin{align}
\label{eq:Qm}
Q^{\rm (m)}_{lm}(\bm{k}) &= O_{lm}(\bm{k}) \ \ {\rm for \  even\ } l, \\
\label{eq:Tm}
T^{\rm (m)}_{lm}(\bm{k}) &=  O_{lm}(\bm{k}) \ \  {\rm for \  odd\ } l.   
\end{align}
The explicit expressions are given by replacing $\bm{r}$ 
with $\bm{k}$ in Table~\ref{tab_Emultipole} in Appendix~\ref{appen:Expressions of electric multipoles} where the odd-rank multipoles in Table~\ref{tab_Emultipole} should be replaced with $T_{lm}^{\rm (m)}(\bm{k})$.

In general, the momentum dependence in crystals has periodicity and is represented by a superposition of trigonometric functions of $\bm{k}$. 
In this case, the momentum form factor consists of the momentum multipoles up to the infinite rank belonging to the same irrep.
For example, we consider the single-band tight-binding model on a simple square lattice under the point group $D_{4{\rm h}}$ with the lattice constant $a$. 
The momentum form factor for the nearest-neighbor bond $f(\bm{k})$ is given by a linear combination of momentum multipoles as
\begin{align}
f(\bm{k}) &= \cos (k_x a) + \cos (k_y a) \\
                 &= 2 -\frac{a^2}{2}(k_x^2+k_y^2)+ \frac{a^4}{24}(k_x^4+k_y^4)+ \cdots \\
                 &= c_1 Q^{(\rm m)}_0(\bm{k}) + c_2 Q^{(\rm m)}_u(\bm{k})+ c_3 Q^{(\rm m)}_4 (\bm{k}) + \cdots, 
\end{align}
where $c_i$ ($i=1, 2, \cdots$) are the expansion coefficients. 
$f(\bm{k})$ clearly consists of the multipoles belonging to the totally symmetric irrep. $A_{1g}$ under $D_{\rm 4h}$.
To specify the type of multipole, we use the lowest-rank multipole in the superscript of $f(\bm{k})$ as the convention is often used in the field of superconductivity~\cite{Nomoto_PhysRevB.94.174513,Sumita_PhysRevResearch.2.033225}. 
In this case, $f(\bm{k})$ is expressed as $f^{Q_0}(\bm{k})$. 

Similarly, the form factors belonging to the other irreps. are also described by the different set of multipoles. 
In the case of $D_{\rm 4h}$, when the sign of the hopping along the $y$ direction is opposite, the form factor is given by 
\begin{align}
\label{eq:B1g_form}
f(\bm{k}) &= \cos (k_x a) - \cos (k_y a) \\
&=-\frac{a^2}{2} (k_x^2-k_y^2)+\frac{a^4}{24} \left(k_x^4-k_y^4\right) + \cdots \\
&=c_1 Q^{(\rm m)}_v(\bm{k})  +   c_2  Q^{(\rm m)}_{4v}(\bm{k}) 
+\cdots, 
\end{align}
which belongs to the irrep. $B_{1g}$. 
In this situation, we denote $f(\bm{k})$ as $f^{Q_v}(\bm{k})$. 

Moreover, when the imaginary hopping appears only in the $x$ direction under $D_{\rm 4h}$, the form factor is given by
\begin{align}
\label{eq:Eu_form}
f(\bm{k}) &= \sin (k_x a) \\
&=a k_x+\frac{a^3}{3} k_x^3- + \cdots \\
&=c_1 T^{(\rm m)}_x(\bm{k})  +   c_2  T^{\alpha(\rm m)}_{x}(\bm{k}) 
+\cdots, 
\end{align}
which belongs to the irrep. $E_{u}$. 
In this situation, we denote $f(\bm{k})$ as $f^{T_x}(\bm{k})$. 
Such form factors in Eqs.~(\ref{eq:B1g_form}) and (\ref{eq:Eu_form}) can appear in the tight-binding Hamiltonian when the system has the sublattice degree of freedom, as the local site symmetry is lowered than the lattice symmetry.

\subsection{Irreducible representation of multipoles in crystal}
\label{sec:Irreducible representation of multipoles}

In the crystal systems, a part of the rotational symmetry and/or inversion symmetry are lost due to the regular and discrete alignment of the ions. As a result, the multipole degrees of freedom belonging to the same irrep. are not distinguished from the symmetry viewpoint. 
In other words, the irrep. of the rotational group split into subgroups according to the point-group irrep. 
For example, some even-parity and odd-parity multipoles belong to the same irrep. in noncentrosymmetric crystals. 

To avoid such confusion, we uniquely assign the multipoles of irrep. as in Sec.~\ref{sec:Magnetic cluster} by the following rules: 
Among the multipoles belonging to the same irrep., we adopt the lowest even-rank electric multipoles for time-reversal even quantities, whereas we adopt the lowest odd-rank magnetic toroidal multipoles for time-reversal odd quantities.
In this convention, the momentum-type odd-rank electric multipoles and even-rank magnetic toroidal multipoles do not appear in the Hamiltonian, as will be clarified in Sec.~\ref{sec:Momentum-dependent spin splitting and band deformation}. 

Following the above rules, the multipoles and the irrep. have a one-to-one correspondence.
We summarize the corresponding multipole notations under 32 point groups in Appendix~\ref{appen:Multipole notations under 11 Laue classes}, where we divide them into eleven Laue classes with the same number of the irreps., in Tables~\ref{tab_multipoles_table_m3m}-\ref{tab_multipoles_table_1}.
The compatibility relation of multipole within the same Laue class is given in the same row in the table. 
On the other hand, for the compatibility relation of multipole between the different Laue classes, we use the group-subgroup compatibility relation and adopt the lower-rank multipole to assign the irrep. 
For example, by the relation between $T_{\rm d}$ and $T$, the electric monopole $Q_0$ belonging to $A_{1}$ and the electric hexadecapole $Q_{6t}$ belonging to $A_{2}$  under the point group $T_{\rm d}$ turn into the same irrep. $A$ under the point group $T$. 
In this case, the multipole belonging to $A$ under $T$ is denoted as $Q_0$. 

Let us remark on the connection of the quantities introduced in the cluster and the lattice. 
Although we assign the multipoles to the electronic degrees of freedom by introducing the magnetic cluster, there is a situation where the lattice symmetry is higher than the cluster symmetry due to the additional operations combined with the translation. 
In such a situation, we replace the irreps. in a cluster with the corresponding ones in a lattice in accordance with the compatibility relation. 
Accordingly, the multipoles in a cluster are mapped onto those in a lattice.
In Sec.~\ref{sec:Application to lattice systems}, we exemplify this by considering the triangular and kagome lattices consisting of the triangle unit where the cluster and lattice symmetries are different with each other.

\section{Momentum-dependent spin splitting and band deformation}
\label{sec:Momentum-dependent spin splitting and band deformation}

By using the multipole notations introduced in Sec.~\ref{sec:Multipole description}, we express the Hamiltonian in terms of multipoles in Sec.~\ref{sec:Hamiltonian}. 
Then, we analyze systematically when and how the spin splitting and antisymmetric deformation in the band structure occur in Sec.~\ref{sec:Band deformation}. 

\subsection{Hamiltonian}
\label{sec:Hamiltonian}

In the absence of the SOC, the single-orbital Hamiltonian consists of the hopping part without the spin dependence and the mean-field part gives rise to the symmetry breaking due to the magnetic ordering. 
The total tight-binding Hamiltonian is generally represented by 
\begin{align}
\label{eq:Ham_mul}
\mathcal{H}=\!\! \sum_{\bm{k} \sigma \sigma' \gamma \gamma'}
c^{\dagger}_{\bm{k}\gamma \sigma} \left[\delta_{\sigma \sigma'}(H_t^Q+H_t^T)^{\gamma \gamma'}+ \delta_{\gamma \gamma'} H_m^{\sigma\sigma'}\right]c_{\bm{k} \gamma'\sigma'}, 
\end{align}
where $c^{\dagger}_{\bm{k}\gamma \sigma}$ ($c_{\bm{k}\gamma \sigma}$) is the creation (annihilation) operator at wave vector $\bm{k}$ and sublattice $\gamma$. 
$H_t^Q$ and $H_t^T$ stand for the real and imaginary hopping matrices, respectively, which are 
represented by a linear combination of the product between bond and momentum multipoles as 
\begin{align}
\label{eq:Ham_bondQ}
H^{Q}_{t}&=\sum_{lm} f^{Q_{lm}}(\bm{k})Q^{\rm (b)}_{lm} \ \ \ (l: {\rm even}), \\
\label{eq:Ham_bondT}
H^{T}_{t}&=\sum_{lm} f^{T_{lm}}(\bm{k})T^{\rm (b)}_{lm} \ \ \ (l: {\rm odd}). 
\end{align}
In Eq.~(\ref{eq:Ham_bondQ}), the electric monopole contribution $f^{Q_{0}}(\bm{k})Q^{\rm (b)}_{0}$ always appears in $H^{Q}_{t}$, while the higher-rank contribution depends on the lattice symmetry. 
On the other hand, $H^{T}_{t}$ in Eq.~(\ref{eq:Ham_bondT}) exists only in the absence of the local inversion symmetry.
It is noted that the cross terms $f^{T_{lm}}(\bm{k})Q^{\rm (b)}_{lm} $ and $f^{Q_{lm}}(\bm{k})T^{\rm (b)}_{lm}$ do not appear in the Hamiltonian due to the time-reversal symmetry. 
The mean-field term $H_m$ is represented by the cluster multipole as already introduced in Eq.~(\ref{eq:Hm}).

\subsection{Band deformation}
\label{sec:Band deformation}

\begin{figure}[h!]
\begin{center}
\includegraphics[width=1.0 \hsize]{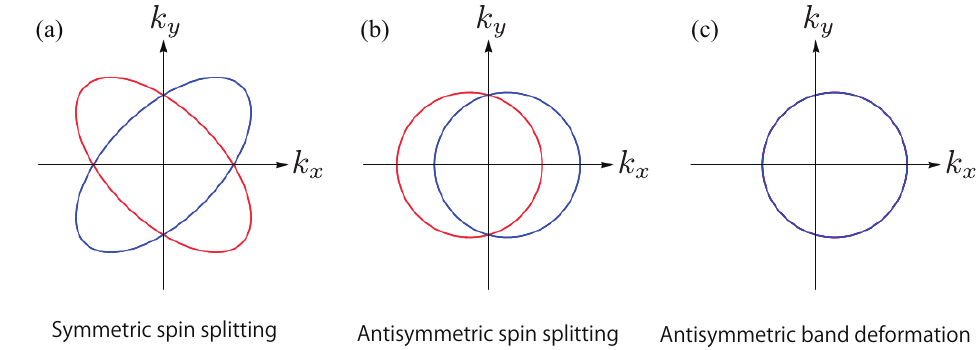} 
\caption{
\label{Fig:band}
Schematic pictures of three types of band deformations; (a) the symmetric spin splitting, (b) the antisymmetric spin splitting, and (c) the antisymmetric band deformation with spin degeneracy.  
In (a) and (b), the red and blue curves represent the up-spin and down-spin polarized bands, respectively. 
}
\end{center}
\end{figure}

Let us first discuss the essential points for the spin-split band structures in AFMs without the SOC. 
We consider three types of band deformations, the symmetric spin splitting, the antisymmetric spin splitting, and the antisymmetric band deformation, which are categorized into the different symmetry classes. 
Note that the time-reversal symmetry is always broken, as we focus on magnetic orderings.  

The first category is the symmetric spin splitting with respect to $\bm{k}$ when there is the spatial inversion symmetry in the system.
In this situation, the spin-dependent band dispersion is described by the product of the even function of $\bm{k}$ and spin $\sigma$. 
This means that the symmetric spin splitting arises through the effective coupling between the momentum electric multipole $Q^{\rm (m)}_{lm}(\bm{k})$ and $\sigma$. 
The lowest-order contribution is given by the rank-0 electric monopole, which merely corresponds to the momentum-independent Zeeman-like spin splitting in the band structure. 
In the following, we mainly focus on the higher-rank contribution for $l \geq 2$. 
The schematic example in the case $Q^{\rm (m)}_{xy}(\bm{k})\sigma \sim k_x k_y \sigma $ is shown in Fig.~\ref{Fig:band}(a). 

The second category is the antisymmetric spin splitting with respect to $\bm{k}$ in the absence of the spatial inversion symmetry and the product symmetry of time-reversal and spatial inversion operations in addition to the breaking of time-reversal symmetry. 
The functional form of the spin splittings is represented by the product of the odd function of $\bm{k}$ and spin $\sigma$. 
Thus, the antisymmetic spin splitting occurs when momentum magnetic toroidal multipole $T^{\rm (m)}_{lm}(\bm{k})$ is coupled with $\sigma$. 
The schematic example in the case of $T^{\rm (m)}_{x}(\bm{k})\sigma \sim k_x \sigma$ is shown in Fig.~\ref{Fig:band}(b). 

The third category is the antisymmetric band deformation with the spin degeneracy in the absence of spatial inversion symmetry, with preserving the product symmetry of time-reversal and spatial inversion operations. 
The band structure becomes asymmetric due to the contribution from the odd function of $\bm{k}$ in addition to the even function of $\bm{k}$. 
In terms of the multipole description, the antisymmetric part of the band deformation is described by the emergence of the momentum magnetic toroidal multipole $T^{\rm (m)}_{lm}(\bm{k})$ solely without spin dependence~\cite{dubovik1990toroid,kopaev2009toroidal,Yanase_JPSJ.83.014703,Hayami_PhysRevB.90.024432}. 
The schematic example in the case of $T^{\rm (m)}_{x}(\bm{k}) \sim k_x$ is shown in Fig.~\ref{Fig:band}(c).

To clarify a necessary condition of the microscopic model parameters for the band deformations in AFM orderings beyond symmetry argument, one can need to know when and how the momentum multipoles, $Q^{\rm (m)}_{lm}(\bm{k})$ and $T^{\rm (m)}_{lm}(\bm{k})$, become active and are coupled with spin $\sigma$. 
To examine such conditions, we introduce the following quantity at wave vector $\bm{k}$ in the magnetic unit cell, 
\begin{align}
\label{eq:expec_spin}
\mathrm{Tr}[e^{-\beta \mathcal{H}_{\bm{k}}} \sigma_\mu]
= \sum_{s} \frac{(-\beta)^{s}}{s!}
g_s^{\mu}(\bm{k}), 
\end{align} 
where $\mu=0,x,y,z$, $\mathcal{H}=\sum_{\bm{k}}\mathcal{H}_{\bm{k}}$ and $\beta$ is the inverse temperature.
By means of a sort of high-temperature expansion, the $s$th order expansion coefficient of the $\mu$-component, $g_s^{\mu}(\bm{k})$, gives the corresponding effective multipole coupling as $g_{s}^{\mu}(\bm{k})\sigma_{\mu}/2$.
As the Hamiltonian in Eq.~(\ref{eq:Ham_mul}) consists of the cluster and bond multipoles in the matrix form, the $s$th order expansion of $e^{-\beta \mathcal{H}_{\bm{k}}}$ can be described by the product of the $s$-tuple of the matrices, $Q^{\rm (b)}_{lm}$, $T^{\rm (b)}_{lm}$, and $Q^{\rm (c)}_{lm}$. 
It is noted that the $\bm{k}$ dependence arises from the momentum multipoles $f^{Q_{lm}}(\bm{k})$ and $f^{T_{lm}}(\bm{k})$, which are always coupled with the bond multipoles in the scalar form as Eqs.~(\ref{eq:Ham_bondQ}) and (\ref{eq:Ham_bondT}). 
This analysis can be applied to not only $\bm{Q}=\bm{0}$ orderings but also finite commensurate $\bm{Q}$ orderings by choosing the appropriate minimal magnetic unit cell.

We present microscopic conditions for the band deformations from a multipole viewpoint in the cases of symmetric spin splitting, antisymmetric spin splitting, and antisymmetric band deformation with spin degeneracy in Secs.~\ref{sec:Symmetric spin splitting}, \ref{sec:Antisymmetric spin splitting}, and \ref{sec:Antisymmetric band deformation without spin splitting}, respectively.

\subsubsection{Symmetric spin splitting}
\label{sec:Symmetric spin splitting}

The symmetric spin splitting, $g_s^{\mu}(\bm{k})=g_s^{\mu}(-\bm{k})$ for $\mu=x,y,z$, occurs under the presence of the spatial inversion symmetry and the absence of the time-reversal symmetry.
The conditions for the symmetric spin splitting are obtained by considering the space-time inversion properties ($\mathcal{P},\mathcal{T}$); the product of $s$-tuple of multipoles, which consists of the coupling between the bond and cluster multipoles, must have the same parities as those of the symmetric spin splitting, i.e., $(\mathcal{P},\mathcal{T})=(+1,-1)$. 

Since the bond multipoles consist of electric multipoles with $(\mathcal{P}, \mathcal{T})=(+1,+1)$ and magnetic toroidal multipoles with $(\mathcal{P}, \mathcal{T})=(-1,-1)$, while the cluster multipoles (electric multipoles) coupled with spin with $(\mathcal{P}, \mathcal{T})=(+1,-1)$, we obtain the following conditions for the product of $s$-tuple of multipoles to realize the symmetric spin splitting: 
\begin{enumerate}
\item Bond electric multipoles or even number of bond magnetic toroidal multipoles are involved. 
\item Odd number of cluster electric multipoles are involved.  
\item Trace of the sublattice degree of freedom (product of cluster multipoles) remains finite. 
\end{enumerate}
The conditions (i) and (ii) are required from the symmetry of the symmetric spin-split band dispersions, as mentioned above.
The condition (ii) indicates that only the symmetric spin splitting occurs in collinear magnets. 
From the condition (iii), one can find that the symmetric spin splitting can occur when $H^{Q}_t$ and $H_m$ contain the same symmetry of electric multipoles, while the term $H^{T}_t$ is not necessary. 
In other words, the momentum electric multipole can be coupled with $\sigma$ through the higher-order coupling between the bond electric multipoles and cluster electric multipoles, which is necessary to yield the symmetric spin splitting. 

Let us look at the example in an AFM with the collinear order parameter $\sum_{lm} h_{lm}^z Q^{\rm (c)}_{lm}\sigma_z$. 
The lowest-order contribution to Eq.~(\ref{eq:expec_spin}) arises from the third order, which is proportional to 
\begin{align}
{\rm Tr}\left[\left\{H_t , \left\{ H_t,  H_m\right\}\right\}\sigma_{\mu}\right] 
\propto
m_{lm}^{\mu}{\rm Tr}\left[\left\{H_t , \left\{ H_t,  Q_{lm}^{\rm (c)}\right\}\right\}\right],
\end{align}
where $\{ \cdots \}$ is the anticommutator and $H_t=H_t^Q + H_t^T$. 
Here, the nonzero anticommutator between $H^{Q}_t$ (or $H^{T}_t$) and $H_m$ is essential to give rise to the spin-split band structure, since it gives a nontrivial coupling between the kinetic motions of electrons and the spin textures instead of the SOC.
We can use the following relations among $Q^{\rm (b)}_{lm}$, $T^{\rm (b)}_{lm}$, and $Q^{\rm (c)}_{lm}$,
\begin{align}
\label{eq:anticommu}
\left\{ Q^{\rm (b)}_{l'm'},  Q^{\rm (c)}_{l''m''} \right\} &= \sum_{lm} c_{lm} Q^{\rm (b)}_{lm}, \nonumber \\
\left\{ T^{\rm (b)}_{l'm'},  Q^{\rm (c)}_{l''m''} \right\} &= \sum_{lm} c_{lm} T^{\rm (b)}_{lm}, \nonumber \\
\left\{ Q^{\rm (b)}_{l'm'},  Q^{\rm (b)}_{l''m''} \right\} &= \sum_{lm} c_{lm} Q^{\rm (b)}_{lm}+\sum_{lm} c'_{lm} Q^{\rm (c)}_{lm}, \nonumber \\
\left\{ T^{\rm (b)}_{l'm'},  T^{\rm (b)}_{l''m''} \right\} &= \sum_{lm} c_{lm} Q^{\rm (b)}_{lm}+\sum_{lm} c'_{lm} Q^{\rm (c)}_{lm}, \nonumber \\
\left\{ Q^{\rm (b)}_{l'm'},  T^{\rm (b)}_{l''m''} \right\} &= \sum_{lm} c_{lm} T^{\rm (b)}_{lm}, 
\end{align}
where $c_{lm}$ and $c'_{lm}$ are expansion coefficients. 
These expressions are obtained from the comparison of the spatial and time-reversal parities of electric and magnetic toroidal multipoles in both sides. 
We omit the indices $l'm'$ and $l''m''$ of $c_{lm}$ and $c'_{lm}$ for notational simplicity.
By using the first and third relations in Eq.~(\ref{eq:anticommu}), one can easily find that
\begin{align*}
&
g_{s}^{\mu}(\bm{k})\sim m_{lm}^{\mu}
f^{Q_{0}}(\bm{k})f^{Q_{l'm'}}(\bm{k})
\cr&\quad\times
{\rm Tr}\left[\left\{Q^{\rm (b)}_{0}, \left\{ Q^{\rm (b)}_{l'm'}, Q^{\rm (c)}_{lm} 
\right\}\right\}\right]
\end{align*}
becomes nonzero only when $H_t$ and $H_m$ contain the same symmetry of multipole, i.e., $l'=l$, $m'=m$. 
Then, the functional form of the spin splitting is given by $f^{Q_{0}}(\bm{k})f^{Q_{lm}}(\bm{k})  \sim Q^{\rm (m)}_{lm}(\bm{k})$. 
In other words, the functional form of the spin splitting is characterized by the higher-rank momentum electric multipole $Q_{lm}^{\rm (m)}(\bm{k})$. 
Note that the bond magnetic toroidal multipoles can also contribute to the spin splitting by the effective coupling as $m_{lm}^{\mu}{\rm Tr}\left[\left\{T^{\rm (b)}_{l'm'}, \left\{ T^{\rm (b)}_{l''m''},  Q^{\rm (c)}_{lm}\right\}\right\}\right]$ when $T^{\rm (b)}_{l'm'}T^{\rm (b)}_{l''m''}$ belongs to the same irrep. as $Q^{\rm (c)}_{lm}$.

\subsubsection{Antisymmetric spin splitting}
\label{sec:Antisymmetric spin splitting}

In contrast to the symmetric spin splitting, the antisymmetric spin splitting, $g_s^{\mu}(\bm{k})=-g_s^{\mu}(-\bm{k})$ for $\mu=x,y,z$, occurs only in noncollinear magnets.  
This is because in collinear magnets without the spin-orbit coupling the spin rotational operation [$(\bm{k}, \sigma) \to (\bm{k},-\sigma)$] with combining the time-reversal operation [$(\bm{k}, \sigma) \to (-\bm{k},-\sigma)$] ensures the spatial inversion symmetry [$(\bm{k}, \sigma) \to (-\bm{k},\sigma)$]~\cite{hayami2019momentum}. 

By similar argument as in the symmetric spin splitting, the conditions for the product of $s$-tuple of multipoles are given as follows:   
\begin{enumerate}
\item Odd number of bond magnetic toroidal multipoles are involved. 
\item At least, two spin components leading to noncollinear spin textures are involved. 
\item Trace of the sublattice degree of freedom (product of cluster multipoles) remains finite. 
\end{enumerate}
The conditions (i) and (ii) are required from the antisymmetric spin-split band dispersions under the breakings of spatial, time-reversal, and their product symmetries, which are obtained from the similar analysis in Sec.~\ref{sec:Symmetric spin splitting}. 
From the above conditions, the emergence of the antisymmetric spin splittings are due to the effective coupling between the bond magnetic toroidal multipoles and cluster electric multipoles. 

Let us look at the example in a noncollinear AFM with the order parameter $\sum_{lm} (h_{lm}^x Q_{lm}^{\rm (c)}\sigma_x + h_{lm}^y Q_{lm}^{\rm (c)}\sigma_y)$. 
One of the contributions comes from the fifth order, which is proportional to 
\begin{align}
\label{eq:ass_5th}
{\rm Tr}[\left\{H_t , \left\{\left\{ H_t,  H_m\right\},\left\{ H_t,  H_m\right\}\right\}\right\}\sigma_{z}]. 
\end{align}
In contrast to Eq.~(\ref{eq:anticommu}) where the spin is simply traced out leaving the anticommutator without spin dependence, the plural $H_{m}$ terms depending on different component of spins are involved in this case. 
All the necessary anticommutator appearing in Eq.~(\ref{eq:expec_spin}) are represented in the form, 
\begin{align}
\label{eq:XY}
\left\{
 X_{lm} \sigma_\mu, Y_{l'm'}\sigma_\nu 
\right\}
&= \left\{ X_{lm}, Y_{l'm'} \right\}\delta_{\mu,\nu} \sigma_0 \nonumber \\
&+ i \left[ X_{lm}, Y_{l'm'}\right] \sum_{\kappa} \varepsilon_{\mu\nu \kappa}\sigma_{\kappa}, 
\end{align}
where $[ \cdots ]$ is the commutator, and $i \left[ X_{lm}, Y_{l'm'}\right]$ is the hermite matrix. 
From the fact that the imaginary unit $i$ represents the time-reversal-odd scalar and the antisymmetric tensor $\varepsilon_{\mu\nu\kappa}$ changes the sign of the spatial parity, the commutation relation is given as follows: 
\begin{align}
\label{eq:anticommu2}
i\left[ Q^{\rm (c)}_{l'm'},  Q^{\rm (c)}_{l''m''} \right] &= 0, \nonumber \\
i\left[ Q^{\rm (b)}_{l'm'},  Q^{\rm (c)}_{l''m''} \right] &= \sum_{lm} c_{lm} T^{\rm (b)}_{lm}, \nonumber \\
i\left[ T^{\rm (b)}_{l'm'},  Q^{\rm (c)}_{l''m''} \right] &= \sum_{lm} c_{lm} Q^{\rm (b)}_{lm}, \nonumber \\
i\left[ Q^{\rm (b)}_{l'm'},  Q^{\rm (b)}_{l''m''} \right] &= \sum_{lm} c_{lm} T^{\rm (b)}_{lm}, \nonumber \\
i\left[ T^{\rm (b)}_{l'm'},  T^{\rm (b)}_{l''m''} \right] &= \sum_{lm} c_{lm} T^{\rm (b)}_{lm}, \nonumber \\
i\left[ Q^{\rm (b)}_{l'm'},  T^{\rm (b)}_{l''m''} \right] &= \sum_{lm} c_{lm} Q^{\rm (b)}_{lm}+\sum_{lm} c'_{lm} Q^{\rm (c)}_{lm}, 
\end{align}
where $c_{lm}$ and $c'_{lm}$ are expansion coefficients where we again omit their indices, $l'm'$ and $l''m''$. 
By using the first and second relations in Eq.~(\ref{eq:anticommu}) and the sixth relation in Eq.~(\ref{eq:anticommu2}), we find that
\begin{align*}
{\rm Tr}\left[\left\{Q^{\rm (b)}_0 , \left\{\left\{ Q^{\rm (b)}_0,  Q^{\rm (c)}_{lm}\sigma_x \right\},\left\{ T^{\rm (b)}_{l'm'},  Q^{\rm (c)}_{l''m''} \sigma_y \right\}\right\}\right\}\sigma_z \right]
\end{align*}
becomes nonzero for the $\sigma_z$ component perpendicular to the coplanar magnetic moments. 
The functional form of the spin splitting is then given by the active magnetic toroidal multipole $f^{T_{l'm'}}(\bm{k}) (f^{Q_{0}}(\bm{k}))^2 \sim T^{\rm (m)}_{l'm'}(\bm{k})$. 
Note that the other multipole coupling can also contribute to the spin splitting, e.g.,
\begin{align*}
{\rm Tr}\left[T^{\rm (b)}_{l'''m'''} , \left\{\left\{ T^{\rm (b)}_{l''''m''''},  Q^{\rm (c)}_{lm}\sigma_x \right\},\left\{ T^{\rm (b)}_{l'm'},  Q^{\rm (c)}_{l''m''} \sigma_y \right\}\right\}\sigma_z \right]
\end{align*}
as long as the quantity remains finite after tracing them out.

\subsubsection{Antisymmetric band deformation with spin degeneracy} 
\label{sec:Antisymmetric band deformation without spin splitting}
Finally, we discuss the antisymmetric band deformation with spin degeneracy, $g_s^{0}(\bm{k})=-g_s^{0}(-\bm{k})$, which occurs in noncoplanar magnets without spatial inversion symmetry. 
The conditions for the effective multipole couplings are given as follows: 
\begin{enumerate}
\item Odd number of bond magnetic toroidal multipoles are involved. 
\item Three spin components, which are necessary to represent noncoplanar spin structures, are involved.  
\item Trace of the sublattice and spin degrees of freedom remains finite. 
\end{enumerate}
The conditions (i) and (ii) are required to satisfy the symmetry for the antisymmetric band deformations. 
The condition (iii) indicates that the spin dependence is not important. 

We show the example in a noncoplanar AFM with the order parameter $\sum_{lm} (h_{lm}^x Q_{lm}^{\rm (c)}\sigma_x + h_{lm}^y Q_{lm}^{\rm (c)}\sigma_y+ h_{lm}^z Q_{lm}^{\rm (c)}\sigma_z)$. 
One of the six-order contributions to the antisymmetric band deformation is given by 
\begin{align}
{\rm Tr}[
\left\{\left\{ H_t,  H_m\right\} , \left\{\left\{ H_t,  H_m\right\},\left\{ H_t,  H_m\right\}\right\}\right\}]
. 
\end{align}
By using Eq.~(\ref{eq:XY}), the contribution
\begin{align*}
{\rm Tr}\left[\left\{\left\{Q^{\rm (b)}_0,  Q^{\rm (c)}_{lm}\sigma_z\right\} , \left\{\left\{ Q^{\rm (b)}_0,  Q^{\rm (c)}_{lm}\sigma_x \right\},\left\{ T^{\rm (b)}_{l'm'},  Q^{\rm (c)}_{l''m''} \sigma_y \right\}\right\}\right\} \right]
\end{align*}
can remain finite. 
Although the noncoplanar magnets are rare as compared to the coplanar magnets, the antisymmetric band deformations can also be realized by applying the magnetic field to the coplanar AFMs without the spatial inversion symmetry in the out-of-plane-moment direction~\cite{Hayami_PhysRevB.101.220403}. 

We summarize the functional form of the band deformations and related magnetic textures in  Table~\ref{tab_band_mp}. 

\begin{table}
\caption{
Three types of band deformations and their functional forms: symmetric spin splitting (SS), antisymmetric spin splitting, and antisymmetric band deformation (BD).  
The necessary conditions of space-time parities of the system and magnetic textures are also shown. 
}
\label{tab_band_mp}
\centering
\begin{tabular}{cccccccc} \hline\hline
Type & form
& $\mathcal{P}$& $\mathcal{PT}$ & magnetic textures \\ \hline
Symmetric SS  & $f^{Q_{lm}}(\bm{k})\sigma_\mu$ & $\circ$ & $\times$ & collinear \\
Antisymmetric SS & $f^{T_{lm}}(\bm{k})\sigma_\mu$ & $\times$ & $\times$ & coplanar\\
Antisymmetric BD& $f^{T_{lm}}(\bm{k})$ & $\times$ & $\circ$ & noncoplanar\\
\hline\hline
\end{tabular}
\end{table}

\section{Application to triangular lattice systems}
\label{sec:Application to lattice systems}

We apply the present scheme to specific lattice systems. 
We take three examples consisting of a triangle cluster: triangular, kagome, and breathing kagome structures. 
After introducing multipole degrees of freedom in the triangle cluster in Sec.~\ref{sec:Triangle cluster}, we show that spin splittings and band deformations are induced by the 120$^{\circ}$ AFM ordering on three specific lattices.
We present the antisymmetric spin splitting on a triangular lattice in Sec.~\ref{sec:Triangular}, the symmetric spin splitting on a kagome lattice in Sec.~\ref{sec:Kagome}, and symmetric and antisymmetric spin splittings on a breathing kagome lattice in Sec.~\ref{sec:Breathing kagome}. 
We also show the effect of an external magnetic field on the noncollinear ordering on a breathing kagome lattice in Sec.~\ref{sec:magnetic field}.

\subsection{Triangle cluster}
\label{sec:Triangle cluster}

\begin{figure}[h!]
\begin{center}
\includegraphics[width=1.0 \hsize]{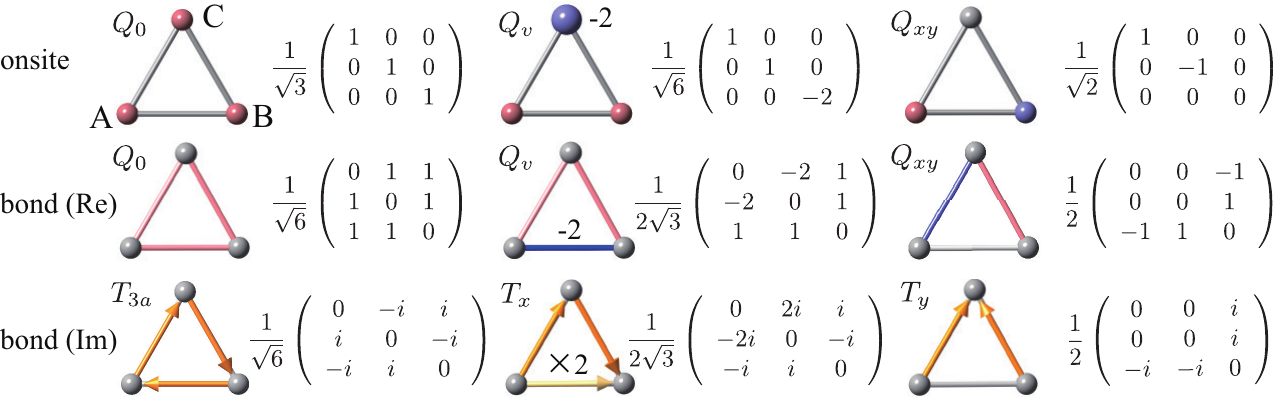} 
\caption{
\label{Fig:triangle}
Cluster and bond multipoles in a triangle cluster~\cite{Hayami_PhysRevB.101.220403}. 
The correspondence between multipoles and matrix elements is shown. 
The red (blue) circles represent the positive (negative) onsite potential, and the red (blue) lines and orange arrows on each bond represent the positive (negative) real and imaginary hoppings, respectively. 
The gray lines represent no hoppings. 
}
\end{center}
\end{figure}

We consider a triangle cluster whose sublattice basis function consists of $(\ket{{\rm A}}, \ket{{\rm B}}, \ket{{\rm C}})$ as shown in Fig.~\ref{Fig:triangle}. 
This cluster belongs to the point group $D_{3{\rm h}}$ and has nine multipole degrees of freedom.
From the irreducible decomposition in Table~\ref{tab_IRREP} and corresponding multipole table in Table~\ref{tab_multipoles_table_6mmm}, three onsite degrees of freedom with $A'^+_1 \oplus E'^+$ correspond to 
$Q_0^{\rm (c)}$, $Q_v^{\rm (c)}$, and $Q_{xy}^{\rm (c)}$, three real bond degrees of freedom with $A'^+_1 \oplus E'^+$ correspond to $Q_0^{\rm (b)}$, $Q_v^{\rm (b)}$, and $Q_{xy}^{\rm (b)}$, and three imaginary bond degrees of freedom with $A'^-_2 \oplus E'^-$ correspond to $T_{3a}^{\rm (b)}$, $T_x^{\rm (b)}$, and $T_{y}^{\rm (b)}$. 
It is noted that there are two settings in choosing the $C'_2$ rotational axis in $D_{3{\rm h}}$. 
Here, we take the $y$ axis as the $C'_2$ rotational axis (See the column $D_{3{\rm h}}$ in Table~\ref{tab_multipoles_table_6mmm}). 

The specific matrix elements for each multipole are shown in Fig.~\ref{Fig:triangle}~\cite{comment_Erep}. 
In the following sections, we assume the noncollinear 120$^{\circ}$ AFM magnetic structure with the form of $Q^{\rm (c)}_{xy}\sigma_x +Q^{\rm (c)}_{v} \sigma_y$ on the triangular, kagome, and breathing kagome structures. 
We implicitly assume that the spin rotational symmetry is spontaneously broken through the phase transition. 

\subsection{Triangular}
\label{sec:Triangular}

\begin{figure}[h!]
\begin{center}
\includegraphics[width=1.0 \hsize]{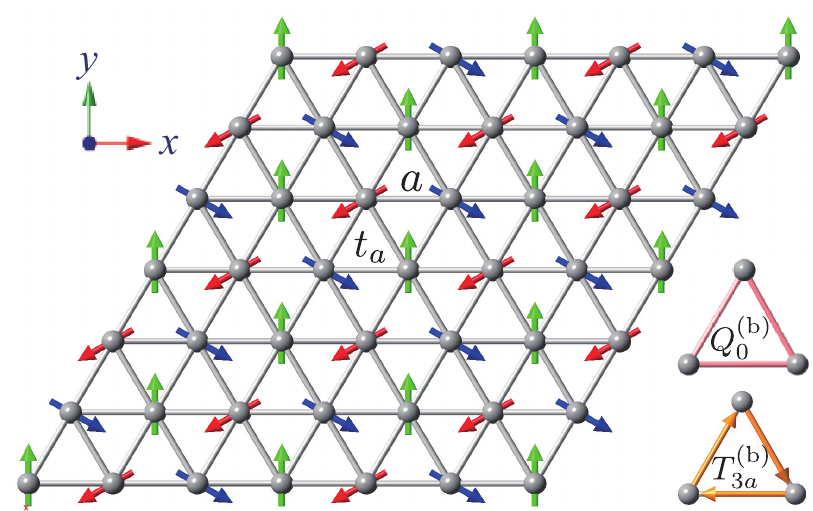} 
\caption{
\label{Fig:TL}
Schematic pictures of the 120$^{\circ}$ AFM on a triangular lattice. 
The active multipoles are also shown. 
}
\end{center}
\end{figure}

\begin{figure}[h!]
\begin{center}
\includegraphics[width=1.0 \hsize]{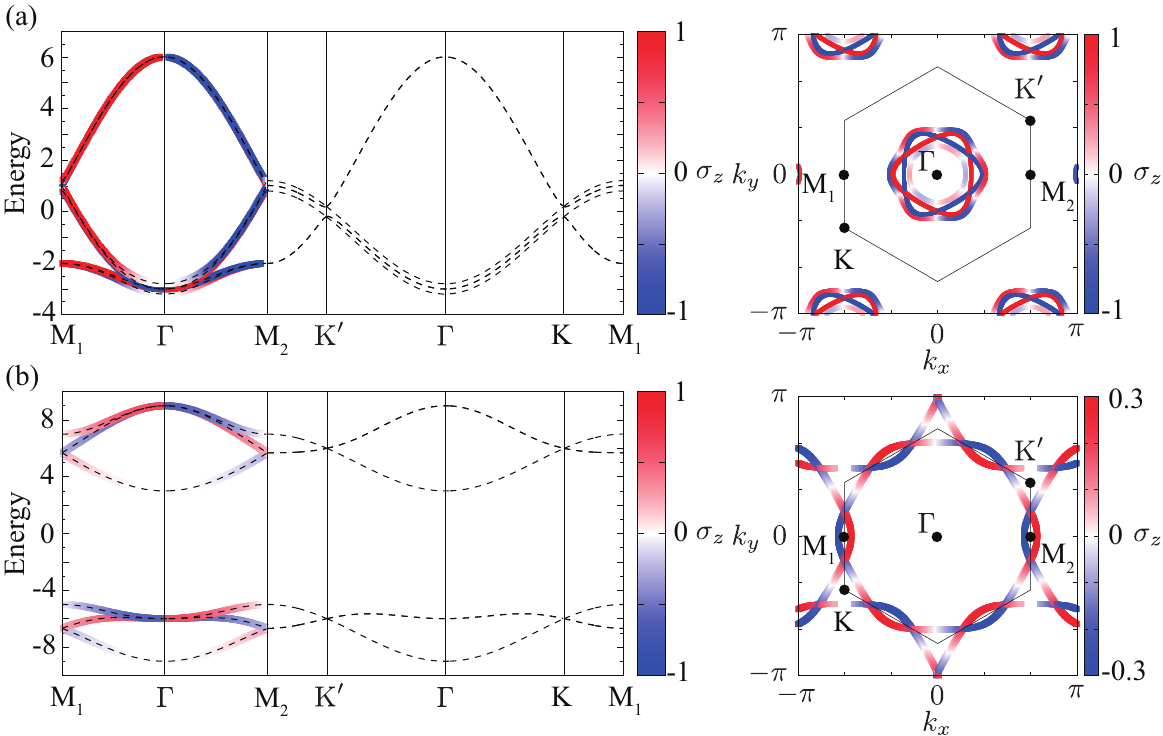} 
\caption{
\label{Fig:TL_band}
(Left panel) The band structure of the model on the triangular lattice at (a) $m = 0.2$ and (b) $m=6$.
The other model parameter is $t_a=1$. 
The dashed lines show the band dispersions and the color map shows the spin polarization of the $z$ component at each wave vector. 
(Right panel) The isoenergy surfaces at $\mu=-2.5$ and $\mu=-6.5$. 
The hexagon in the right panel represents the magnetic first Brillouin zone. 
}
\end{center}
\end{figure}

First, we consider the triangular lattice with the lattice constant $a$, as shown in Fig.~\ref{Fig:TL}. 
It is noted that the symmetry of the triangular lattice $D_{6 {\rm h}}$ is different from that of the triangle cluster $D_{3 {\rm h}}$, both of which belong to the same Laue class $6/mmm$, as shown in Table~\ref{tab_multipoles_table_6mmm}. 
In this case, from compatibility relation from $D_{\rm 3h}$ to $D_{\rm 6h}$, the irrep. should be replaced as $A'^+_1 \to A^+_{1g}$, $A''^-_{1} \to A_{1u}^-$, and so on. 
Meanwhile, by looking the correspondence between the multipoles and the irrep. in the same row in Table~\ref{tab_multipoles_table_6mmm}, one can find that the same multipole notations, e.g., $Q_{4a}$, $Q_{zx}$, are used for cluster and lattice systems.

The matrices of the hopping and mean-field Hamiltonians in the three-sublattice triangular system are given by 
\begin{align}
&H^{Q}_{t}=
f^{Q_{0}}(\bm{k})Q^{\rm (b)}_{0},
\cr&
H^{T}_{t}=
f^{T_{3a}}(\bm{k})T^{\rm (b)}_{3a},
\cr&
H_{m}=-m (Q^{\rm (c)}_{xy} \sigma_x +Q^{\rm (c)}_{v} \sigma_y), 
\label{eq:ham_TL}
\end{align}
where the form factors for the nearest-neighbor site are represented by 
\begin{align}
f^{Q_{0}}(\bm{k})&=\sqrt{6}t_{a}(\cos k_{x}a+2\cos\tilde{k}_{x}a\cos\tilde{k}_{y}a), \nonumber \\
f^{T_{3a}}(\bm{k})&=-\sqrt{6}t_a(\sin k_xa-2\sin\tilde{k}_{x}a\cos\tilde{k}_{y}a), 
\end{align}
with the hopping amplitude $t_a$. 
Here and hereafter, we use the abbreviated notations $\tilde{k}_{x}=k_{x}/2$ and $\tilde{k}_{y}=\sqrt{3}k_{y}/2$. 
We consider the first-neighbor hopping in Eq.~(\ref{eq:ham_TL}), which is expressed by the electric monopole and magnetic toroidal octupole degrees of freedom. 
The presence of magnetic toroidal multipole, $T^{\rm (b)}_{3a}$, is attributed to the introduction of the sublattice degree of freedom by taking into account the magnetic unit cell, and it does not exist in the case of a single-site unit cell.
As we will show below, $T^{\rm (b)}_{3a}$ plays an important role for the emergent spin splitting as a result from the coupling with the noncollinear three-sublattice magnetic structures.
The mean-field matrix $H_m$ consists of two spin components to express the $120^{\circ}$ noncollinear magnetic order with the amplitude $m$. 

It is noted that the active bond multipoles appearing in the hopping matrices, $H_t^Q$ and $H_t^T$ depend on the nature of hopping and the choice of the magnetic unit cell. 
For example, the further neighbor hoppings may bring about the other types of electric and magnetic toroidal multipoles, as shown in Table~\ref{tab_IRREP}. 
Nevertheless, in the present triangular-lattice case, the further neighbor hoppings do not give rise to the other multipoles due to the lattice symmetry. 
Thus, the symmetric spin splitting does not appear even by taking account of further neighbor hoppings due to the lack of higher-rank electric multipoles in $H_t^Q$.
On the other hand, the antisymmetric spin splitting can occur according to the conditions given in Sec.~\ref{sec:Antisymmetric spin splitting}.
The lowest-order contribution is given by 
\begin{align}
\label{eq:spinsplit_asym_TL}
g^{z}_5(\bm{k}) =
-\frac{1}{3} \sqrt{\frac{2}{3}} m^2 f^{T_{3a}}(\bm{k}) \left[ (f^{T_{3a}}(\bm{k}))^2-3  (f^{Q_0}(\bm{k}))^2\right].   
\end{align}
As $f^{T_{3a}}(\bm{k}) \propto k_x(k_x^2-3k_y^2)$ and $f^{Q_0}(\bm{k}) \propto 1$ in the $\bm{k}\to \bm{0}$ limit, the essential anisotropy is given by 
\begin{align}
\label{eq:spinsplit_asym_TL2}
g^{z}_5(\bm{k}) &\simeq
 \sqrt{\frac{2}{3}} m^2 (f^{Q_0}(\bm{k}))^2 f^{T_{3a}}(\bm{k})  \nonumber \\
 &=24 m^2 t_a^3 \sin \tilde{k}_x  \left(\cos \tilde{k}_y -\cos \tilde{k}_x \right) \nonumber \\
 &  \ \ \  \times \left(2 \cos \tilde{k}_x \cos \tilde{k}_y+\cos k_x\right)^2 \nonumber \\
 &\simeq  \frac{27}{2} m^2 t_a^3  k_x \left(k_x^2-3 k_y^2\right)a^3. 
\end{align} 
In this way, the functional form of the antisymmetric spin splitting satisfying the magnetic space group symmetry is obtained from the effective multipole coupling. 
Moreover, one can obtain the model parameter dependence for the spin splitting.
As is consistent with the discussion in Sec.~\ref{sec:Antisymmetric spin splitting}, the expressions in Eq.~(\ref{eq:spinsplit_asym_TL2}) contain the product of the even number of order parameters as $m^2$ and the bond magnetic toroidal multipole $T_{3a}^{\rm (m)}(\bm{k})$.
The opposite spin alignment is realized by reversing the vector spin chirality; the sign of one spin components in $H_{m}$ is reversed as $H_{m}=-m (Q^{\rm (c)}_{xy} \sigma_x -Q^{\rm (c)}_{v} \sigma_y)$.
This is consistent with the analysis in Eq.~(\ref{eq:ass_5th}), which results in opposite sign to Eq.~(\ref{eq:spinsplit_asym_TL2}).  

The effective multipole coupling leads to physical phenomena related with the inversion symmetry breaking~\cite{Hayami_PhysRevB.101.220403}. 
For example, the active magnetic toroidal multipoles in the form of $T^{(\rm m)}_{3a}(\bm{k})\sigma_{z}\sim k_x (k_x^2-3 k_y^2)\sigma_{z}$ in Eq.~(\ref{eq:spinsplit_asym_TL2}), implies that a spontaneous threefold rotational nonreciprocity is induced by a magnetic field along the $z$ direction if one divides it as $k_{x}(k_{x}^{2}-3k_{y}^{2})\times \sigma_{z}$.

The emergent antisymmetric spin splitting is confirmed by diagonalizing the Hamiltonian. 
We show the electronic band structure in Figs.~\ref{Fig:TL_band}(a) and \ref{Fig:TL_band}(b). 
The result clearly shows the spin splitting along the $M_1$-$\Gamma$-$M_2$ line, while there is no spin splitting along the $K$-$\Gamma$-$K'$ line irrespective of the value of $m$, which is consistent with Eq.~(\ref{eq:spinsplit_asym_TL2}).

\subsection{Kagome}
\label{sec:Kagome}

\begin{figure}[h!]
\begin{center}
\includegraphics[width=1.0 \hsize]{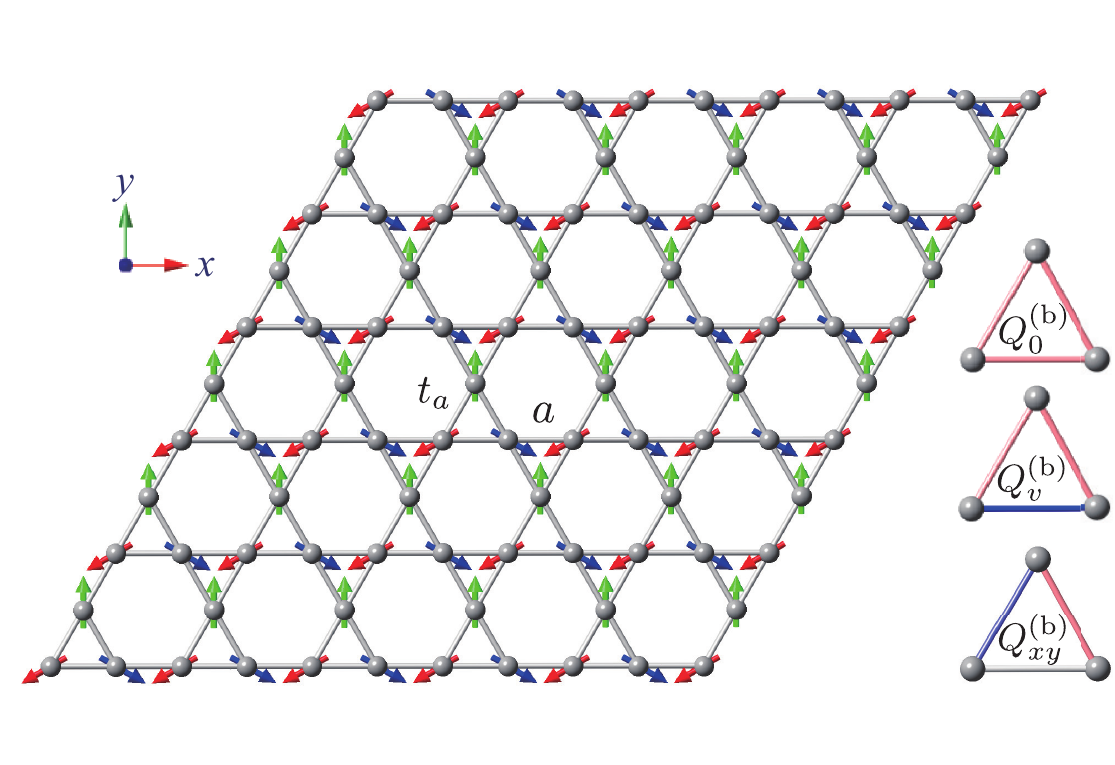} 
\caption{
\label{Fig:KL}
Schematic pictures of the 120$^{\circ}$ AFM on a kagome lattice. 
The active multipoles are also shown. 
}
\end{center}
\end{figure}

\begin{figure}[h!]
\begin{center}
\includegraphics[width=1.0 \hsize]{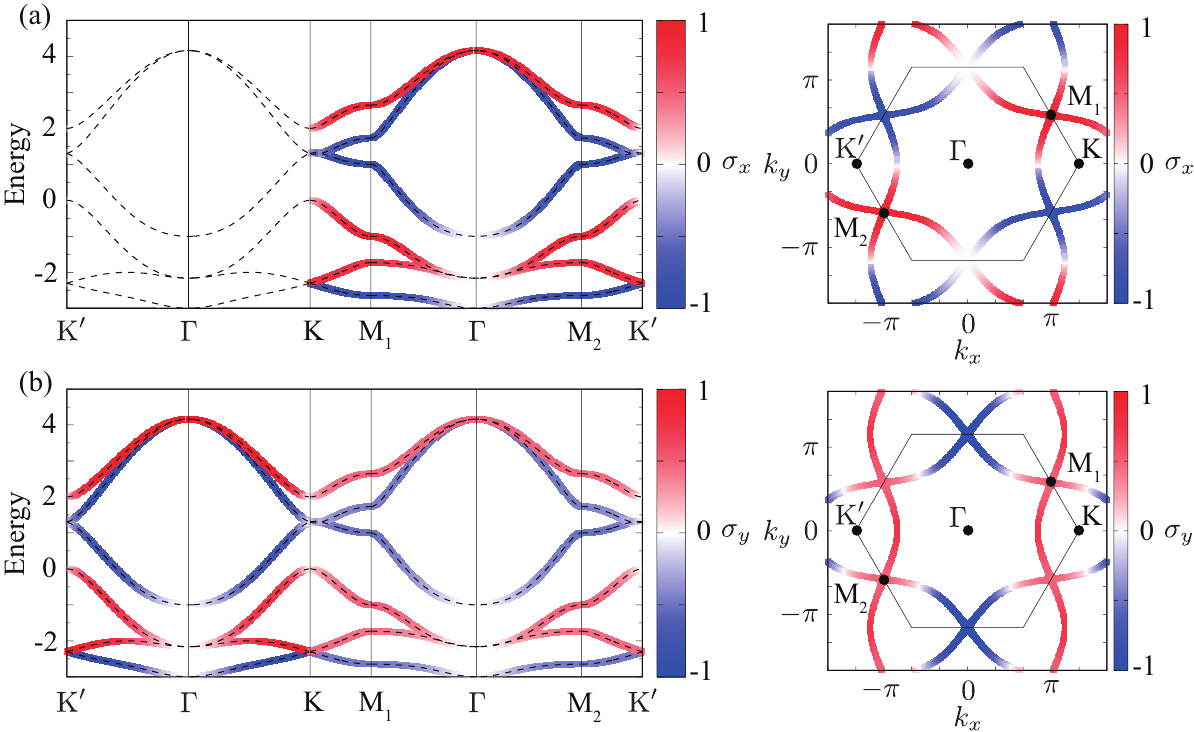} 
\caption{
\label{Fig:KL_band}
(Left panel) The band structure of the model on the kagome lattice at $t_a=1$ and $m = 1$.
The dashed lines show the band dispersions and the color map shows the spin polarization of the (a) $x$ and (b) $y$ components at each wave vector. 
(Right panel) The isoenergy surfaces at $\mu=-1$. 
The hexagon in the right panel represents the magnetic first Brillouin zone. 
}
\end{center}
\end{figure}

Next, we consider the 120$^{\circ}$ AFM on the kagome lattice with the lattice constant $2a$, as shown in Fig.~\ref{Fig:KL}. 
The point group of the kagome structure is $D_{6{\rm h}}$, which is the same as that of the triangular lattice. Owing to the different lattice geometry from the triangular lattice in the previous section, the different bond and momentum multipoles appear, as will be shown below.

The matrices of the hopping and mean-field Hamiltonians in the three-sublattice kagome system are given by 
\begin{align}
&H^{Q}_{t}=
f^{Q_{0}}(\bm{k})Q^{\rm (b)}_{0}+f^{Q_{v}}(\bm{k})Q^{\rm (b)}_{v}+f^{Q_{xy}}(\bm{k})Q^{\rm (b)}_{xy},
\cr&
H^{T}_{t}= 0,
\cr&
H_{m}=-m (Q^{\rm (c)}_{xy}\sigma_x +Q^{\rm (c)}_{v} \sigma_y), 
\label{eq:ham_KL}
\end{align}
where the form factors for the nearest-neighbor site are represented by 
\begin{align}
f^{Q_{0}}(\bm{k})&=2\sqrt{\frac{2}{3}}t_{a}(\cos k_{x}a+2\cos\tilde{k}_{x}a\cos\tilde{k}_{y}a), \nonumber \\ f^{Q_{v}}(\bm{k})&=\frac{4}{\sqrt{3}}t_{a}(\cos\tilde{k}_{x}a\cos\tilde{k}_{y}a-\cos k_{x}a), \nonumber \\ 
f^{Q_{xy}}(\bm{k})&=4t_{a}\sin\tilde{k}_{x}a\sin\tilde{k}_{y}a,
\end{align}
with the hopping amplitude $t_a$. 
There are two differences from the case in the triangular lattice in Sec.~\ref{sec:Triangular}: 
One is that the higher-rank electric multipoles are present, which indicates that the symmetric spin splitting can occur. 
The other is that there are no magnetic toroidal multipoles in the hopping matrix, since all the sublattice sites have the local inversion symmetry. 
Thus, any antisymmetric band deformations do not occur within the three-sublattice ordering in the kagome structure. 
We consider the 120$^{\circ}$AFM structure where the mean-field matrix $H_m$ is the same as that in the case of triangular lattice. 

The symmetric spin splitting due to the presence of $Q^{\rm (b)}_{v}$ and $Q^{\rm (b)}_{xy}$ is given by
\begin{align}
\label{eq:spinsplit_sym_KL_gx}
g^{x}_3(\bm{k}) &=m \left(2 f^{Q_0}(\bm{k}) f^{Q_{xy}}(\bm{k}) + \sqrt{2} f^{Q_{v}}(\bm{k}) f^{Q_{xy}}(\bm{k}) \right),\\
\label{eq:spinsplit_sym_KL_gy}
g^{y}_3(\bm{k}) &= m \left[2 f^{Q_0}(\bm{k}) f^{Q_{v}}(\bm{k}) \right. \nonumber \\
&\ \ \ \left.-\frac{1}{\sqrt{2}} \left\{(f^{Q_{v}}(\bm{k}))^2 - (f^{Q_{xy}}(\bm{k}))^2\right\}\right].   
\end{align}
As $f^{Q_0}(\bm{k}) \propto 1$, $f^{Q_v}(\bm{k}) \propto k_x^2-k_y^2$, and $f^{Q_{xy}}(\bm{k}) \propto k_x k_y$ in the $\bm{k}\to \bm{0}$ limit, the essential anisotropy is given by 
\begin{align}
\label{eq:spinsplit_asym_KL2_gx}
g^{x}_3(\bm{k}) &\simeq
2 m  f^{Q_0}(\bm{k}) f^{Q_{xy}}(\bm{k})  \simeq  12 \sqrt{2}  m t_a^2 k_x k_y a^2. \\
\label{eq:spinsplit_asym_KL2_gy}
g^{y}_3(\bm{k}) &\simeq
2 m f^{Q_0}(\bm{k}) f^{Q_{v}}(\bm{k}) \simeq   6 \sqrt{2}   m t_a^2 \left(k_x^2-k_y^2\right) a^2. 
\end{align} 
In contrast to the antisymmetric spin splitting in Eq.~(\ref{eq:spinsplit_asym_TL2}), $g^{x}_3(\bm{k})$ and $g^{y}_3(\bm{k})$ are proportional to $m$, which implies that the one spin component, i.e., the collinear spin structure, is sufficient to realize the symmetric spin splitting, as discussed in Sec.~\ref{sec:Symmetric spin splitting}. 
In fact, when we switch off one of the order parameters $m Q^{\rm (c)}_{v}=0$ or $m Q^{\rm (c)}_{xy}=0$, $g^{x}_3(\bm{k})$ or $g^{y}_3(\bm{k})$ remains finite, i.e., the symmetric spin splittings for $x$ and $y$ spin components are independent with each other. 
Moreover, one can find the opposite direction of the AFM moment results in the opposite spin splittings.  

The symmetric spin splitting as a result of the effective multipole coupling affects physical response tensors~\cite{Hayami_PhysRevB.98.165110}. 
For example, from the coupling between $\sigma_x$ and $Q_{xy}(\bm{k})\sim k_x k_y$ in Eq.~(\ref{eq:spinsplit_sym_KL_gx}), we can expect the magneto-elastic effect where a spontaneous $xy$-type shear stress is induced by a magnetic field along the $x$ direction or the spin-current generation where the spin current along the $x$ direction with the $x$-spin component is generated by an electric field along the $y$ direction~\cite{naka2019spin,hayami2019momentum}. 

The above analysis for the symmetric spin splitting is confirmed by calculating the electronic band structure.  Figures~\ref{Fig:KL_band}(a) and (b) show the band structure at $t_a=1$ and $m=1$ where the color map shows the spin polarization for the $x$ and $y$ components, respectively. 
The result clearly shows that the spin splittings are symmetric with respect to $\bm{k}$ and their functional forms are characterized by $Q^{\rm (m)}_{xy}(\bm{k}) \sigma_x$ in Fig.~\ref{Fig:KL_band}(a) and $Q^{\rm (m)}_{v}(\bm{k}) \sigma_y$ in Fig.~\ref{Fig:KL_band}(b).

\subsection{Breathing kagome}
\label{sec:Breathing kagome}

\begin{figure}[h!]
\begin{center}
\includegraphics[width=1.0 \hsize]{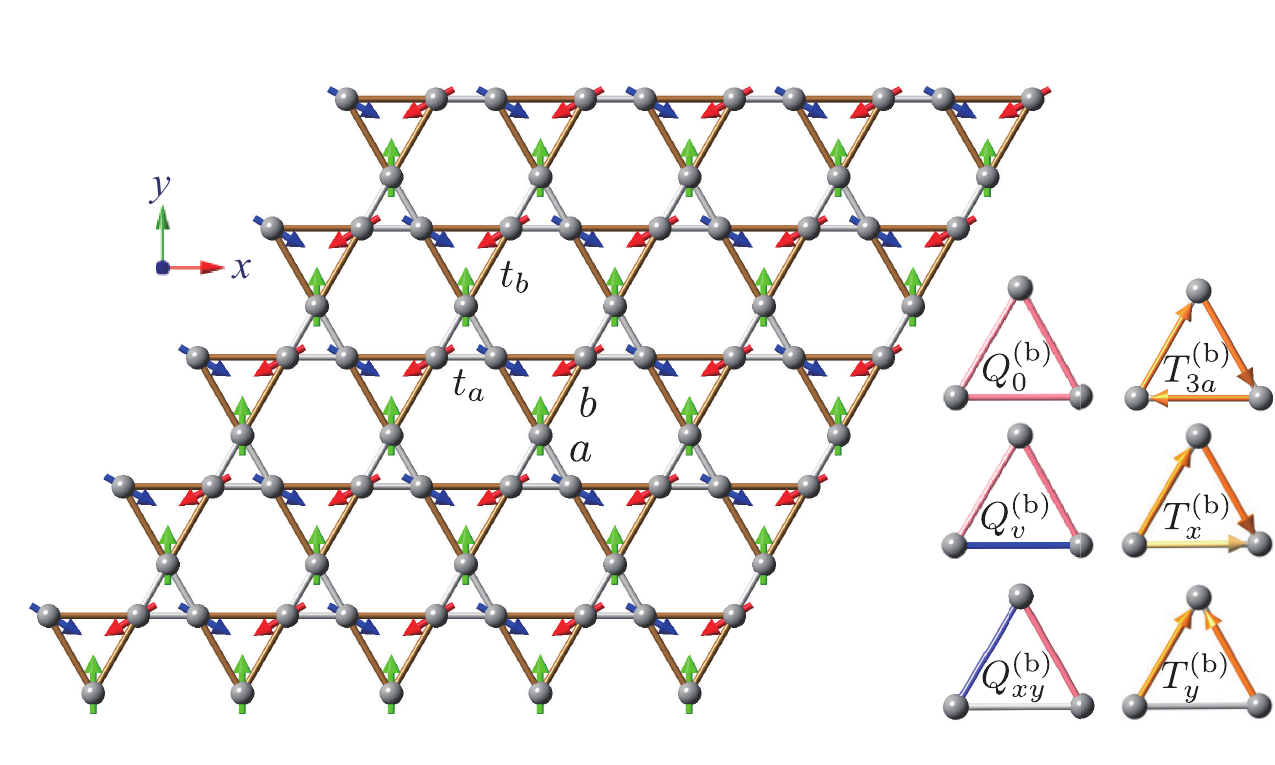} 
\caption{
\label{Fig:BKL}
Schematic pictures of the 120$^{\circ}$ AFM on a breathing kagome lattice. 
The active multipoles are also shown. 
}
\end{center}
\end{figure}

\begin{figure}[h!]
\begin{center}
\includegraphics[width=1.0 \hsize]{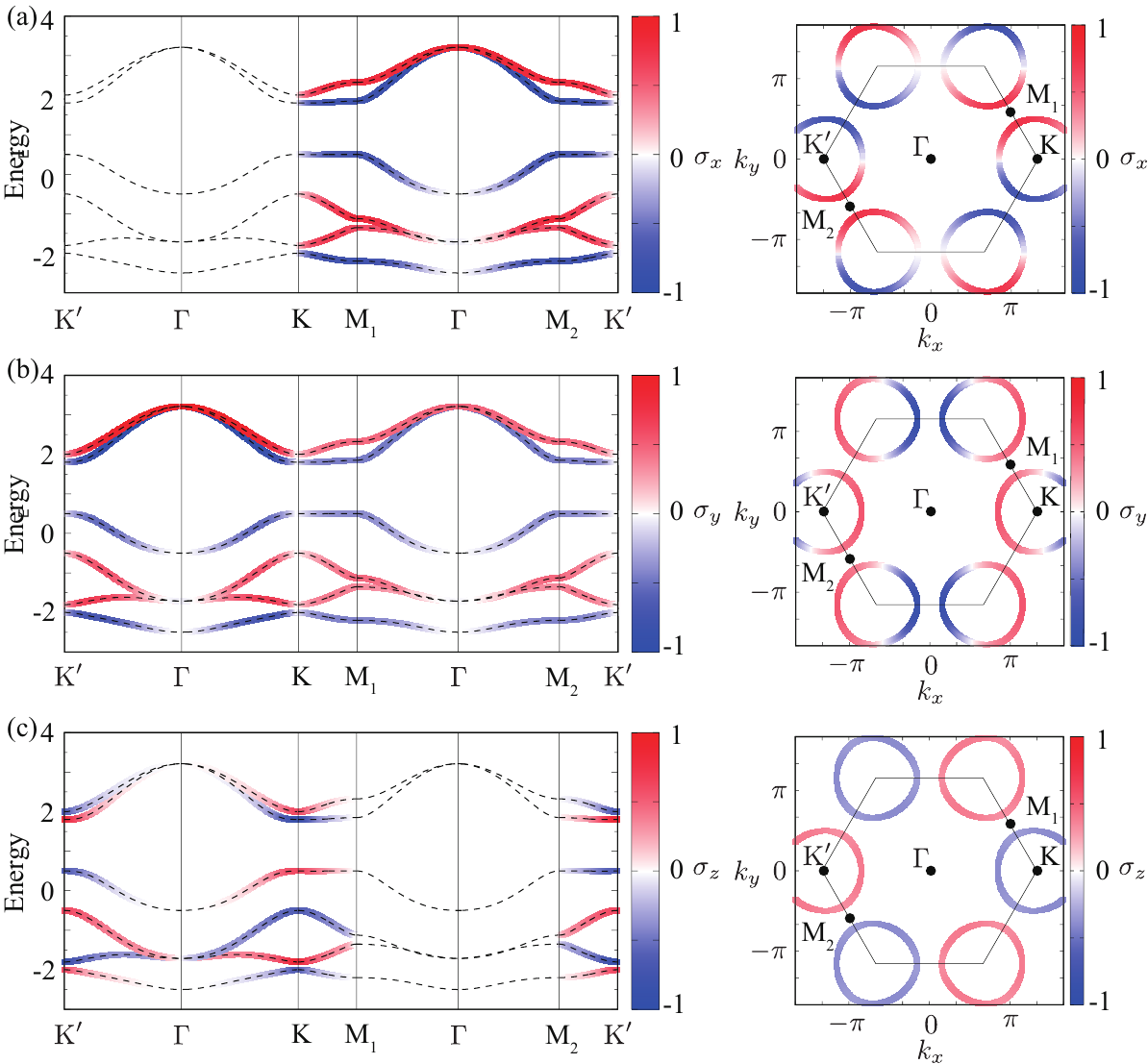} 
\caption{
\label{Fig:BKL_band}
(Left panel) The band structure of the model on the breathing kagome lattice at $t_a=1$, $t_b=0.5$, and $m = 1$.
The dashed lines show the band dispersions and the color map shows the spin polarization of the (a) $x$, (b) $y$, and $z$ components at each wave vector. 
(Right panel) The isoenergy surfaces at $\mu=-1$. 
The hexagon in the right panel represents the magnetic first Brillouin zone. 
}
\end{center}
\end{figure}

The last example is the 120$^{\circ}$ AFM on the breathing kagome lattice with the lattice constant $a+b$, as shown in Fig.~\ref{Fig:BKL}. 
The point group of the breathing kagome structure is $D_{3{\rm h}}$, which is the same as that of the triangular cluster. 

The matrices of the hopping and mean-field Hamiltonians in the three-sublattice breathing kagome system are given by 
\begin{align}
&H^{Q}_{t}=
f^{Q_{0}}(\bm{k})Q_{0}^{\rm (b)}+f^{Q_{v}}(\bm{k})Q_{v}^{\rm (b)}+f^{Q_{xy}}(\bm{k})Q_{xy}^{\rm (b)},
\cr&
H^{T}_{t}=f^{T_{3a}}(\bm{k})T_{3a}^{\rm (b)}+f^{T_{x}}(\bm{k})T_{x}^{\rm (b)}+f^{T_{y}}(\bm{k})T_{y}^{\rm (b)},
\cr&
H_{m}=-m (Q_{xy}^{\rm (c)}\sigma_x +Q_{v}^{\rm (c)}\sigma_y), 
\label{eq:Ham_BKL}
\end{align}
where the form factors are represented by 
\begin{align}
f^{Q_{0}}(\bm{k})&=\sqrt{\frac{2}{3}}\sum_{\eta}t_{\eta}(\cos k_{x}\eta+2\cos\tilde{k}_{x}\eta\cos\tilde{k}_{y}\eta), \nonumber \\ 
f^{Q_{v}}(\bm{k})&=\frac{2}{\sqrt{3}}\sum_{\eta}t_{\eta}(\cos\tilde{k}_{x}\eta\cos\tilde{k}_{y}\eta-\cos k_{x}\eta), \nonumber \\ 
f^{Q_{xy}}(\bm{k})&= 2\sum_{\eta}t_{\eta}\sin\tilde{k}_{x}\eta\sin\tilde{k}_{y}\eta, \nonumber \\ 
f^{T_{3a}}(\bm{k})&=-\sqrt{\frac{2}{3}}\sum_{\eta}p_{\eta}(\sin k_x\eta-2\sin\tilde{k}_{x}\eta\cos\tilde{k}_{y}\eta), \nonumber \\
f^{T_{x}}(\bm{k})&=\frac{2}{\sqrt{3}}\sum_{\eta}p_{\eta}(\sin k_{x}\eta+\sin\tilde{k}_{x}\eta\cos\tilde{k}_{y}\eta), \nonumber \\ 
f^{T_{y}}(\bm{k})&=2\sum_{\eta}p_{\eta}\cos\tilde{k}_{x}\eta\sin\tilde{k}_{y}\eta, 
\end{align}
for $\eta=a,b$, $p_a=t_a$, and $p_b=-t_b$.  
The hopping amplitudes are defined as $t_a$ within upward triangles and $t_b$ within downward triangles. 
The mean-field matrix $H_m$ is the same as that in Eqs.~(\ref{eq:ham_TL}) and (\ref{eq:ham_KL}). 

Owing to the presence of bond electric and magnetic toroidal multipoles for $l \geq 1$, both the symmetric and antisymmetric spin splittings can occur.
The lowest-order contribution to the symmetric spin splitting arises at $s=3$ in Eq.~(\ref{eq:expec_spin}) as  
\begin{align}
\label{eq:spinsplit_sym_BKL}
g^{x}_3(\bm{k}) &=m \left[2 f^{Q_0}(\bm{k}) f^{Q_{xy}}(\bm{k})+ \sqrt{2} f^{Q_v}(\bm{k}) f^{Q_{xy}}(\bm{k}) \right. \nonumber \\ 
& \ \ \ \left.
-2  f^{T_{3a}}(\bm{k}) f^{T_{y}}(\bm{k})- \sqrt{2} f^{T_x}(\bm{k}) f^{T_y}(\bm{k})\right], \\
g^{y}_3(\bm{k}) &=  m \left[2 f^{Q_0}(\bm{k}) f^{Q_v}(\bm{k}) -\frac{1}{\sqrt{2}} \left\{ (f^{Q_v}(\bm{k}))^2 - (f^{Q_{xy}}(\bm{k}))^2\right\} \right. \nonumber \\ 
& \ \ \ \left.
+2  f^{T_{3a}}(\bm{k}) f^{T_x}(\bm{k})-\frac{1}{\sqrt{2}} \left\{ (f^{T_x}(\bm{k}))^2-  (f^{T_y}(\bm{k}))^2\right\}\right].   
\end{align}
It is easily confirmed that the effective multipole couplings with electric multipoles are the same as those in Eqs.~(\ref{eq:spinsplit_sym_KL_gx}) and (\ref{eq:spinsplit_sym_KL_gy}). 
There are additional effective multipole couplings with magnetic toroidal multipoles. 
When the limit of $\bm{k}\to \bm{0}$ is taken, the essential anisotropy is given by 
\begin{align}
g^{x}_3(\bm{k}) &\simeq m \left[2 f^{Q_0}(\bm{k}) f^{Q_{xy}}(\bm{k}) - \sqrt{2} f^{T_x}(\bm{k}) f^{T_y}(\bm{k})\right] \nonumber \\ 
&\simeq \frac{6}{\sqrt{2}} m t_a t_b (a+b)^2  k_x k_y , \\
g^{y}_3(\bm{k}) &\simeq 
m \left[2 f^{Q_0}(\bm{k}) f^{Q_v}(\bm{k})  \right. \nonumber \\ 
& \ \ \ \left.
-\frac{1}{\sqrt{2}} \left\{ (f^{T_x}(\bm{k}))^2-  (f^{T_y}(\bm{k}))^2\right\}\right], \nonumber \\
&\simeq \frac{3}{\sqrt{2}} m t_a t_b (a+b)^2 \left(k_x^2 -k_y^2 \right).  
\end{align}
Also in this case, the functional forms are similar to those in the case of the kagome lattice in Eqs.~(\ref{eq:spinsplit_asym_KL2_gx}) and (\ref{eq:spinsplit_asym_KL2_gy}). 
In fact, $g^{x}_3(\bm{k})$ and $g^{y}_3(\bm{k})$ for the kagome and breathing kagome lattices are identical when we regard as $a+b \to 2a$ and $t_b \to t_a$. 

There are the contributions to the antisymmetric spin splitting in the $z$-component due to the presence of bond magnetic toroidal multipoles. 
The lowest-order contribution is obtained at the fifth order in Eq.~(\ref{eq:expec_spin}) as
\begin{align}
\label{eq:spinsplit_asym}
&g_{5}^{z}(\bm{k})
=\frac{m^2}{6 \sqrt{3}} \left(6 \sqrt{2} f^{Q_0}(\bm{k})^2 f^{T_{3a}}(\bm{k})
\right. \nonumber \\ 
& \ \ \ \left.
+6 \sqrt{2} f^{Q_0}(\bm{k}) (f^{Q_{xy}}(\bm{k}) f^{T_y}(\bm{k})-f^{Q_v}(\bm{k}) f^{T_x}(\bm{k}))
\right. \nonumber \\ 
& \ \ \ \left.
-3 \sqrt{2}  \left[f^{Q_v}(\bm{k})^2 +
  f^{Q_{xy}}(\bm{k})^2 \right]f^{T_{3a}}(\bm{k})
-2 \sqrt{2} f^{T_{3a}}(\bm{k})^3
\right. \nonumber \\ 
& \ \ \ \left.
 -6 \left[f^{Q_v}(\bm{k})^2 
- f^{Q_{xy}}(\bm{k})^2\right] f^{T_x}(\bm{k})
\right. \nonumber \\
& \ \ \ \left.
-12 f^{Q_v}(\bm{k}) f^{Q_{xy}}(\bm{k}) f^{T_y}(\bm{k})
\right. \nonumber \\ 
& \ \ \ \left.
+3 \sqrt{2} f^{T_{3a}}(\bm{k}) \left[f^{T_x}(\bm{k})^2+ f^{T_y}(\bm{k})^2 \right]
\right. \nonumber \\ 
& \ \ \ \left.
+2f^{T_x}(\bm{k}) \left[  f^{T_x}(\bm{k})^2-3  f^{T_y}(\bm{k})^2 \right]\right). 
\end{align}
All the terms contain the odd number of magnetic toroidal multipoles, as shown in Eq.~(\ref{eq:spinsplit_asym_TL}).
In the $\bm{k}\to \bm{0}$ limit, $g_{5}^{z}(\bm{k})$ becomes 
\begin{align}
\label{eq:BKL_gz}
&g_{5}^{z}(\bm{k})\simeq \sqrt{\frac{2}{3}}m^2 \biggl\{f^{Q_0}(\bm{k}) [f^{Q_{xy}}(\bm{k}) f^{T_y}(\bm{k})-f^{Q_v}(\bm{k}) f^{T_x}(\bm{k})]
\cr&\quad
+ f^{Q_0}(\bm{k})^2 f^{T_{3a}}(\bm{k}) + \frac{1}{3\sqrt{2}}f^{T_x}(\bm{k})[f^{T_x}(\bm{k})^2-3 f^{T_y}(\bm{k})^2]\biggr\} \nonumber \\
&=8 m^2  t_a t_b (t_a-t_b) \sin k'_x  \left(\cos k'_x-\cos k'_y\right) \nonumber \\
&\simeq -\frac{1}{2}(a+b)^3  m^2 t_a t_b (t_a-t_b) k_x  (k_x^2-3 k_y^2),  
\end{align}
where $k'_x=k_x (a+b)/2$ and $k'_y=k_y (a+b)\sqrt{3}/2$. 
From the expression in Eq.~(\ref{eq:BKL_gz}), the antisymmetric spin splitting occurs for $t_a \neq 0$, $t_b \neq 0$, and $t_a \neq t_b$, i.e., the breathing structure is important.

The above analysis for the spin splittings is also confirmed by calculating explicitly the electronic band structure.  
Figures~\ref{Fig:BKL_band}(a) and (b) show the band structure at $t_a=1$, $t_b=0.5$, and $m=1$ where the color map shows the spin polarization for the $x$ and $y$ components, respectively. 
The result is similar to that in the kagome case in Fig.~\ref{Fig:KL_band}; the symmetric spin splittings are characterized by $Q^{\rm (m)}_{xy}(\bm{k}) \sigma_x$ in Fig.~\ref{Fig:BKL_band}(a) and $Q^{\rm (m)}_{v}(\bm{k}) \sigma_x$ in Fig.~\ref{Fig:BKL_band}(b). 
In contrast to the result in the kagome system, the antisymmetric spin splitting occurs, as shown in Fig.~\ref{Fig:BKL_band}(c), which is similar to that in the triangular case in Fig.~\ref{Fig:TL_band}. 

\subsection{Effect of external magnetic field}
\label{sec:magnetic field}

We investigate the effect of an external magnetic field on the breathing kagome AFM. 
To this end, we add the Zeeman coupling term, $- \bm{H}\cdot \sum_{i\sigma\sigma'}c_{i\sigma}^{\dagger} \bm{\sigma}_{\sigma \sigma'}c_{i\sigma'}^{}$, to the Hamiltonian in Eq.~(\ref{eq:Ham_BKL}), which is given in the multipole notation as 
\begin{align}
H_{\rm mag} = - Q_{0}^{\rm (c)} \bm{H}\cdot  \bm{\sigma}. 
\end{align}
With this Zeeman term, the expansion procedure leads to the additional multipole couplings according to the symmetry reduction. 
There are mainly two types of additional couplings.  
One is the term proportional to the odd order of $\bm{H}$, and the other is the term proportional to the even order of $\bm{H}$. 

For $\bm{H}\parallel [100]$, the antisymmetric contributions proportional to $H_x$ are 
given by 
\begin{align}
g^{0}(\bm{k}) &\sim m H_x Q^{\rm (m)}_{xy}(\bm{k}) \sim k^2 \sin 2 \phi, \\
g^{x}(\bm{k}) &\sim m^2 H_x Q^{\rm (m)}_{v}(\bm{k}) \sim k^2 \cos 2 \phi, \\
g^{y}(\bm{k}) &\sim m^2 H_x Q^{\rm (m)}_{xy}(\bm{k}) \sim k^2 \sin 2 \phi, \\
g^{z}(\bm{k}) &\sim m^3 H_x Q^{\rm (m)}_{xy}(\bm{k})T_{3a}^{\rm (m)}(\bm{k})\sim k^5 \sin \phi, 
\end{align}
where $\bm{k}=k( \cos \phi,  \sin \phi)$ and we omit the subscript of $g^{\mu}_s(\bm{k})$. 
Thus, one can expect the following additional effects under the [100] magnetic field: the $xy$-type shear stress in the form of $Q^{\rm (m)}_{xy}(\bm{k})$, the symmetric spin splitting in forms of $Q^{\rm (m)}_{v}(\bm{k})$ and $Q^{\rm (m)}_{xy}(\bm{k})$ for $\sigma_x$ and $\sigma_y$ spin components, respectively, and the antisymmetric spin splitting in the form of $Q^{\rm (m)}_{xy}(\bm{k})T^{\rm (m)}_{3a}(\bm{k})$. 
Especially, the last additional antisymmetric spin splitting is related to the magnetoelectric effect, since $Q^{\rm (m)}_{xy}(\bm{k})T_{3a}^{\rm (m)}(\bm{k})$ has the same symmetry as the electric dipole $Q_x$~\cite{Hayami_PhysRevB.98.165110}.

The symmetric contributions proportional to $H_x^{2}$ are given by 
\begin{align}
g^{0}(\bm{k}) &\sim m^2 H^2_x Q^{\rm (m)}_{v}(\bm{k}) \sim k^2 \cos 2 \phi, \\
g^{x}(\bm{k}) &\sim m H^2_x Q^{\rm (m)}_{xy}(\bm{k}) \sim k^2 \sin 2 \phi, \\
g^{y}(\bm{k}) &\sim m^3 H^2_x Q^{\rm (m)}_0(\bm{k}) \sim 1, \\
g^{z}(\bm{k}) &\sim m^4 H^2_x Q^{\rm (m)}_{v}(\bm{k})T^{\rm (m)}_{3a}(\bm{k}) \sim k^5 \cos \phi. 
\end{align}
The obtained expressions indicate that magnetization in the $y$ component is spontaneously induced by applying the magnetic field even along the [100] direction. 
In this way, the multipole couplings under the magnetic field are obtained systematically. 
We summarize some of active multipoles under the magnetic field along various directions in Tables~\ref{tab_multipoles_Hxy}, \ref{tab_multipoles_Hzx}, and \ref{tab_multipoles_Hzy}.

\begin{table}[t!]
\caption{
Some of active momentum multipoles when the magnetic field is applied in the $xy$ plane, $\bm{H}=H(\cos \theta, \sin \theta,0)$. 
The superscript $\rm (m)$ and $(\bm{k})$ in the multipoles $Q^{\rm (m)}_{lm}(\bm{k})$ and $T^{\rm (m)}_{lm}(\bm{k})$ are omitted for notational simplicity. 
}
\label{tab_multipoles_Hxy}
\centering
\begin{tabular}{cccccccc} \hline\hline
$H(\cos \theta, \sin \theta,0)$ & $\sigma_0$ & $\sigma_x$ & $\sigma_y$ & $\sigma_z$  \\  \hline
$H\cos \theta$ & $m Q_{xy}$ & $m^2 Q_{v}$ & $m^2 Q_{xy}$ & $ m^3 Q_{xy} T_{3a}$  \\ 
$H\sin \theta$ & $m Q_v$ & $m^2 Q_{xy}$  & $m^2 Q_v$  & $m^3  Q_v T_{3a}$  \\
$H^2\cos 2\theta$ & $m^2 Q_v$  & $m  Q_{xy}$ & $ m^3 ,   m  Q_v$  & $m^4  Q_v T_{3a}$ \\
$H^2\sin 2\theta$ & $m^2 Q_{xy}$  & $ m^3,   m  Q_v$ & $m Q_{xy}$  & $ m^4 Q_{xy} T_{3a}$  \\
\hline\hline
\end{tabular}
\end{table}

\begin{table}[t!]
\caption{
Some of active momentum multipoles when the magnetic field is applied in the $zx$ plane, $\bm{H}=H(\sin \theta, 0, \cos \theta)$. 
The superscript $\rm (m)$ and $(\bm{k})$ in the multipoles $Q^{\rm (m)}_{lm}(\bm{k})$ and $T^{\rm (m)}_{lm}(\bm{k})$ are omitted for notational simplicity. 
}
\label{tab_multipoles_Hzx}
\centering
\begin{tabular}{cccccccc} \hline\hline
$H(\sin \theta, 0, \cos \theta)$ & $\sigma_0$ & $\sigma_x$ & $\sigma_y$ & $\sigma_z$  \\  \hline
$H\cos \theta$ & $m^2  T_{3a}$ & $m^3  Q_{xy} T_{3a}$  &$ m^3  Q_v T_{3a}$ & $m^2 H^2 Q_v$ \\ 
$H\sin \theta$ & $m Q_{xy} $ & $m^2 Q_v$  & $m^2 Q_{xy}$  & $m^3 Q_{xy} T_{3a}$  \\
$H^2\cos 2\theta$ & $m^2 Q_v$ & $m Q_{xy}$ & $m^3 , m^3 Q_v$  & $m^2 T_{3a}$ \\
$H^2\sin 2\theta$ & $m^3 Q_{xy} T_{3a}$   & $m^2 T_{3a}$ &$ m^4  Q_{xy} T_{3a}$ &$m Q_{xy}$  \\
\hline\hline
\end{tabular}
\end{table}

\begin{table}[t!]
\caption{
Some of active momentum multipoles when the magnetic field is applied in the $zy$ plane, $\bm{H}=H(0,\sin \theta, \cos \theta)$. 
The superscript $\rm (m)$ and $(\bm{k})$ in the multipoles $Q^{\rm (m)}_{lm}(\bm{k})$ and $T^{\rm (m)}_{lm}(\bm{k})$ are omitted for notational simplicity. 
}
\label{tab_multipoles_Hzy}
\centering
\begin{tabular}{cccccccc} \hline\hline
$H(0,\sin \theta, \cos \theta)$ & $\sigma_0$ & $\sigma_x$ & $\sigma_y$ & $\sigma_z$  \\ \hline
$H\cos \theta$ & $m^2 T_{3a}$  & $m^3  Q_{xy} T_{3a}$  & $ m^3 Q_v T_{3a}$ & $m^2 H^2 Q_v$ \\ 
$H\sin \theta$ & $m Q_v$  & $m^2 Q_{xy}$  & $ m^2 Q_v$  & $m^3 Q_v T_{3a}$  \\
$H^2\cos 2\theta$ & $m^2 Q_v$  & $m^3 Q_{xy}$  &  $m^3 , m  Q_v$  & $m^2 T_{3a}$ \\
$H^2\sin 2\theta$ & $m^3 Q_v T_{3a}$  & $ m^4  Q_{xy} T_{3a}$  & $m^2  T_{3a}$ & $m  Q_v$  \\
\hline\hline
\end{tabular}
\end{table}

\section{Discussion on Materials and Summary}
\label{sec:Summary}

The spin splittings and band deformations by the effective multipole-spin couplings irrespective of the SOC are ubiquitously found in various structures of magnetic materials with a variety of chemical compositions. 
The symmetric spin splitting in collinear AFM state has been studied for an organic $\kappa$-(BETD-TTF)$_2$Cu[N(CN)$_2$]Cl~\cite{naka2019spin,hayami2019momentum,Hayami2020b,Naka_PhysRevB.102.075112}, transition metal oxide RuO$_2$~\cite{Berlijn_PhysRevLett.118.077201,Ahn_PhysRevB.99.184432}, and transition metal fluoride MnF$_2$~\cite{Yuan_PhysRevB.102.014422}.
Moreover, the antisymmetric spin splitting and band deformation have been studied for a transition metal oxide Ba$_{3}$MnNb$_{2}$O$_{9}$~\cite{Lee_PhysRevB.90.224402,Hayami_PhysRevB.101.220403}. 
A similar analysis can be applied to the exchange Hamiltonian for insulating noncollinear magnets where an emergent Dzyaloshinskii-Moriya interaction without the SOC arises through the multipole couplings~\cite{cardias2020dzyaloshinskii}. 

With the knowledge of the multipole couplings, we list up the candidate materials that could exhibit spin-split band structures at the onset of the AFM phase transition having ordering vector $\bm{Q}=\bm{0}$ in Tables~\ref{tab:sym} and \ref{tab:asym}, which are obtained in accordance with MAGNDATA, magnetic structure database~\cite{gallego2016magndata}. 
It is noted that the AFM materials with a finite ordering vector $\bm{Q}$ are also candidates, as exemplified for Ba$_{3}$MnNb$_{2}$O$_{9}$~\cite{Hayami_PhysRevB.101.220403}.
In the candidate materials, the symmetric spin splitting emerges in the magnetic-ordered-moment direction, while the antisymmetric spin splitting emerges in the direction perpendicular to the coplanar magnetic structure, when the present mechanism dominates over ordinary SOC one.
Since the present mechanism of the spin splitting and band deformation does not rely on the presence of the SOC, we can explore a variety class of materials including simple compounds with lighter elements.
The multipole-spin couplings can be flexibly tuned by temperature, pressure, and magnetic fields as discussed in this paper.
The effect of the splitting and deformation is expected to be large as compared with those by the spin-orbit coupling origin since the magnitude of the effective coupling is characterized by the kinetic energy and the molecular field of AFM, which is the order of the Coulomb interaction.
These advantages further promote the efficient engineering of spin-orbit-coupling free materials exhibiting a giant spin-dependent and/or nonreciprocal transport, magneto- electric and elastic responses.

To summarize, we have clarified the efficient bottom-up design procedure of electronic band structures in AFMs without the spin-orbit coupling. 
Our microscopic guideline to engineer the spin and momentum dependent band structures was established by introducing the concept of augmented multipoles consisting of the electric and magnetic toroidal multipoles. 
We showed that arbitrary Hamiltonians in the tight-binding model are decomposed into a scalar-product form in terms of electric and magnetic toroidal multipoles. 
The hopping Hamiltonian is expressed as a linear combination of products between bond and momentum multipoles, while the mean-field Hamiltonian is expressed as a linear combination of products between cluster multipoles and spins. 
By using such multipole degrees of freedom, we demonstrated that the spin-split and reshaped electronic band structures are caused by the effective multipole couplings. 
The symmetric spin splitting emerges in the presence of the coupling between cluster and bond multipoles with the same symmetry in collinear AFMs, whereas the antisymmetric one is induced by the coupling including a bond-type magnetic toroidal multipole in noncollinear AFMs. 
Furthermore, we found that the antisymmetric band deformation with spin degeneracy is realized in noncoplanar AFMs.  
We analyzed the band deformations under the AFM orderings by exemplifying three lattice systems consisting of a triangle cluster, triangular, kagome, and breathing kagome structures. 
Lastly, we listed candidate materials showing intrinsic band deformations driven by the magnetic orderings by referring MAGNDATA, magnetic structures database, which would be useful to unveil unexplored fascinating functional materials.

\begin{longtable*}[c]{cllcccc}
\caption{
Symmetric spin-splitting materials listed in Ref.~\onlinecite{gallego2016magndata}.
SG, MSG, and MPG represent space group, magnetic space group, and magnetic point group, respectively. 
$\mathcal{P}$ stands for the presence ($\circ$) or absence ($\times$) of the spatial inversion symmetry. 
\# represents the serial number of space group. 
The symbol * shows that there are different magnetic patterns. 
}
\label{tab:sym}
\\
\hline \hline
Crystal systems & Materials & \# & SG & MSG & MPG & $\mathcal{P}$ \\ \hline
Monoclinic&LiFeP$_{2}$O$_{7}$   &     $4$  &                 $P2_{1}$  &                 $P2_{1}$  &                 $2$ & $\times$    \\ \relax
&*CaFe$_{5}$O$_{7}$   &    $11$  &               $P2_{1}/m$  &               $P2_{1}/m$  &               $2/m$  & $\circ$  \\ \relax
&*CaFe$_{5}$O$_{7}$   &    $11$  &               $P2_{1}/m$  &             $P2_{1}'/m'$  &             $2'/m'$  & $\circ$    \\ \relax
&Nd$_{2}$NaRuO$_{6}$   &    $14$  &               $P2_{1}/n$  &               $P2_{1}/c$  &               $2/m$ & $\circ$    \\ \relax
&LiFe(SO$_{4}$)$_{2}$   &    $14$  &               $P2_{1}/c$  &               $P2_{1}/c$  &               $2/m$ & $\circ$    \\ \relax
&Li$_{2}$Co(SO$_{4}$)$_{2}$   &    $14$  &               $P2_{1}/c$  &             $P2_{1}'/c'$  &             $2'/m'$ & $\circ$   \\ \relax
&Li$_{2}$Mn(SO$_{4}$)$_{2}$   &    $14$  &               $P2_{1}/c$  &               $P2_{1}/c$  &               $2/m$ & $\circ$    \\ \relax
&La$_{2}$LiRuO$_{6}$   &    $14$  &               $P2_{1}/n$  &               $P2_{1}/c$  &               $2/m$ & $\circ$   \\ \relax
&Y$_{2}$MnCoO$_{6}$   &    $14$  &               $P2_{1}/c$  &             $P2_{1}'/c'$  &             $2'/m'$  & $\circ$  \\ \relax
&FeCl$_{5}$D$_{2}$O(ND$_{4}$)$_{2}$   &    $14$  &               $P2_{1}/c$  &                $P2_{1}'$  &                $2'$  & $\times$   \\ \relax
&Ca$_{2}$MnReO$_{6}$   &    $14$  &               $P2_{1}/c$  &               $P2_{1}/c$  &               $2/m$ & $\circ$   \\ \relax
&Sr$_{2}$MnReO$_{6}$   &    $14$  &               $P2_{1}/c$  &             $P2_{1}'/c'$  &             $2'/m'$ & $\circ$   \\ \relax
&Li$_{3}$Fe$_{2}$(PO$_{4}$)$_{3}$   &    $14$  &               $P2_{1}/n$  &             $P2_{1}'/c'$  &             $2'/m'$  & $\circ$  \\ \relax
&*Tb$_{2}$MnNiO$_{6}$   &    $14$  &               $P2_{1}/c$  &             $P2_{1}'/c'$  &             $2'/m'$& $\circ$    \\ \relax
&*Tb$_{2}$MnNiO$_{6}$   &    $14$  &               $P2_{1}/c$  &             $P2_{1}'/c'$  &             $2'/m'$ & $\circ$   \\ \relax
&*Tb$_{2}$MnNiO$_{6}$   &    $14$  &               $P2_{1}/c$  &             $P2_{1}'/c'$  &             $2'/m'$  & $\circ$  \\ \relax
&*Tb$_{2}$MnNiO$_{6}$   &    $14$  &               $P2_{1}/c$  &               $P2_{1}/c$  &               $2/m$ & $\circ$   \\ \relax
&Tl$_{2}$NiMnO$_{6}$   &    $14$  &               $P2_{1}/c$  &               $P2_{1}/c$  &               $2/m$ & $\circ$   \\ \relax
&*Cu$_{2}$(OD)$_{3}$Cl   &    $14$  &               $P2_{1}/c$  &               $P2_{1}/c$  &               $2/m$ & $\circ$   \\ \relax
&*Cu$_{2}$(OD)$_{3}$Cl   &    $14$  &               $P2_{1}/c$  &               $P2_{1}/c$  &               $2/m$ & $\circ$   \\ \relax
&Sr$_{2}$CoTeO$_{6}$   &    $14$  &               $P2_{1}/n$  &               $P2_{1}/c$  &               $2/m$  & $\circ$  \\ \relax
&Sr$_{2}$Co$_{0.9}$Mg$_{0.1}$TeO$_{6}$   &    $14$  &               $P2_{1}/n$  &               $P2_{1}/c$  &               $2/m$ & $\circ$   \\ \relax
&Ho$_{2}$CoMnO$_{6}$   &    $14$  &               $P2_{1}/c$  &             $P2_{1}'/c'$  &             $2'/m'$ & $\circ$   \\ \relax
&*Tm$_{2}$CoMnO$_{6}$   &    $14$  &               $P2_{1}/c$  &             $P2_{1}'/c'$  &             $2'/m'$  & $\circ$  \\ \relax
&*Tm$_{2}$CoMnO$_{6}$   &    $14$  &               $P2_{1}/c$  &             $P2_{1}'/c'$  &             $2'/m'$  & $\circ$  \\ \relax
&Cu$_{1.94}$Mn$_{1.06}$BO$_{5}$   &    $14$  &               $P2_{1}/c$  &             $P2_{1}'/c'$  &             $2'/m'$  & $\circ$  \\ \relax
&KMnF$_{4}$   &    $14$  &               $P2_{1}/a$  &             $P2_{1}'/c'$  &             $2'/m'$ & $\circ$   \\ \relax
&RbMnF$_{4}$   &    $14$  &               $P2_{1}/a$  &               $P\bar{1}$  &           $\bar{1}$ & $\circ$   \\ \relax
&Li$_{2}$FeP$_{2}$O$_{7}$   &    $14$  &               $P2_{1}/c$  &               $P2_{1}/c$  &               $2/m$ & $\circ$    \\ \relax
&Co$_{4}$(OH)$_{2}$(C$_{10}$H$_{16}$O$_{4}$)$_{3}$   &    $14$  &               $P2_{1}/c$  &             $P2_{1}'/c'$  &             $2'/m'$ & $\circ$    \\ \relax
&Mn$_{2}$ScSbO$_{6}$   &    $14$  &               $P2_{1}/n$  &               $P2_{1}/c$  &               $2/m$   & $\circ$  \\ \relax
&[CH$_{3}$NH$_{3}$]
[Co(COOH)$_{3}$]  &    $14$  &               $P2_{1}/n$  &             $P2_{1}'/c'$  &             $2'/m'$  & $\circ$  \\ \relax
&La$_{2}$CoIrO$_{6}$   &    $14$  &               $P2_{1}/n$  &               $P2_{1}/c$  &               $2/m$  & $\circ$  \\ \relax
&Fe$_{3}$(PO$_{4}$)$_{2}$(OH)$_{2}$   &    $14$  &               $P2_{1}/c$  &               $P2_{1}/c$  &               $2/m$ & $\circ$   \\ \relax
&Cs$_{2}$FeCl$_{5}
\cdot $D$_{2}$O   &    $15$  &                   $C2/c$  &                     $C2$  &                 $2$   & $\times$  \\ \relax
&*BiCrO$_{3}$   &    $15$  &                   $C2/c$  &               $P\bar{1}$  &           $\bar{1}$ & $\circ$   \\ \relax
&*BiCrO$_{3}$   &    $15$  &                   $C2/c$  &                   $C2/c$  &               $2/m$   & $\circ$  \\ \relax
&FeSO$_{4}$F   &    $15$  &                   $C2/c$  &                 $C2'/c'$  &             $2'/m'$  & $\circ$  \\ \relax
&Sr$_{2}$CoOsO$_{6}$   &    $15$  &                   $B2/n$  &                   $C2/c$  &               $2/m$ & $\circ$   \\ \relax
&NaCrGe$_{2}$O$_{6}$   &    $15$  &                   $C2/c$  &                 $C2'/c'$  &             $2'/m'$& $\circ$    \\ \relax
&Na$_{2}$BaFe(VO$_{4}$)$_{2}$   &    $15$  &                   $C2/c$  &                 $C2'/c'$  &             $2'/m'$  & $\circ$  \\ \hline \relax
Orthorhombic &SrMn(VO$_{4}$)(OH)   &    $19$  &       $P2_{1}2_{1}2_{1}$  &                 $P2_{1}$  &                 $2$  & $\times$   \\ \relax
&BaCrF$_{5}$   &    $19$  &       $P2_{1}2_{1}2_{1}$  &     $P2_{1}'2_{1}'2_{1}$  &             $2'2'2$  & $\times$   \\ \relax
&GaFeO$_{3}$   &    $33$  &               $Pna2_{1}$  &             $Pna'2_{1}'$  &             $m'm2'$  & $\times$    \\ \relax
&*Fe$_{2}$O$_{3}$   &    $33$  &               $Pna2_{1}$  &             $Pna'2_{1}'$  &             $m'm2'$   & $\times$  \\ \relax
&*Fe$_{2}$O$_{3}$   &    $33$  &               $Pna2_{1}$  &             $Pna'2_{1}'$  &             $m'm2'$   & $\times$  \\ \relax
&*[C(ND$_{2}$)$_{3}$]Cu(DCOO)$_{3}$   &    $33$  &               $Pna2_{1}$  &               $Pna2_{1}$  &               $mm2$   & $\times$  \\ \relax
&*[C(ND$_{2}$)$_{3}$]Cu(DCOO)$_{3}$   &    $33$  &                    
$Pna2_{1}$ &             $Pn'a'2_{1}$  &             $m'm'2$  & $\times$   \\ \relax
&Y$_{2}$Cu$_{2}$O$_{5}$   &    $33$  &               $Pna2_{1}$  &               $Pna2_{1}$  &               $mm2$  & $\times$   \\ \relax
&BaCuF$_{4}$   &    $36$  &               $Cmc2_{1}$  &             $Cm'c'2_{1}$  &             $m'm'2$  & $\times$   \\ \relax
&Ca$_{3}$Mn$_{2}$O$_{7}$   &    $36$  &               $Cmc2_{1}$  &             $Cm'c2_{1}'$  &             $m'm2'$    & $\times$ \\ \relax
&ErGe$_{1.83}$   &    $36$  &               $Cmc2_{1}$  &               $Cmc2_{1}$  &               $mm2$  & $\times$   \\ \relax
&Cu$_{2}$V$_{2}$O$_{7}$   &    $43$  &                   $Fdd2$  &                 $Fd'd'2$  &             $m'm'2$   & $\times$  \\ \relax
&BiFe$_{0.5}$Sc$_{0.5}$O$_{3}$   &    $46$  &                   $Ima2$  &                 $Im'a2'$  &             $m'm2'$   & $\times$  \\ \relax
&[C(ND$_{2}$)$_{3}$]Mn(DCOO)$_{3}$   &    $52$  &                   $Pnna$  &                 $Pn'n'a$  &             $m'm'm$  & $\circ$  \\ \relax
&[C(ND$_{2}$)$_{3}$]Co(DCOO)$_{3}$   &    $52$  &                   $Pnna$  &                 $Pn'na'$  &             $m'm'm$   & $\circ$ \\ \relax
&Fe$_{1.5}$Mn$_{1.5}$BO$_{5}$   &    $55$  &                   $Pbam$  &                   $Pbam$  &               $mmm$ & $\circ$   \\ \relax
&Fe(N(CN$_{2}$))$_{2}$   &    $58$  &                   $Pnnm$  &                 $Pnn'm'$  &             $m'm'm$   & $\circ$ \\ \relax
&KCo$_{4}$(PO$_{4}$)$_{3}$   &    $58$  &                   $Pnnm$  &                 $Pnn'm'$  &             $m'm'm$  & $\circ$  \\ \relax
&Mn(N(CN$_{2}$))$_{2}$   &    $58$  &                   $Pnnm$  &                 $Pnn'm'$  &             $m'm'm$  & $\circ$  \\ \relax
&*TmMn$_{3}$O$_{6}$   &    $59$  &                   $Pmmn$  &                 $Pm'm'n$  &             $m'm'm$  & $\circ$  \\ \relax
&*TmMn$_{3}$O$_{6}$   &    $59$  &                   $Pmmn$  &                 $Pmm'n'$  &             $m'm'm$ & $\circ$   \\ \relax
&*$\alpha$-Mn$_{2}$O$_{3}$
&    $61$  &                   $Pbca$  &                   $Pbca$  &               $mmm$  & $\circ$  \\ \relax
&*$\alpha$-Mn$_{2}$O$_{3}$
&    $61$  &                   $Pbca$  &                   $Pbca$  &               $mmm$  & $\circ$  \\ \relax
&Ca$_{2}$RuO$_{4}$   &    $61$  &                   $Pbca$  &                   $Pbca$  &               $mmm$ & $\circ$   \\ \relax
&CuFePO$_{5}$   &    $62$  &                   $Pnma$  &                   $Pnma$  &               $mmm$  & $\circ$  \\ \relax
&NiFePO$_{5}$   &    $62$  &                   $Pnma$  &                   $Pnma$  &               $mmm$  & $\circ$  \\ \relax
&CoFePO$_{5}$   &    $62$  &                   $Pnma$  &                 $Pnm'a'$  &             $m'm'm$  & $\circ$  \\ \relax
&Fe$_{2}$PO$_{5}$   &    $62$  &                   $Pnma$  &                   $Pnma$  &               $mmm$ & $\circ$    \\ \relax
&CoSO$_{4}$   &    $62$  &                   $Pnma$  &                   $Pnma$  &               $mmm$ & $\circ$    \\ \relax
&YCr$_{0.5}$Mn$_{0.5}$O$_{3}$   &    $62$  &                   $Pnma$  &                 $Pn'ma'$  &             $m'm'm$  & $\circ$  \\ \relax
&*Mn$_{2}$GeO$_{4}$   &    $62$  &                   $Pnma$  &                 $Pn'm'a$  &             $m'm'm$ & $\circ$   \\ \relax
&*Mn$_{2}$GeO$_{4}$   &    $62$  &                   $Pnma$  &                   $Pnma$  &               $mmm$  & $\circ$  \\ \relax
&*Mn$_{2}$GeO$_{4}$   &    $62$  &                   $Pnma$  &               $P2_{1}/c$  &               $2/m$ & $\circ$   \\ \relax
&NH$_{4}$Fe$_{2}$O$_{6}$   &    $62$  &                   $Pnma$  &                   $Pnma$  &               $mmm$  & $\circ$  \\ \relax
&*NdMnO$_{3}$   &    $62$  &                   $Pnma$  &                 $Pn'ma'$  &             $m'm'm$  & $\circ$  \\ \relax
&*NdMnO$_{3}$   &    $62$  &                   $Pnma$  &                 $Pn'ma'$  &             $m'm'm$ & $\circ$   \\ \relax
&ErVO$_{3}$   &    $62$  &                   $Pbnm$  &             $P2_{1}'/m'$  &             $2'/m'$ & $\circ$   \\ \relax
&NiTe$_{2}$O$_{5}$   &    $62$  &                   $Pnma$  &                   $Pnma$  &               $mmm$ & $\circ$   \\ \relax
&(Tm$_{0.7}$Mn$_{0.3}$)MnO$_{3}$   &    $62$  &                   $Pnma$  &                 $Pnm'a'$  &             $m'm'm$ & $\circ$   \\ \relax
&Cu$_{4}$(OD)$_{6}$FBr   &    $62$  &                   $Pnma$  &                 $Pn'm'a$  &             $m'm'm$ & $\circ$   \\ \relax
&Nd$_{5}$Ge$_{4}$   &    $62$  &                   $Pnma$  &                 $Pnm'a'$  &             $m'm'm$ & $\circ$   \\ \relax
&ErVO$_{3}$   &    $62$  &                   $Pbnm$  &               $P2_{1}/c$  &               $2/m$  & $\circ$  \\ \relax
&RbFe$_{2}$F$_{6}$   &    $62$  &                   $Pnma$  &                   $Pnma$  &               $mmm$ & $\circ$   \\ \relax
&Ca$_{2}$PrCr$_{2}$NbO$_{9}$   &    $62$  &                   $Pnma$  &                 $Pn'm'a$  &             $m'm'm$  & $\circ$  \\ \relax
&Ca$_{2}$PrCr$_{2}$TaO$_{9}$   &    $62$  &                   $Pnma$  &                 $Pn'm'a$  &             $m'm'm$  & $\circ$  \\ \relax
&DyVO$_{3}$   &    $62$  &                   $Pbnm$  &             $P2_{1}'/m'$  &             $2'/m'$  & $\circ$  \\ \relax
&NaOsO$_{3}$   &    $62$  &                   $Pnma$  &                 $Pn'ma'$  &             $m'm'm$& $\circ$    \\ \relax
&Ca$_{2}$Fe$_{0.875}$Cr$_{0.125}$GaO$_{5}$   &    $62$  &                   $Pnma$  &                 $Pn'm'a$  &             $m'm'm$  & $\circ$  \\ \relax
&La$_{0.5}$Sr$_{0.5}$FeO$_{2.5}$F$_{0.5}$   &    $62$  &                   $Pnma$  &                 $Pn'ma'$  &             $m'm'm$   & $\circ$ \\ \relax
&ScCrO$_{3}$   &    $62$  &                   $Pnma$  &                   $Pnma$  &               $mmm$  & $\circ$  \\ \relax
&InCrO$_{3}$   &    $62$  &                   $Pnma$  &                   $Pnma$  &               $mmm$  & $\circ$  \\ \relax
&TlCrO$_{3}$   &    $62$  &                   $Pnma$  &                   $Pnma$  &               $mmm$  & $\circ$  \\ \relax
&*Co$_{2}$SiO$_{4}$   &    $62$  &                   $Pnma$  &                   $Pnma$  &               $mmm$  & $\circ$  \\ \relax
&*Co$_{2}$SiO$_{4}$   &    $62$  &                   $Pnma$  &                   $Pnma$  &               $mmm$  & $\circ$  \\ \relax
&Mn$_{2}$SiO$_{4}$   &    $62$  &                   $Pnma$  &                 $Pn'm'a$  &             $m'm'm$  & $\circ$  \\ \relax
&Fe$_{2}$SiO$_{4}$   &    $62$  &                   $Pnma$  &                   $Pnma$  &               $mmm$  & $\circ$  \\ \relax
&DyFeO$_{3}$   &    $62$  &                   $Pnma$  &             $Pn'a'2_{1}$  &             $m'm'2$  & $\times$   \\ \relax
&LaCrO$_{3}$   &    $62$  &                   $Pnma$  &                   $Pnma$  &               $mmm$  & $\circ$  \\ \relax
&BiFe$_{0.5}$Sc$_{0.5}$O$_{3}$   &    $62$  &                   $Pnma$  &                 $Pn'm'a$  &             $m'm'm$ & $\circ$   \\ \relax
&*NdFeO$_{3}$   &    $62$  &                   $Pnma$  &                 $Pn'ma'$  &             $m'm'm$ & $\circ$   \\ \relax
&*NdFeO$_{3}$   &    $62$  &                   $Pnma$  &             $P2_{1}'/c'$  &             $2'/m'$  & $\circ$  \\ \relax
&*TbFeO$_{3}$   &    $62$  &                   $Pbnm$  &                 $Pn'ma'$  &             $m'm'm$ & $\circ$   \\ \relax
&*TbFeO$_{3}$   &    $62$  &                   $Pbnm$  &                 $Pn'm'a$  &             $m'm'm$  & $\circ$  \\ \relax
&TbCrO$_{3}$   &    $62$  &                   $Pbnm$  &                 $Pn'm'a$  &             $m'm'm$  & $\circ$  \\ \relax
&TbPt$_{0.8}$Cu$_{0.2}$   &    $62$  &                   $Pnma$  &                 $Pn'm'a$  &             $m'm'm$  & $\circ$  \\ \relax
&NdNi$_{0.6}$Cu$_{0.4}$   &    $62$  &                   $Pnma$  &                 $Pnm'a'$  &             $m'm'm$  & $\circ$  \\ \relax
&[CH$_{3}$NH$_{3}$][Co(COOH)$_{3}$]   &    $62$  &                   $Pnma$  &                 $Pn'ma'$  &             $m'm'm$  & $\circ$   \\ \relax
&LaMnO$_{3}$   &    $62$  &                   $Pnma$  &                 $Pn'ma'$  &             $m'm'm$  & $\circ$  \\ \relax
&*NdMnO$_{3}$   &    $62$  &                   $Pnma$  &                 $Pn'ma'$  &             $m'm'm$ & $\circ$   \\ \relax
&*NdMnO$_{3}$   &    $62$  &                   $Pnma$  &                 $Pn'ma'$  &             $m'm'm$& $\circ$    \\ \relax
&La$_{0.75}$Bi$_{0.25}$Fe$_{0.5}$Cr$_{0.5}$O$_{3}$   &    $62$  &                   $Pnma$  &                   $Pnma$  &               $mmm$ & $\circ$   \\ \relax
&*Rb$_{2}$Fe$_{2}$O(AsO$_{4}$)$_{2}$   &    $62$  &                   $Pnma$  &                   $Pnma$  &               $mmm$ & $\circ$   \\ \relax
&*SmFeO$_{3}$   &    $62$  &                   $Pbnm$  &                 $Pn'm'a$  &             $m'm'm$  & $\circ$  \\ \relax
&*SmFeO$_{3}$   &    $62$  &                   $Pnma$  &                 $Pn'ma'$  &             $m'm'm$  & $\circ$  \\ \relax
&*Rb$_{2}$Fe$_{2}$O(AsO$_{4}$)$_{2}$   &    $62$  &                   $Pnma$  &                 $Pn'ma'$  &             $m'm'm$  & $\circ$   \\ \relax
&Ca$_{2}$Fe$_{2}$O$_{5}$   &    $62$  &                   $Pcmn$  &                 $Pcm'n'$  &             $m'm'm$ & $\circ$   \\ \relax
&TeNiO$_{3}$   &    $62$  &                   $Pnma$  &                 $Pn'm'a$  &             $m'm'm$   & $\circ$ \\ \relax
&NdSi   &    $62$  &                   $Pnma$  &                 $Pn'm'a$  &             $m'm'm$ & $\circ$   \\ \relax
&PrSi   &    $62$  &                   $Pnma$  &                 $Pnm'a'$  &             $m'm'm$  & $\circ$  \\ \relax
&TmNi   &    $62$  &                   $Pnma$  &                 $Pn'm'a$  &             $m'm'm$  & $\circ$  \\ \relax
&Y$_{3}$Co$_{3.25}$Al$_{0.75}$   &    $63$  &                   $Cmcm$  &                 $Cm'cm'$  &             $m'm'm$   & $\circ$ \\ \relax
&CaIrO$_{3}$   &    $63$  &                   $Cmcm$  &                 $Cm'cm'$  &             $m'm'm$  & $\circ$  \\ \relax
&LaCaFeO$_{4}$   &    $64$  &                   $Cmce$  &                 $Cm'c'a$  &             $m'm'm$  & $\circ$  \\ \relax
&Gd$_{2}$CuO$_{4}$   &    $64$  &                   $Aeam$  &                 $Cm'ca'$  &             $m'm'm$ & $\circ$   \\ \relax
&Sr$_{4}$Fe$_{4}$O$_{11}$   &    $65$  &                   $Cmmm$  &                 $Cmm'm'$  &             $m'm'm$  & $\circ$  \\ \relax
&YNi$_{4}$Si   &    $65$  &                   $Cmmm$  &                 $Cmm'm'$  &             $m'm'm$  & $\circ$  \\ \relax
&*Y$_{2}$SrCu$_{0.6}$Co$_{1.4}$O$_{6.5}$   &    $72$  &                   $Ibam$  &                 $Ib'a'm$  &             $m'm'm$  & $\circ$  \\ \relax
&*Y$_{2}$SrCu$_{0.6}$Co$_{1.4}$O$_{6.5}$   &    $72$  &                   $Ibam$  &                 $Ib'a'm$  &             $m'm'm$  & $\circ$  \\ \relax
&*YBaMn$_{2}$O$_{5.5}$   &    $72$  &                   $Icam$  &                   $C2/m$  &               $2/m$  & $\circ$  \\ \relax
&*YBaMn$_{2}$O$_{5.5}$   &    $72$  &                   $Icam$  &                 $Ib'a'm$  &             $m'm'm$  & $\circ$  \\ \relax
& Pr$_{0.5}$Sr$_{0.5}$CoO$_{3}$   &    $74$  &                   $Imma$  &                 $Im'm'a$  &             $m'm'm$ & $\circ$   \\ \hline \relax
Tetragonal &MnPrMnSbO$_{6}$   &    $86$  &               $P4_{2}/n$  &               $P4_{2}/n$  &               $4/m$  & $\circ$  \\ \relax
&MnLaMnSbO$_{6}$   &    $86$  &               $P4_{2}/n$  &                 $P2'/c'$  &             $2'/m'$  & $\circ$  \\ \relax
&K$_{y}$Fe$_{2-}$$_{x}$Se$_{2}$   &    $87$  &                   $I4/m$  &                 $C2'/m'$  &             $2'/m'$ & $\circ$   \\ \relax
&TlFe$_{1.6}$Se$_{2}$   &    $87$  &                   $I4/m$  &                   $I4/m$  &               $4/m$ & $\circ$   \\ \relax
&Rb$_{y}$Fe$_{2-}$$_{x}$Se$_{2}$   &    $87$  &                   $I4/m$  &                 $C2'/m'$  &             $2'/m'$  & $\circ$  \\ \relax
&MnV$_{2}$O$_{4}$   &    $88$  &               $I4_{1}/a$  &               $I4_{1}/a$  &               $4/m$  & $\circ$  \\ \relax
&SrMn$_{2}$V$_{2}$O$_{8}$   &   $110$  &               $I4_{1}cd$  &                 $Ib'a2'$  &             $m'm2'$   & $\times$  \\ \relax
&Ba$_{2}$MnSi$_{2}$O$_{7}$   &   $113$  &         $P\bar{4}2_{1}m$  &         $P\bar{4}2_{1}m$  &         $\bar{4}2m$  & $\times$   \\ \relax
&Ba$_{2}$CoGe$_{2}$O$_{7}$   &   $113$  &         $P\bar{4}2_{1}m$  &                 $Cm'm2'$  &             $m'm2'$  & $\times$   \\ \relax
&Ca$_{2}$CoSi$_{2}$O$_{7}$   &   $113$  &         $P\bar{4}2_{1}m$  &         $P2_{1}2_{1}'2'$  &             $2'2'2$    & $\times$ \\ \relax
&CsCoF$_{4}$   &   $120$  &             $I\bar{4}c2$  &              $I\bar{4}'$  &          $\bar{4}'$   & $\times$  \\ \relax
&CeMn$_{2}$Ge$_{4}$O$_{12}$   &   $125$  &                 $P4/nbm$  &               $P4'/nbm'$  &           $4'/mm'm$  & $\circ$  \\ \relax
&CeMnCoGe$_{4}$O$_{12}$   &   $125$  &                 $P4/nbm$  &                 $Pb'an'$  &             $m'm'm$  & $\circ$  \\ \relax
&ZrCo$_{2}$Ge$_{4}$O$_{12}$   &   $125$  &                 $P4/nbm$  &                 $Pb'an'$  &             $m'm'm$  & $\circ$  \\ \relax
&ZrMn$_{2}$Ge$_{4}$O$_{12}$   &   $125$  &                 $P4/nbm$  &               $P4'/nbm'$  &           $4'/mm'm$  & $\circ$  \\ \relax
&CsMnF$_{4}$   &   $129$  &                 $P4/nmm$  &                 $Pmm'n'$  &             $m'm'm$  & $\circ$  \\ \relax
&MnF$_{2}$   &   $136$  &             $P4_{2}/mnm$  &           $P4_{2}'/mnm'$  &           $4'/mm'm$ & $\circ$   \\ \relax
&NiF$_{2}$   &   $136$  &             $P4_{2}/mnm$  &                 $Pnn'm'$  &             $m'm'm$  & $\circ$  \\ \relax
&CoF$_{2}$   &   $136$  &             $P4_{2}/mnm$  &           $P4_{2}'/mnm'$  &           $4'/mm'm$  & $\circ$  \\ \relax
&Nd$_{2}$NiO$_{4.11}$   &   $138$  &             $P4_{2}/ncm$  &           $P4_{2}/nc'm'$  &           $4/mm'm'$  & $\circ$  \\ \relax
&*Nd$_{2}$NiO$_{4}$   &   $138$  &             $P4_{2}/ncm$  &           $P4_{2}/nc'm'$  &           $4/mm'm'$ & $\circ$   \\ \relax
&*La$_{2}$NiO$_{4}$   &   $138$  &             $P4_{2}/ncm$  &                 $Pc'c'n$  &             $m'm'm$  & $\circ$  \\ \relax
&Sr$_{2}$Mn$_{2}$CuAs$_{2}$O$_{2}$   &   $139$  &                 $I4/mmm$  &               $I4/mm'm'$  &           $4/mm'm'$ & $\circ$   \\ \relax
&Mn$_{2.85}$Ga$_{1.15}$   &   $139$  &                 $I4/mmm$  &               $I4/mm'm'$  &           $4/mm'm'$ & $\circ$   \\ \relax
&EuCr$_{2}$As$_{2}$   &   $139$  &                 $I4/mmm$  &           $I\bar{4}m'2'$  &       $\bar{4}2'm'$  & $\times$   \\ \relax
&CaFe$_{4}$Al$_{8}$   &   $139$  &                 $I4/mmm$  &               $I4'/mmm'$  &           $4'/mm'm$  & $\circ$  \\ \relax
&Pr$_{0.5}$Sr$_{0.5}$CoO$_{3}$   &   $140$  &                 $I4/mcm$  &                 $Fm'm'm$  &             $m'm'm$ & $\circ$   \\ \relax
&NiCr$_{2}$O$_{4}$   &   $141$  &             $I4_{1}/amd$  &                 $Fd'd'd$  &             $m'm'm$  & $\circ$  \\ \relax
&Sr$_{2}$Ir$_{0.92}$Sn$_{0.08}$O$_{4}$   &   $142$  &             $I4_{1}/acd$  &                 $Ib'c'a$  &             $m'm'm$  & $\circ$  \\ \hline \relax
Trigonal& Mn$_{2}$ScSbO$_{6}$   &   $146$  &                     $R3$  &                     $P1$  &                 $1$  & $\times$   \\ \relax
&Mn$_{2}$FeMoO$_{6}$   &   $146$  &                     $R3$  &                     $R3$  &                 $3$   & $\times$  \\ \relax
&Mn$_{2}$FeSbO$_{6}$   &   $148$  &               $R\bar{3}$  &               $P\bar{1}$  &           $\bar{1}$ & $\circ$   \\ \relax
&NiN$_{2}$O$_{6}$   &   $148$  &               $R\bar{3}$  &               $R\bar{3}$  &           $\bar{3}$  & $\circ$  \\ \relax
&Li$_{3}$Fe$_{2}$(PO$_{4}$)$_{3}$   &   $148$  &               $R\bar{3}$  &               $R\bar{3}$  &           $\bar{3}$ & $\circ$   \\ \relax
&Cr$_{2}$S$_{3}$   &   $148$  &               $R\bar{3}$  &               $P\bar{1}$  &           $\bar{1}$ & $\circ$   \\ \relax
&NaMnFeF$_{6}$   &   $150$  &                   $P321$  &                  $P32'1$  &               $32'$  & $\times$   \\ \relax
&GaFeO$_{3}$   &   $161$  &                    $R3c$  &                    $Cc'$  &                $m'$   & $\times$  \\ \relax
&ScFeO$_{3}$   &   $161$  &                    $R3c$  &                    $Cc'$  &                $m'$   & $\times$  \\ \relax
&MnTiO$_{3}$   &   $161$  &                    $R3c$  &                    $Cc'$  &                $m'$  & $\times$   \\ \relax
&PbNiO$_{3}$   &   $161$  &                    $R3c$  &                    $R3c$  &                $3m$   & $\times$  \\ \relax
&[NH$_{2}$(CH$_{3}$)$_{2}$][FeCo(HCOO)$_{6}$]   &   $163$  &             $P\bar{3}c1$  &                 $C2'/c'$  &             $2'/m'$  & $\circ$  \\ \relax
&[NH$_{2}$(CH$_{3}$)$_{2}$][FeMn(HCOO)$_{6}$]   &   $163$  &             $P\bar{3}c1$  &                 $C2'/c'$  &             $2'/m'$  & $\circ$  \\ \relax
&Mn$_{3}$Si$_{2}$Te$_{6}$   &   $163$  &             $P\bar{3}1c$  &                 $C2'/c'$  &             $2'/m'$ & $\circ$   \\ \relax
&Mn$_{3}$Ti$_{2}$Te$_{6}$   &   $163$  &             $P\bar{3}1c$  &                 $C2'/c'$  &             $2'/m'$  & $\circ$  \\ \relax
&Na$_{2}$BaCo(VO$_{4}$)$_{2}$   &   $164$  &             $P\bar{3}m1$  &            $P\bar{3}m'1$  &         $\bar{3}m'$  & $\circ$  \\ \relax
&Nd$_{3}$Sb$_{3}$Mg$_{2}$O$_{14}$   &   $166$  &              $R\bar{3}m$  &             $R\bar{3}m'$  &         $\bar{3}m'$ & $\circ$   \\ \relax
&NiCO$_{3}$   &   $167$  &              $R\bar{3}c$  &                   $C2/c$  &               $2/m$ & $\circ$   \\ \relax
&CoF$_{3}$   &   $167$  &              $R\bar{3}c$  &              $R\bar{3}c$  &          $\bar{3}m$ & $\circ$   \\ \relax
&FeF$_{3}$   &   $167$  &              $R\bar{3}c$  &                 $C2'/c'$  &             $2'/m'$  & $\circ$  \\ \relax
&CoCO$_{3}$   &   $167$  &              $R\bar{3}c$  &                   $C2/c$  &               $2/m$  & $\circ$  \\ \relax
&Sr$_{3}$LiRuO$_{6}$   &   $167$  &              $R\bar{3}c$  &                 $C2'/c'$  &             $2'/m'$  & $\circ$  \\ \relax
&MnCO$_{3}$   &   $167$  &              $R\bar{3}c$  &                   $C2/c$  &               $2/m$  & $\circ$  \\ \relax
&FeCO$_{3}$   &   $167$  &              $R\bar{3}c$  &              $R\bar{3}c$  &          $\bar{3}m$ & $\circ$   \\ \relax
&FeBO$_{3}$   &   $167$  &              $R\bar{3}c$  &                 $C2'/c'$  &             $2'/m'$  & $\circ$  \\ \relax
&Ca$_{3}$Co$_{2-}$$_{x}$Mn$_{x}$O$_{6}$   &   $167$  &              $R\bar{3}c$  &                    $R3c$  &                $3m$   & $\times$  \\ \relax
&Ca$_{3}$LiOsO$_{6}$   &   $167$  &              $R\bar{3}c$  &                 $C2'/c'$  &             $2'/m'$ & $\circ$   \\ \relax
&[NH$_{2}$(CH$_{3}$)$_{2}$]$_{n}$[Fe$^{\mathrm{III}}$Fe$^{\mathrm{II}}$(HCOO)$_{6}$]$_{n}$   &   $167$  &              $R\bar{3}c$  &             $R\bar{3}c'$  &         $\bar{3}m'$  & $\circ$  \\ \relax
&Sr$_{3}$NaRuO$_{6}$   &   $167$  &              $R\bar{3}c$  &                 $C2'/c'$  &             $2'/m'$  & $\circ$  \\ \relax
&Ca$_{3}$LiRuO$_{6}$   &   $167$  &              $R\bar{3}c$  &                 $C2'/c'$  &             $2'/m'$  & $\circ$  \\ \relax
&*$\alpha$-Fe$_{2}$O$_{3}$
&   $167$  &              $R\bar{3}c$  &                 $C2'/c'$  &             $2'/m'$ & $\circ$   \\ \relax
&*$\alpha$-Fe$_{2}$O$_{3}$
&   $167$  &              $R\bar{3}c$  &               $P\bar{1}$  &           $\bar{1}$  & $\circ$  \\ \hline \relax
Hexagonal &Cu$_{4}$(OH)$_{6}$FBr   &   $176$  &                  $P63/m$  &             $P2_{1}'/m'$  &             $2'/m'$ & $\circ$   \\ \relax
& Fe$_{2}$Mo$_{3}$O$_{8}$   &   $186$  &               $P6_{3}mc$  &             $P6_{3}'m'c$  &             $6'mm'$    & $\times$ \\ \relax
&*Co$_{2}$Mo$_{3}$O$_{8}$   &   $186$  &               $P6_{3}mc$  &             $P6_{3}'m'c$  &             $6'mm'$   & $\times$  \\ \relax
&Mn$_{2}$Mo$_{3}$O$_{8}$   &   $186$  &               $P6_{3}mc$  &             $P6_{3}m'c'$  &             $6m'm'$  & $\times$   \\ \relax
&*Co$_{2}$Mo$_{3}$O$_{8}$   &   $186$  &               $P6_{3}mc$  &             $P6_{3}'m'c$  &             $6'mm'$   & $\times$  \\ \relax
&Mn$_{5}$Ge$_{3}$   &   $193$  &             $P6_{3}/mcm$  &           $P6_{3}/mc'm'$  &           $6/mm'm'$  & $\circ$  \\ \relax
&*Mn$_{3}$Sn   &   $194$  &             $P6_{3}/mmc$  &                 $Cmc'm'$  &             $m'm'm$  & $\circ$  \\ \relax
&*Mn$_{3}$As   &   $194$  &             $P6_{3}/mmc$  &                 $Cmc'm'$  &             $m'm'm$  & $\circ$  \\ \relax
&*Mn$_{3}$As   &   $194$  &             $P6_{3}/mmc$  &                 $Cm'cm'$  &             $m'm'm$  & $\circ$  \\ \relax
&*MnPtGa   &   $194$  &             $P6_{3}/mmc$  &                 $Cm'c'm$  &             $m'm'm$  & $\circ$  \\ \relax
&*MnPtGa   &   $194$  &             $P6_{3}/mmc$  &                 $Cm'c'm$  &             $m'm'm$  & $\circ$  \\ \relax
&*Mn$_{3}$Sn   &   $194$  &             $P6_{3}/mmc$  &                 $Cm'cm'$  &             $m'm'm$ & $\circ$   \\ \relax
&*Mn$_{3}$Ge   &   $194$  &             $P6_{3}/mmc$  &                 $Cm'cm'$  &             $m'm'm$  & $\circ$  \\ \relax
&*Mn$_{3}$Ge   &   $194$  &             $P6_{3}/mmc$  &                 $C2'/m'$  &             $2'/m'$  & $\circ$  \\ \relax
&Ba$_{5}$Co$_{5}$ClO$_{13}$   &   $194$  &             $P6_{3}/mmc$  &          $P6_{3}'/m'm'c$  &          $6'/m'mm'$  & $\circ$  \\ \relax
&*Pr$_{3}$Ru$_{4}$Al$_{12}$   &   $194$  &             $P6_{3}/mmc$  &                 $Cm'c'm$  &             $m'm'm$  & $\circ$  \\ \relax
&*Pr$_{3}$Ru$_{4}$Al$_{12}$   &   $194$  &             $P6_{3}/mmc$  &                 $C2'/c'$  &             $2'/m'$  & $\circ$  \\ \relax
&Nd$_{3}$Ru$_{4}$Al$_{12}$   &   $194$  &             $P6_{3}/mmc$  &                 $Cm'c'm$  &             $m'm'm$  & $\circ$  \\ \relax
&Mn$_{2.85}$Ga$_{1.15}$   &   $194$  &             $P6_{3}/mmc$  &          $P6_{3}'/m'm'c$  &          $6'/m'mm'$  & $\circ$  \\ \hline \relax
Cubic&Cu$_{2}$OSeO$_{3}$   &   $198$  &                $P2_{1}3$  &                     $R3$  &                 $3$   & $\times$  \\ \relax
&Na$_{3}$Co(CO$_{3}$)$_{2}$Cl   &   $203$  &              $Fd\bar{3}$  &              $Fd\bar{3}$  &          $m\bar{3}$ & $\circ$   \\ \relax
&MnTe$_{2}$   &   $205$  &              $Pa\bar{3}$  &              $Pa\bar{3}$  &          $m\bar{3}$ & $\circ$   \\ \relax
&NiS$_{2}$   &   $205$  &              $Pa\bar{3}$  &              $Pa\bar{3}$  &          $m\bar{3}$  & $\circ$  \\ \relax
&Tb$_{2}$C$_{3}$   &   $220$  &             $I\bar{4}3d$  &                 $Fd'd2'$  &             $m'm2'$  & $\times$   \\ \relax
&Mn$_{3}$Cu$_{0.5}$Ge$_{0.5}$N   &   $221$  &             $Pm\bar{3}m$  &              $R\bar{3}m$  &          $\bar{3}m$  & $\circ$  \\ \relax
&*Mn$_{3}$NiN   &   $221$  &             $Pm\bar{3}m$  &               $R\bar{3}$  &           $\bar{3}$ & $\circ$  \\ \relax
&*Mn$_{3}$NiN   &   $221$  &             $Pm\bar{3}m$  &               $R\bar{3}$  &           $\bar{3}$   & $\circ$ \\ \relax
&Mn$_{3}$Ir   &   $221$  &             $Pm\bar{3}m$  &             $R\bar{3}m'$  &         $\bar{3}m'$ & $\circ$   \\ \relax
&Mn$_{3}$Pt   &   $221$  &             $Pm\bar{3}m$  &             $R\bar{3}m'$  &         $\bar{3}m'$ & $\circ$   \\ \relax
&Mn$_{3}$GaN   &   $221$  &             $Pm\bar{3}m$  &              $R\bar{3}m$  &          $\bar{3}m$ & $\circ$   \\ \relax
&Mn$_{3}$ZnN   &   $221$  &             $Pm\bar{3}m$  &              $R\bar{3}m$  &          $\bar{3}m$ & $\circ$   \\ \relax
&*Mn$_{3}$AlN   &   $221$  &             $Pm\bar{3}m$  &             $R\bar{3}m'$  &         $\bar{3}m'$  & $\circ$  \\ \relax
&*Mn$_{3}$AlN   &   $221$  &             $Pm\bar{3}m$  &                 $Cmm'm'$  &             $m'm'm$  & $\circ$  \\ \relax
&Mn$_{4}$N   &   $221$  &             $Pm\bar{3}m$  &             $R\bar{3}m'$  &         $\bar{3}m'$ & $\circ$   \\ \relax
&Mn$_{3}$(Co$_{0.61}$Mn$_{0.39}$)N   &   $221$  &             $Pm\bar{3}m$  &               $R\bar{3}$  &           $\bar{3}$  & $\circ$  \\ \relax
&Ho$_{2}$CrSbO$_{7}$   &   $227$  &             $Fd\bar{3}m$  &           $I4_{1}/am'd'$  &           $4/mm'm'$ & $\circ$   \\ \relax
&Bi$_{2}$RuMnO$_{7}$   &   $227$  &             $Fd\bar{3}m$  &                 $Fd'd'd$  &             $m'm'm$   & $\circ$ \\ \relax
&Gd$_{2}$Sn$_{2}$O$_{7}$   &   $227$  &             $Fd\bar{3}m$  &           $I4_{1}'/amd'$  &           $4'/mm'm$ & $\circ$   \\ \relax
&Tb$_{2}$Ti$_{2}$O$_{7}$   &   $227$  &             $Fd\bar{3}m$  &             $R\bar{3}m'$  &         $\bar{3}m'$   & $\circ$ \\ \relax
&Tb$_{2}$Sn$_{2}$O$_{7}$   &   $227$  &             $Fd\bar{3}m$  &           $I4_{1}/am'd'$  &           $4/mm'm'$ & $\circ$   \\ \relax
&Nd$_{2}$Hf$_{2}$O$_{7}$   &   $227$  &             $Fd\bar{3}m$  &            $Fd\bar{3}m'$  &        $m\bar{3}m'$  &  $\circ$   \\ \relax
&Nd$_{2}$Zr$_{2}$O$_{7}$   &   $227$  &             $Fd\bar{3}m$  &            $Fd\bar{3}m'$  &        $m\bar{3}m'$ &  $\circ$    \\ \relax
&*Ho$_{2}$Ru$_{2}$O$_{7}$   &   $227$  &             $Fd\bar{3}m$  &           $I4_{1}/am'd'$  &           $4/mm'm'$ & $\circ$   \\ \relax
&Er$_{2}$Sn$_{2}$O$_{7}$   &   $227$  &             $Fd\bar{3}m$  &           $I4_{1}'/amd'$  &           $4'/mm'm$ &  $\circ$    \\ \relax
&Er$_{2}$Pt$_{2}$O$_{7}$   &   $227$  &             $Fd\bar{3}m$  &           $I4_{1}'/amd'$  &           $4'/mm'm$  &  $\circ$   \\ \relax
&Er$_{2}$Ti$_{2}$O$_{7}$   &   $227$  &             $Fd\bar{3}m$  &           $I4_{1}'/am'd$  &           $4'/mm'm$ & $\circ$   \\ \relax
&Tm$_{2}$Mn$_{2}$O$_{7}$   &   $227$  &             $Fd\bar{3}m$  &           $I4_{1}/am'd'$  &           $4/mm'm'$  & $\circ$  \\ \relax
&Er$_{2}$Ru$_{2}$O$_{7}$   &   $227$  &             $Fd\bar{3}m$  &           $I4_{1}'/am'd$  &           $4'/mm'm$  & $\circ$  \\ \relax
&Yb$_{2}$Sn$_{2}$O$_{7}$   &   $227$  &             $Fd\bar{3}m$  &           $I4_{1}/am'd'$  &           $4/mm'm'$ & $\circ$   \\ \relax
&Yb$_{2}$Ti$_{2}$O$_{7}$   &   $227$  &             $Fd\bar{3}m$  &           $I4_{1}/am'd'$  &           $4/mm'm'$  & $\circ$  \\ \relax
&*Ho$_{2}$Ru$_{2}$O$_{7}$   &   $227$  &             $Fd\bar{3}m$  &           $I4_{1}/am'd'$  &           $4/mm'm'$ &  $\circ$    \\ \relax
&Cd$_{2}$Os$_{2}$O$_{7}$   &   $227$  &             $Fd\bar{3}m$  &            $Fd\bar{3}m'$  &        $m\bar{3}m'$& $\circ$    \\ \relax
&CdYb$_{2}$S$_{4}$   &   $227$  &             $Fd\bar{3}m$  &             $I4_{1}/amd$  &             $4/mmm$   &  $\circ$  \\ \relax
&CdYb$_{2}$Se$_{4}$   &   $227$  &             $Fd\bar{3}m$  &             $I4_{1}/amd$  &             $4/mmm$   &  $\circ$  \\ \relax
&Nd$_{2}$Sn$_{2}$O$_{7}$   &   $227$  &             $Fd\bar{3}m$  &            $Fd\bar{3}m'$  &        $m\bar{3}m'$ &  $\circ$    \\ \relax
&*Nd$_{0.5}$Tb$_{0.5}$Co$_{2}$   &   $227$  &             $Fd\bar{3}m$  &                 $C2'/m'$  &             $2'/m'$   & $\circ$ \\ \relax
&*Nd$_{0.5}$Tb$_{0.5}$Co$_{2}$   &   $227$  &             $Fd\bar{3}m$  &                 $C2'/m'$  &             $2'/m'$   & $\circ$ \\ \relax
&*NdCo$_{2}$   &   $227$  &             $Fd\bar{3}m$  &                 $Imm'a'$  &             $m'm'm$ &  $\circ$   \\ \relax
&*NdCo$_{2}$   &   $227$  &             $Fd\bar{3}m$  &                 $C2'/c'$  &             $2'/m'$  & $\circ$  \\ \relax
&*NdCo$_{2}$   &   $227$  &             $Fd\bar{3}m$  &           $I4_{1}/am'd'$  &           $4/mm'm'$ &  $\circ$  \\ \relax
&TbCo$_{2}$   &   $227$  &             $Fd\bar{3}m$  &             $R\bar{3}m'$  &         $\bar{3}m'$   & $\circ$ \\ \relax
&Dy$_{3}$Al$_{5}$O$_{12}$   &   $230$  &             $Ia\bar{3}d$  &            $Ia\bar{3}d'$  &        $m\bar{3}m'$ &  $\circ$    \\  
\hline \hline
\end{longtable*}

\begin{longtable*}[c]{lllcccc}
\caption{
Antisymmetric spin-splitting materials listed in Ref.~\onlinecite{gallego2016magndata}.
The notations are the same as those in Table~\ref{tab:asym}.
}
\label{tab:asym}
\\
\hline \hline
Crystal systems & Materials & \# & SG & MSG & MPG & $\mathcal{P}$ \\ \hline
Monoclinic& *Tb$_{2}$MnNiO$_{6}$   &    $14$  &               $P2_{1}/c$  &                $P2_{1}'$  &                $2'$  & $\times$    \\ \relax
&SrCo(VO$_{4}$)(OH)   &    $19$  &       $P2_{1}2_{1}2_{1}$  &       $P2_{1}2_{1}2_{1}$  &               $222$   & $\times$  \\ \hline \relax
Orthorhombic&Mn$_{3}$B$_{7}$O$_{13}$I   &    $29$  &               $Pca2_{1}$  &             $Pc'a2_{1}'$  &             $m'm2'$   & $\times$  \\ \relax
&Ni$_{3}$B$_{7}$O$_{13}$Br   &    $29$  &               $Pca2_{1}$  &             $Pc'a2_{1}'$  &             $m'm2'$    & $\times$ \\ \relax
&Ni$_{3}$B$_{7}$O$_{13}$Cl   &    $29$  &               $Pca2_{1}$  &             $Pc'a2_{1}'$  &             $m'm2'$    & $\times$  \\ \relax
&Co$_{3}$B$_{7}$O$_{13}$Br   &    $29$  &               $Pca2_{1}$  &             $Pc'a2_{1}'$  &             $m'm2'$   & $\times$  \\ \relax
&Tm$_{2}$Cu$_{2}$O$_{5}$   &    $33$  &               $Pna2_{1}$  &             $Pn'a'2_{1}$  &             $m'm'2$    & $\times$ \\ \relax
&CaBaCo$_{4}$O$_{7}$   &    $33$  &               $Pbn2_{1}$  &             $Pna'2_{1}'$  &             $m'm2'$   & $\times$  \\ \relax
&DyCrWO$_{6}$  &    $33$  &               $Pna2_{1}$  &                 $P2_{1}$  &                 $2$   & $\times$  \\ \relax
&Er$_{2}$Cu$_{2}$O$_{5}$   &    $33$  &               $Pna2_{1}$  &               $Pna2_{1}$  &               $mm2$  & $\times$   \\ \relax
&Tb$_{3}$Ge$_{5}$   &    $43$  &                   $Fdd2$  &                   $Fdd2$  &               $mm2$   & $\times$  \\ \relax
&DyFeO$_{3}$   &    $62$  &                   $Pnma$  &       $P2_{1}2_{1}2_{1}$  &               $222$   & $\times$  \\ \relax
&TbFeO$_{3}$   &    $62$  &                   $Pbnm$  &     $P2_{1}'2_{1}'2_{1}$  &             $2'2'2$  & $\times$    \\ \relax
&*Cu$_{3}$Mo$_{2}$O$_{9}$   &    $62$  &                   $Pnma$  &     $P2_{1}'2_{1}'2_{1}$  &             $2'2'2$  & $\times$   \\ \relax
&*Cu$_{3}$Mo$_{2}$O$_{9}$   &    $62$  &                   $Pnma$  &             $Pm'c2_{1}'$  &             $m'm2'$   & $\times$  \\ \relax
&FePO$_{4}$   &    $62$  &                   $Pnma$  &       $P2_{1}2_{1}2_{1}$  &               $222$   & $\times$  \\ \relax
&Fe$_{3}$BO$_{5}$   &    $62$  &                   $Pnma$  &             $Pm'c2_{1}'$  &             $m'm2'$  & $\times$   \\ \hline \relax
Tetragonal &U$_{3}$Al$_{2}$Si$_{3}$   &    $79$  &                     $I4$  &                    $C2'$  &                $2'$   & $\times$  \\ \relax
&Nd$_{5}$Si$_{4}$   &    $92$  &           $P4_{1}2_{1}2$  &         $P4_{1}2_{1}'2'$  &             $42'2'$   & $\times$  \\ \relax
&Ho$_{2}$Ge$_{2}$O$_{7}$  &    $92$  &           $P4_{1}2_{1}2$  &           $P4_{1}2_{1}2$  &               $422$  & $\times$   \\ \relax
&KMnFeF$_{6}$   &   $106$  &               $P4_{2}bc$  &                 $Pb'a2'$  &             $m'm2'$   & $\times$  \\ \relax
&FeSb$_{2}$O$_{4}$   &   $135$  &             $P4_{2}/mbc$  &               $Pmc2_{1}$  &               $mm2$   & $\times$  \\  \relax
&FePbBiO$_{4}$   &   $135$  &             $P4_{2}/mbc$  &               $Pmc2_{1}$  &               $mm2$   & $\times$   \\  \hline \relax
Hexagonal &Cu$_{0.82}$Mn$_{1.18}$As   &   $174$  &               $P\bar{6}$  &              $P\bar{6}'$  &          $\bar{6}'$   & $\times$  \\ \relax
&*HoMnO$_{3}$   &   $185$  &               $P6_{3}cm$  &               $P6_{3}cm$  &               $6mm$  & $\times$   \\ \relax
&*HoMnO$_{3}$   &   $185$  &               $P6_{3}cm$  &               $P6_{3}cm$  &               $6mm$  & $\times$   \\ \relax
&*HoMnO$_{3}$   &   $185$  &               $P6_{3}cm$  &             $P6_{3}'c'm$  &             $6'mm'$    & $\times$ \\ \relax
&*HoMnO$_{3}$   &   $185$  &               $P6_{3}cm$  &             $P6_{3}'cm'$  &             $6'mm'$    & $\times$ \\ \relax
&*HoMnO$_{3}$   &   $185$  &               $P6_{3}cm$  &             $P6_{3}c'm'$  &             $6m'm'$    & $\times$   \\ \relax
&*YMnO$_{3}$   &   $185$  &               $P6_{3}cm$  &                $P6_{3}'$  &                $6'$   & $\times$  \\ \relax
&*YMnO$_{3}$   &   $185$  &               $P6_{3}cm$  &               $P6_{3}cm$  &               $6mm$   & $\times$  \\ \relax
&*ScMnO$_{3}$   &   $185$  &               $P6_{3}cm$  &                 $P6_{3}$  &                 $6$  & $\times$   \\ \relax
&*ScMnO$_{3}$   &   $185$  &               $P6_{3}cm$  &             $P6_{3}c'm'$  &             $6m'm'$ & $\times$     \\ \relax
&LuFeO$_{3}$   &   $185$  &               $P6_{3}cm$  &             $P6_{3}c'm'$  &             $6m'm'$   & $\times$  \\ \relax
&YbMnO$_{3}$   &   $185$  &               $P6_{3}cm$  &             $P6_{3}'c'm$  &             $6'mm'$   & $\times$  \\ \relax
&Co$_{6}$(OH)$_{3}$(TeO$_{3}$)$_{4}$(OH)$\sim$$0.9$(H$_{2}$O) &   $186$  &               $P6_{3}mc$  &             $P6_{3}'mc'$  &             $6'mm'$   & $\times$  \\ \relax
&Nd$_{15}$Ge$_{9}$C$_{0.39}$   &   $186$  &               $P6_{3}mc$  &             $P6_{3}m'c'$  &             $6m'm'$   & $\times$  \\ \relax
&TmAgGe   &   $189$  &             $P\bar{6}2m$  &                 $Am'm'2$  &             $m'm'2$   & $\times$  \\ \hline \relax
Cubic&U$_{3}$P$_{4}$   &   $220$  &             $I\bar{4}3d$  &                   $R3c'$  &               $3m'$    & $\times$ \\ \relax
&U$_{3}$As$_{4}$   &   $220$  &             $I\bar{4}3d$  &                   $R3c'$  &               $3m'$   & $\times$  \\  
\hline \hline
\end{longtable*}

\appendix

\section{Expressions of electric multipoles}
\label{appen:Expressions of electric multipoles}

In this appendix, we show the multipole expressions by using the cubic and hexagonal harmonics up to rank $4$ in Table~\ref{tab_Emultipole}. 

\begin{table*}[htb!]
\caption{
The correspondence between electric multipoles and cubic and hexagonal harmonics up to rank $4$. 
The expressions for rank-0-2 harmonics are common. 
$(lm)$ and $(lm)'$ stand for $(-1)^l (O_{lm}+O^*_{lm})/\sqrt{2}$ and $(-1)^l 
(O_{lm}-O^*_{lm})/\sqrt{2}i$, respectively. 
}
\label{tab_Emultipole}
\centering
\begingroup
\renewcommand{\arraystretch}{1.0}
 \begin{tabular}{ccccc}
 \hline \hline
\multicolumn{4}{c}{Cubic harmonics} \\
rank & symbol & Definition & correspondence \\ \hline
$0$  & $Q_0$ & $1$ & $(00)$ \\
\hline
 $1$  & $Q_x$, $Q_y$, $Q_z$ & $x$, $y$, $z$ & $(11)$, $(11)'$, $(10)$ \\ \hline
$2$   & $Q_{u}$, $Q_{v}$ & $\frac{1}{2}(3z^2-r^2)$, $ \frac{\sqrt{3}}{2}(x^2-y^2)$ & $(20)$, $(22)$ \\
     &  $Q_{yz}$, $Q_{zx}$, $Q_{xy}$ & $  \sqrt{3}y z$, $\sqrt{3}z x$, $\sqrt{3}x y$ 
     & $(21)'$, $(21)$, $(22)'$\\
\hline
3   & $Q_{xyz}$ & $\sqrt{15}x y z$ & $(32)'$ \\
  & $Q_{x}^{\alpha}$ & $  \frac{1}{2}x(5x^{2}-3r^{2})$ & $  \frac{1}{2\sqrt{2}}[\sqrt{5}(33)-\sqrt{3}(31)]$\\
  &  $Q_{y}^{\alpha}$ & $  \frac{1}{2}y(5y^{2}-3r^{2})$ & $  -\frac{1}{2\sqrt{2}}[\sqrt{5}(33)'+\sqrt{3}(31)']$\\
     &$Q_{z}^{\alpha}$ & $ \frac{1}{2}z(5z^{2}-3r^{2})$ & $(30)$ \\
 &  $Q_{x}^{\beta}$ & $  \frac{\sqrt{15}}{2}x(y^{2}-z^{2})$ & $ -\frac{1}{2\sqrt{2}}[\sqrt{3}(33)+\sqrt{5}(31)]$\\
 &  $Q_{y}^{\beta}$ & $ \frac{\sqrt{15}}{2}y(z^{2}-x^{2})$ & $ \frac{1}{2\sqrt{2}}[-\sqrt{3}(33)'+\sqrt{5}(31)']$\\
 &  $Q_{z}^{\beta}$ & $ \frac{\sqrt{15}}{2}z(x^{2}-y^{2})$ & $(32)$ \\
\hline
$4$  & $Q_{4}$ & $  \frac{5\sqrt{21}}{12}\left(x^4+y^4+z^4-\frac{3}{5} r^4 \right)$ & $  (4) \equiv \frac{1}{2\sqrt{3}} [\sqrt{5} (44)+\sqrt{7}(40)]$ \\
       & $Q_{4u}$& $ \frac{7\sqrt{15}}{6}\left[z^4-\frac{x^4+y^4}{2}-\frac{3}{7} r^2(3z^2-r^2)\right]$ & $  -\frac{1}{2\sqrt{3}}[\sqrt{7}(44)-\sqrt{5}(40)]$ \\ 
      & $Q_{4v}$ & $ \frac{7\sqrt{5}}{4}\left[x^4-y^4-\frac{6}{7}r^2 (x^2-y^2)\right]$ & $-(42)$ \\
      & $Q^{\alpha}_{4x}$ & $ \frac{\sqrt{35}}{2}y z(y^2-z^2)$ & $ -\frac{1}{2\sqrt{2}}[(43)'+\sqrt{7}(41)']$ \\
       & $Q^{\alpha}_{4y}$ & $ \frac{\sqrt{35}}{2}z x(z^2-x^2)$ & $ -\frac{1}{2\sqrt{2}}[(43)-\sqrt{7}(41)]$\\
       & $Q^{\alpha}_{4z}$ & $ \frac{\sqrt{35}}{2}x y(x^2-y^2)$ & $(44)'$ \\
      & $Q^{\beta}_{4x}$ & $ \frac{\sqrt{5}}{2}yz(7x^2-r^2)$ & $ \frac{1}{2\sqrt{2}}[\sqrt{7}(43)'-(41)']$\\
      & $Q^{\beta}_{4y}$ & $ \frac{\sqrt{5}}{2}zx(7y^2-r^2)$ &$ -\frac{1}{2\sqrt{2}}[\sqrt{7}(43)+(41)]$ \\
      & $Q^{\beta}_{4z}$ & $ \frac{\sqrt{5}}{2}xy(7z^2-r^2)$ & $(42)'$ \\
\hline
\hline
\multicolumn{4}{c}{Hexagonal harmonics} \\
 rank & symbol & Definition & correspondence \\ \hline
3   & $Q_z^\alpha$  & $ \frac{1}{2}z (5z^2-3r^2)$ & $(30)$  \\
  & $Q_{3a}$  & $ \frac{\sqrt{10}}{4}x(x^2-3y^2)$ & $(33)$ \\
  & $Q_{3b}$  & $ \frac{\sqrt{10}}{4}y(3x^2-y^2)$ & $(33)'$ \\
  & $Q_{3u}$, $Q_{3v}$ & $ \frac{\sqrt{6}}{4}x(5z^2-r^2)$, $ \frac{\sqrt{6}}{4}y(5z^2-r^2)$ & $(31)$, $(31)'$\\
  & $Q_z^\beta$, $Q_{xyz}$ & $ \frac{\sqrt{15}}{2}z(x^2-y^2)$, $ \sqrt{15}xyz$ & $(32)$, $(32)'$ \\
\hline
$4$  & $Q_{40}$ & $ \frac{1}{8}(35z^4-30 z^2 r^2 + 3 r^4)$ & $(40)$ \\
       & $Q_{4a}$ & $ \frac{\sqrt{70}}{4}yz (3x^2-y^2)$ & $(43)'$ \\
       & $Q_{4b}$ & $ \frac{\sqrt{70}}{4}zx (x^2-3y^2)$ & $(43)$ \\
      & $Q_{4u}^{\alpha},Q_{4v}^{\alpha}$ & $ \frac{\sqrt{10}}{4} zx (7z^2-3r^2)$, $ \frac{\sqrt{10}}{4} yz (7z^2-3r^2)$ & $(41)$, $(41)'$ \\
      & $Q_{4u}^{\beta1},Q_{4v}^{\beta1}$ & $ \frac{\sqrt{35}}{8}(x^4-6x^2y^2 +y^4)$, $ \frac{\sqrt{35}}{2}xy (x^2-y^2)$ & $(44)$, $(44)'$ \\
      & $Q_{4u}^{\beta2},Q_{4v}^{\beta2}$ & $ \frac{\sqrt{5}}{4} (x^2-y^2)(7z^2-r^2)$, $ \frac{\sqrt{5}}{2} xy (7z^2-r^2)$ & $(42)$, $(42)'$ \\
\hline
\hline
\end{tabular}
\endgroup
\end{table*}

\section{Multipole notations under 11 Laue classes}
\label{appen:Multipole notations under 11 Laue classes}

We show the multipole notations per each Laue class in Tables~\ref{tab_multipoles_table_m3m}--\ref{tab_multipoles_table_1}. 

\begin{table}[htb!]
\caption{
Multipoles under Laue class $m\bar{3}m$. 
The upper and lower columns represent even-parity electric and odd-parity magnetic toroidal multipoles, respectively. 
We omit the numerical coefficients of the basis functions.
}
\label{tab_multipoles_table_m3m}
\centering
\begin{tabular}{ccccccc|ccccccc|ccc|ccc} \hline\hline
$O_{\rm h}$ & $O$ & $T_{\rm d}$ &
MP & basis functions &  
\\ \hline
$A^+_{1g}$ & $A^+_{1}$ & $A^+_{1}$  &
$Q_{0}$ & $1$ & 
\\
$A^+_{2g}$ & $A^+_{2}$ & $A^+_{2}$  &
$Q_{6t}$ & $(k_{y}^{2}-k_{z}^{2})(k_{z}^{2}-k_{x}^{2})(k_{x}^{2}-k_{y}^{2})$  &  
\\
$E^+_{g}$ & $E^+$ & $E^+$  &
$Q_{u}$ & $\frac{1}{\sqrt{3}}(3k_z^2-k^2)$ & 
\\
& & &  
$Q_{v}$ & $k^2_x-k^2_y$ & 
\\
$T^+_{1g}$ & $T^+_{1}$ & $T^+_{1}$  &
$Q_{4x}^{\alpha}$ & $k_y k_z (k_y^2-k_z^2)$   & 
\\
& & &  
$Q_{4y}^{\alpha}$ & $k_z k_x (k_z^2-k_x^2)$ &  
\\
& & &  
$Q_{4z}^{\alpha}$ & $k_x k_y (k_x^2-k_y^2)$ &  
\\
$T^+_{2g}$ & $T^+_{2}$ & $T^+_{2}$  &
$Q_{yz}$ &  $k_y k_z$ & 
\\
& & &  
$Q_{zx}$ & $k_z k_x$ &  
\\
& & &  
$Q_{xy}$ & $k_x k_y$ &  
\\ \hline
$A^-_{1u}$ & $A^-_{1}$ & $A^-_{2}$  &
 $T_{9u}$ & $k_x k_y k_z (k_x^2-k_y^2)(k_y^2-k_z^2)(k_z^2-k_x^2) $  & 
\\
$A^-_{2u}$ & $A^-_{2}$ & $A^-_{1}$  &
$T_{xyz}$ & $k_x k_y k_z$ &  
\\
$E^-_{u}$ & $E^-$ & $E^-$  &
$T_{5u}$ & $\sqrt{3} k_x k_y k_z (k_x^2-k_y^2)$ &  
\\
& & &  
$T_{5v}$ & $-k_x k_y k_z (3k_z^2-k^2)$ & 
\\
$T^-_{1u}$ & $T^-_{1}$ & $T^-_{2}$  &
$T_{x}$ & $k_x$ &  
\\
& & &  
$T_{y}$ & $k_y$  &  
\\
& & &  
$T_{z}$ & $k_z$ &  
\\
$T^-_{2u}$ & $T^-_{2}$ & $T^-_{1}$  &
$T_{x}^{\beta}$ & $k_x (k_y^2-k_z^2)$  &  
\\
& & &  
$T_{y}^{\beta}$ & $k_y (k_z^2-k_x^2)$ &  
\\
& & &  
$T_{z}^{\beta}$ & $k_z (k_x^2-k_y^2)$ &  
\\
\hline\hline
\end{tabular}
\end{table}

\begin{table}[t!]
\caption{
Multipoles under Laue class $m\bar{3}$. 
}
\label{tab_multipoles_table_m3}
\centering
\begin{tabular}{ccccccc|ccccccc|ccc|ccc} \hline\hline
$T_{\rm h}$ & $T$ &
MP & basis functions &  
\\ \hline
$A^+_{g}$ & $A^+$ &
$Q_{0}$ & $1$ & 
 \\
$E^+_{g}$ & $E^+$ &
$Q_{u}-i Q_{v}$ & $\frac{1}{\sqrt{3}}(3k_z^2-k^2)-i (k^2_x-k^2_y) $ & 
 \\
& &
$Q_{u}+i Q_{v}$ & $\frac{1}{\sqrt{3}}(3k_z^2-k^2)+i (k^2_x-k^2_y) $ & 
 \\
$T^+_{g}$ & $T^+$ &
$Q_{yz}$ &  $k_y k_z$ & 
 \\
& &
$Q_{zx}$ & $k_z k_x$ &  
 \\
& &
$Q_{xy}$ & $k_x k_y$ &  
 \\ \hline
$A^-_{u}$ & $A^-$ &
$T_{xyz}$ & $k_x k_y k_z$ &  
 \\
$E^-_{u}$ & $E^-$ &
$T_{5u}-i T_{5v}$ & $\sqrt{3} k_x k_y k_z (k_x^2-k_y^2)+ i k_x k_y k_z (3k_z^2-k^2)$ &  
 \\
& &
$T_{5u}+i T_{5v}$ & $\sqrt{3} k_x k_y k_z (k_x^2-k_y^2)- i k_x k_y k_z (3k_z^2-k^2)$ & 
 \\
$T^-_{u}$ & $T^-$ &
$T_{x}$ & $k_x$ &  
 \\
& &
$T_{y}$ & $k_y$  &  
 \\
& &
$T_{z}$ & $k_z$ &  
 \\
\hline\hline
\end{tabular}
\end{table}

\begin{table}[t!]
\caption{
Multipoles under Laue class $4/mmm$. 
We take the $x$ ($[110]$) axis as the $C_{2}'$ rotation axis for $D_{\rm 2d}$ ($D_{\rm 2d}'$). 
}
\label{tab_multipoles_table_4mmm}
\centering
\begin{tabular}{ccccccccc|ccc|ccc} \hline\hline
$D_{\rm 4h}$ & $D_{4}$ & $D_{\rm 2d}$ & $D'_{\rm 2d}$ & $C_{\rm 4v}$ &
MP & basis functions &  
\\ \hline
$A^+_{1g}$ & $A^+_{1}$ & $A^+_{1}$ & $A^+_{1}$ & $A^+_{1}$ &
$Q_{0}$ & $1$ & 
 \\
$A^+_{2g}$ & $A^+_{2}$ & $A^+_{2}$ & $A^+_{2}$ & $A^+_{2}$ &
$Q_{4z}^{\alpha}$ & $k_x k_y (k_x^2-k_y^2)$ &  
 \\
$B^+_{1g}$ & $B^+_{1}$ & $B^+_{1}$ & $B^+_{2}$ & $B^+_{1}$  &
$Q_{v}$ & $k^2_x-k^2_y$ & 
\\
$B^+_{2g}$ & $B^+_{2}$ & $B^+_{2}$ & $B^+_{1}$ & $B^+_{2}$  &
$Q_{xy}$ & $k_x k_y$ &  
\\ 
$E^+_{g}$ & $E^+$ & $E^+$ & $E^+$ & $E^+$  &
$Q_{yz}$ &  $k_y k_z$ & 
\\
& & & &  &
$Q_{zx}$ & $k_z k_x$ &  
\\
\hline
$A^-_{1u}$ & $A^-_{1}$ & $B^-_{1}$ & $B^-_{1}$ & $A^-_{2}$ &
$T_{5u}$ & $k_x k_y k_z (k_x^2-k_y^2)$ &
 \\
$A^-_{2u}$ & $A^-_{2}$ & $B^-_{2}$ & $B^-_{2}$ & $A^-_{1}$ &
$T_{z}$ & $k_z$ &  
 \\
$B^-_{1u}$ & $B^-_{1}$ & $A^-_{1}$ & $A^-_{2}$ & $B^-_{2}$ &
$T_{xyz}$ & $k_x k_y k_z$ &  
 \\
$B^-_{2u}$ & $B^-_{2}$ & $A^-_{2}$ & $A^-_{1}$ & $B^-_{1}$ &
$T_{z}^{\beta}$ & $k_z (k_x^2-k_y^2)$ &  
 \\
$E^-_{u}$ & $E^-$ & $E^-$ & $E^-$ & $E^-$  &
$T_{x}$ & $k_x$ &  
\\
& & & & &
$T_{y}$ & $k_y$  &  
 \\
\hline\hline
\end{tabular}
\end{table}

\begin{table}[t!]
\caption{
Multipoles under Laue class $4/m$. 
}
\label{tab_multipoles_table_4m}
\centering
\begin{tabular}{ccccccccc|ccc|ccc} \hline\hline
$C_{\rm 4h}$ & $C_{4}$ & $S_{4}$ &
MP & basis functions &  
\\ \hline
$A^+_{g}$ & $A^+$ & $A^+$ &
$Q_{0}$ & $1$ & 
 \\
\multirow{2}{*}{$B^+_{g}$}
 & \multirow{2}{*}{$B^+$} & \multirow{2}{*}{$B^+$} &
$Q_{v}$ & $k^2_x-k^2_y$ & 
 \\
 & &  &
$Q_{xy}$ & $k_x k_y$ &  
\\ 
$E^+_{g}$ & $E^+$ & $E^+$ &
$Q_{yz}-i Q_{zx}$ & $k_y k_z-i k_z k_x$ &  
 \\
& & &
$Q_{yz}+i Q_{zx}$ &  $k_y k_z+i k_z k_x$ & 
 \\
\hline
$A^-_{u}$ & $A^-$ & $B^-$ &
$T_{z}$ & $k_z$ &  
 \\
\multirow{2}{*}{$B^-_{u}$}
 & \multirow{2}{*}{$B^-$} & \multirow{2}{*}{$A^-$} &
$T_{xyz}$ & $k_x k_y k_z$ &  
 \\
 & &  &
$T_{z}^{\beta}$ & $k_z (k_x^2-k_y^2)$ &  
 \\
$E^-_{u}$ & $E^-$ & $E^-$ &
$T_{x}+i T_{y}$ & $k_x+i k_y$ &  
 \\
& & &
$T_{x}-i T_{y}$ & $k_x-i k_y$  &  
 \\
\hline\hline
\end{tabular}
\end{table}

\begin{table}[t!]
\caption{
Multipoles under Laue class $mmm$. 
}
\label{tab_multipoles_table_mmm}
\centering
\begin{tabular}{ccccccccc|ccc|ccc} \hline\hline
$D_{\rm 2h}$ & $D_{2}$ & $C_{\rm 2v}$ &
MP & basis functions &  
\\ \hline
$A^+_{g}$ & $A^+$ & $A^+_{1}$ &
$Q_{0}$ & $1$ & 
 \\
$B^+_{1g}$ & $B^+_{1}$ & $A^+_{2}$ &
$Q_{xy}$ & $k_x k_y$ &  
 \\ 
$B^+_{2g}$ & $B^+_{2}$ & $B^+_{1}$ &
$Q_{zx}$ & $k_z k_x$ &  
 \\
$B^+_{3g}$ & $B^+_{3}$ & $B^+_{2}$ &
$Q_{yz}$ &  $k_y k_z$ & 
 \\
\hline
$A^-_{u}$ & $A^-$ & $A^-_{2}$ &
$T_{xyz}$ & $k_x k_y k_z$ &  
 \\
$B^-_{1u}$ & $B^-_{1}$ & $A^-_{1}$ &
$T_{z}$ & $k_z$ &  
 \\
$B^-_{2u}$ & $B^-_{2}$ & $B^-_{2}$ &
$T_{y}$ & $k_y$  &  
 \\
$B^-_{3u}$ & $B^-_{3}$ & $B^-_{1}$ &
$T_{x}$ & $k_x$ &  
 \\
\hline\hline
\end{tabular}
\end{table}

\begin{table}[t!]
\caption{
Multipoles under Laue class $2/m$. 
}
\label{tab_multipoles_table_2m}
\centering
\begin{tabular}{ccccccccc|ccc|ccc} \hline\hline
$C_{\rm 2h}$ & $C_{\rm 2}$ & $C_{\rm s}$&
MP & basis functions &  
\\ \hline
$A^+_{g}$ & $A^+$ & $A'^+$ &
$Q_{0}$ & $1$ & 
\\
\multirow{1}{*}{$B^+_{g}$} & \multirow{1}{*}{$B^+$} & \multirow{1}{*}{$A''^+$} &
$Q_{zx}$ & $k_z k_x$ &  
\\
\multirow{1}{*}{$B^+_{g}$} & \multirow{1}{*}{$B^+$} & \multirow{1}{*}{$A''^+$} &
$Q_{yz}$ &  $k_y k_z$ & 
\\
\hline
$A^-_{u}$ & $A^-$ & $A''^-$ &
$T_{z}$ & $k_z$ &  
\\
\multirow{1}{*}{$B^-_{u}$} & \multirow{1}{*}{$B^-$} & \multirow{1}{*}{$A'^-$} &
$T_{y}$ & $k_y$  &  
\\
\multirow{1}{*}{$B^-_{u}$} & \multirow{1}{*}{$B^-$} & \multirow{1}{*}{$A'^-$} &
$T_{x}$ & $k_x$ &  
\\
\hline\hline
\end{tabular}
\end{table}

\begin{table}
\caption{
Multipoles under Laue class $6/mmm$. 
For $D_{\rm 6h}$, we take the $y$ and $x$ axes as the $C_{2}'$ and $C_{2}''$ rotation axes, respectively~\cite{comment_D6h, aroyo2006bilbao1, aroyo2006bilbao2}. 
We take the $x$ ($y$) axis as the $C_{2}'$ rotation axis for $D_{\rm 3h}'$ ($D_{\rm 3h}$). 
The sign and coefficient in two dimensional irrep. are chosen so as to satisfy the mutual relationship between two components.
}
\label{tab_multipoles_table_6mmm}
\centering
\begin{tabular}{ccccccccc|ccccc} \hline\hline
$D_{\rm 6h}$ & $D_{6}$ & $C_{\rm 6v}$ & $D_{\rm 3h}$ & $D_{\rm 3h}'$
 & 
MP & basis functions &
\\ \hline
$A^+_{1g}$ & $A^+_{1}$ & $A^+_{1}$ & $A'^+_{1}$ & $A'^+_{1}$ & 
$Q_{0}$ &  $1$  &
\\
$A^+_{2g}$ & $A^+_{2}$ & $A^+_{2}$ & $A'^+_{2}$ & $A'^+_{2}$ & 
$Q_{6s}$ & $k_x k_y (3k_x^2-k_y^2)(k_x^2-3k_y^2) $ &
\\
$B^+_{1g}$ & $B^+_{1}$ & $B^+_{2}$ & $A''^+_{1}$ & $A''^+_{2}$ & 
$Q_{4b}$ & $k_z k_x (k_x^2-3k_y^2)$ &
\\
$B^+_{2g}$ & $B^+_{2}$ & $B^+_{1}$ & $A''^+_{2}$ & $A''^+_{1}$ & 
$Q_{4a}$ & $k_y k_z (3k_x^2-k_y^2)$  &
\\
$E^+_{1g}$ & $E^+_{1}$ & $E^+_{1}$ & $E''^+$ & $E''^+$ &
$Q_{zx}$ & $k_z k_x$  &
\\
& & & & &
$Q_{yz}$ & $k_y k_z$  &
\\
$E^+_{2g}$ & $E^+_{2}$  & $E^+_{2}$ & $E'^+$ & $E'^+$ & 
$Q_{v}$ & $\frac{1}{2} (k_x^2-k_y^2)$  &
\\
& & & & &
$Q_{xy}$ & 
$-k_{x}k_{y}$
\\ \hline
$A^-_{1u}$ & $A^-_{1}$ & $A^-_{2}$  & $A''^-_{1}$  & $A''^-_{1}$ & 
$T_{7u}$ & $k_x k_y k_z (3k_x^2-k_y^2)(k_x^2-3k_y^2)$ &
\\
$A^-_{2u}$ & $A^-_{2}$ & $A^-_{1}$  & $A''^-_{2}$  & $A''^-_{2}$ & 
$T_{z}$ & $k_z$ &
\\
$B^-_{1u}$ & $B^-_{1}$ & $B^-_{1}$  & $A'^-_{1}$  & $A'^-_{2}$ & 
$T_{3b}$ & $k_y (3k_x^2-k_y^2)$  &
\\
$B^-_{2u}$ & $B^-_{2}$ & $B^-_{2}$  & $A'^-_{2}$  & $A'^-_{1}$ & 
$T_{3a}$ & $k_x (k_x^2-3k_y^2)$  &
\\
$E^-_{1u}$ & $E^-_{1}$ & $E^-_{1}$  & $E'^-$ & $E'^-$ & 
$T_{x}$ & $k_x$  &
\\
& & & & &
$T_{y}$ &  $k_y$ &
\\
$E^-_{2u}$ & $E^-_{2}$  & $E^-_{2}$  & $E''^-$ & $E''^-$ & 
$T_{z}^{\beta}$  &  
$\frac{1}{2}k_{z}(k_{x}^{2}-k_{y}^{2})$
\\
& & & & &
$T_{xyz}$ & 
$-k_{x}k_{y}k_{z}$
\\
\hline\hline
\end{tabular}
\end{table}

\begin{table}
\caption{
Multipoles under Laue class $6/m$. 
}
\label{tab_multipoles_table_6m}
\centering
\begin{tabular}{cccccccccccccc} \hline\hline
$C_{\rm 6h}$ & $C_{6}$ & $C_{\rm 3h}$ &  
MP & basis functions &\\ \hline
$A^+_{g}$ & $A^+$ & $A'^+$ &
$Q_{0}$ &  $1$  &
\\
\multirow{2}{*}{$B^+_{g}$} & \multirow{2}{*}{$B^+$} & \multirow{2}{*}{$A''^+$} &
$Q_{4a}$ & $k_y k_z (3k_x^2-k_y^2)$  &
\\
 &  &  &
$Q_{4b}$ & $k_z k_x (k_x^2-3k_y^2)$ &
\\
$E^+_{1g}$ & $E^+_{1}$ & $E''^+$ &
$Q_{zx}+i Q_{yz}$ & $k_z k_x+ i k_y k_z$  &
\\
& & &
$Q_{zx}-i Q_{yz}$ & $k_z k_x - i k_y k_z$  &
\\
$E^+_{2g}$ & $E^+_{2}$ & $E'^+$ &
$Q_{v}+i Q_{xy}$ & $\frac{1}{2} (k_x^2-k_y^2)+i k_x k_y$  &
\\
& & & 
$Q_{v}-i Q_{xy}$ & $\frac{1}{2} (k_x^2-k_y^2)-i k_x k_y$  &
\\ \hline
$A^-_{u}$ & $A^-$  & $A''^-$ &
$T_{z}$ & $k_z$ &
\\
\multirow{2}{*}{$B^-_{u}$} & \multirow{2}{*}{$B^-$}  & \multirow{2}{*}{$A'^-$} &
$T_{3a}$ & $k_x (k_x^2-3k_y^2)$  &
\\
& & &
$T_{3b}$ & $k_y (3k_x^2-k_y^2)$  &
\\
$E^-_{1u}$ & $E^-_{1}$ & $E'^-$ &
$T_{x}+i T_{y}$ & $k_x+i k_y$  &
\\
& & & 
$T_{x}-i T_{y}$ &  $k_x-i k_y$ &
\\
$E^-_{2u}$ & $E^-_{2}$  & $E''^-$ &
$T_{z}^{\beta}+i T_{xyz}$ & $\frac{1}{2}k_z(k_x^2-k_y^2)+i k_x k_y k_z$  &
\\
& & & 
$T_{z}^{\beta}-i T_{xyz}$  &  $\frac{1}{2}k_z(k_x^2-k_y^2)-i k_x k_y k_z$ &
\\
\hline\hline
\end{tabular}
\end{table}

\begin{table}
\caption{
Multipoles under Laue class $\bar{3}m$. 
We take the $x$ ($y$) axis as the $C_{2}'$ rotation axis for $D_{\rm 3d}'$ and $D_{\rm 3}'$ ($D_{\rm 3d}$ and $D_{\rm 3}$). 
For $D'_{3 {\rm d}}$ and $C_{3 {\rm v}}$ ($D_{3 {\rm d}}$ and $C'_{3 {\rm v}}$), we take the $yz$ ($xz$) plane as the $\sigma_v$ or $\sigma_{d}$ mirror plane.
The sign and coefficient in two dimensional irrep. are chosen so as to satisfy the mutual relationship between two components. 
}
\label{tab_multipoles_table_3m}
\centering
\begin{tabular}{cccccccccccccc} \hline\hline
$D_{\rm 3d}$ & $D_{\rm 3d}'$ & $D_{3}$ & $D_{3}'$ & $C_{\rm 3v}$ & $C_{\rm 3v}'$ 
&
MP & basis functions &
\\ \hline
$A^+_{1g}$ & $A^+_{1g}$ & $A^+_{1}$ & $A^+_{1}$ & $A^+_{1}$  & $A^+_{1}$ &
$Q_{0}$ &  $1$  &
\\
$A^+_{1g}$ & $A^+_{2g}$ & $A^+_{1}$ & $A^+_{2}$ & $A^+_{2}$ & $A^+_{1}$ &
$Q_{4b}$ & $k_z k_x (k_x^2-3k_y^2)$ &
\\
$A_{2g}^+$ & $A_{1g}^+$ & $A_{2}^+$ & $A_{1}^+$ & $A_{1}^+$ & $A_{2}^+$ &
$Q_{4a}$ & $k_y k_z (3k_x^2-k_y^2)$  &
\\
$E^+_{g}$ & $E^+_{g}$ & $E^+$ & $E^+$ & $E^+$ & $E^+$ &
$Q_{zx}$ & $k_z k_x$  &
\\
& & & & & &
$Q_{yz}$ & $k_y k_z$  &
\\
$E^+_{g}$ & $E^+_{g}$ & $E^+$ & $E^+$ & $E^+$ & $E^+$ &
$Q_{v}$ & $\frac{1}{2} (k_x^2-k_y^2)$  &
\\
& & & & & &
$Q_{xy}$ & $-k_x k_y$  &
\\ \hline
$A^-_{1u}$ & $A^-_{2u}$ & $A^-_{1}$ & $A^-_{2}$ & $A^-_{1}$ & $A^-_{2}$ &
$T_{3b}$ & $k_y (3k_x^2-k_y^2)$  &
\\
$A^-_{2u}$ & $A^-_{1u}$ & $A^-_{2}$ & $A^-_{1}$ & $A^-_{2}$ & $A^-_{1}$ &
$T_{3a}$ & $k_x (k_x^2-3k_y^2)$  &
\\
$A^-_{2u}$ & $A^-_{2u}$ & $A^-_{2}$ & $A^-_{2}$ & $A^-_{1}$ & $A^-_{1}$ &
$T_{z}$ & $k_z$ &
\\
$E^-_{u}$ & $E^-_{u}$ & $E^-$ & $E^-$ & $E^-$ & $E^-$ &
$T_{x}$ & $k_x$  &
\\
& & & & & &
$T_{y}$ &  $k_y$ &
\\
\hline\hline
\end{tabular}
\end{table}

\begin{table}
\caption{
Multipoles under Laue class $3$. 
$C_{\rm 3i}=S_{6}$. 
}
\label{tab_multipoles_table_3}
\centering
\begin{tabular}{cccccccccccccc} \hline\hline
$C_{\rm 3i}$ & $C_{3}$ &
MP & basis functions &
\\ \hline
$A^+_{g}$ & $A^+$ &
$Q_{0}$ &  $1$  &
\\
\multirow{1}{*}{$E^+_{g}$} & \multirow{1}{*}{$E^+$} &
$Q_{zx}+i Q_{yz}$ & $k_z k_x+i k_y k_z$  &
\\
& &
$Q_{zx}-i Q_{yz}$ & $k_z k_x-i k_y k_z$  &
\\
\multirow{1}{*}{$E^+_{g}$} & \multirow{1}{*}{$E^+$} &
$Q_{v}- i Q_{xy}$ & $\frac{1}{2} (k_x^2-k_y^2)- i k_x k_y$  &
\\
& &
$Q_{v}+ i Q_{xy}$ & $\frac{1}{2} (k_x^2-k_y^2)+ i k_x k_y$  &
\\ \hline
$A^-_{u}$ & $A^-$ &
$T_{z}$ & $k_z$ &
\\
$E^-_{u}$ & $E^-$ &
$T_{x}+ i T_y$ & $k_x+i k_y$  &
\\
 & &
$T_{x}- i T_y$ &  $k_x-i k_y$ &
\\
\hline\hline
\end{tabular}
\end{table}

\begin{table}
\caption{
Multipoles under Laue class $\bar{1}$. 
}
\label{tab_multipoles_table_1}
\centering
\begin{tabular}{cccccccccccccc} \hline\hline
$C_{\rm i}$ & $C$ &
MP & basis functions &
\\ \hline
$A^+_{g}$ & $A^+$ &
$Q_{0}$ &  $1$  &
\\ \hline
$A^-_{u}$  & $A^-$ &
$T_{x}$ & $k_x$ & \\
$A^-_{u}$  & $A^-$ &
$T_{y}$ & $k_y$ & \\
$A^-_{u}$  & $A^-$ &
$T_{z}$ & $k_z$ &
\\
\hline\hline
\end{tabular}
\end{table}

\begin{acknowledgments}
This research was supported by JSPS KAKENHI Grants Numbers JP15H05885, JP18H04296 (J-Physics), JP18K13488, JP19K03752, JP19H01834, and JP20K05299. 
\end{acknowledgments}

\bibliographystyle{apsrev}
\bibliography{ref}

\begin{thebibliography}{59}
\bibitem[{com({\natexlab{a}})}]{comment_Erep}
\bibinfo{note}{It is noted that the basis function belonging to the $E'^+$
  representation is expressed as $(Q_v, -Q_{xy})$ so as to satisfy the mutual
  relationship between two components, although we adopt $(Q_v, Q_{xy})$
  following in Ref.~\onlinecite{Hayami_PhysRevB.101.220403}.}
\bibitem[{com({\natexlab{b}})}]{comment_D6h}
\bibinfo{note}{When we take the $x$ and $y$ axes as the $C_{2}'$ and $C_{2}''$
  rotation axes, respectively, the irreps. $(B_{1g}, B_{2g}, B_{1u}, B_{2u})$
  are replaced with $(B_{2g}, B_{1g}, B_{2u}, B_{1u})$}.
\end{thebibliography}

\end{document}